\documentclass[11pt]{article}
\usepackage{amsmath,amssymb,amsfonts}
\usepackage[latin1]{inputenc}

\usepackage{amsfonts}
\usepackage{color}
\usepackage{epsfig}
 \usepackage{dsfont}
\usepackage{amsmath}
\usepackage{hyperref}

\usepackage{graphicx} 

\usepackage{booktabs} 
\usepackage{array} 
\usepackage{paralist} 
\usepackage{verbatim} 
\usepackage{subfig} 
\usepackage{url}
\usepackage{hyperref}
\usepackage{float}

\usepackage{epsf}
\usepackage{psfrag} 
\usepackage{epsfig,float,psfrag}
\hypersetup{pdfborder={0 0 0},colorlinks,urlcolor=blue}  

\numberwithin{equation}{section}
\newtheorem{thm}{Theorem}[section]

\newtheorem{alem}[thm]{Lemma}
\newtheorem{aprop}[thm]{Proposition}

\newtheorem{arem}[thm]{Remark}
\newenvironment{adem}[1][]%
   {\ \\ {\bf Proof #1~: }}%
   {\hfill\mbox{\rule{2 true mm}{3 true mm}}\vskip 2 ex\noindent}
   {\ \\ {\bf Example #1~: }}%
   {\hfill\mbox{\rule{2 true mm}{3 true mm}}\vskip 2 ex\noindent}

\title{Ninomiya-Victoir scheme: strong convergence, antithetic version and application to multilevel estimators }
\author{A. Al Gerbi, B. Jourdain\thanks{Universit\'e Paris-Est, Cermics (ENPC), INRIA, F-77455, Marne-la-Vall\'ee, France
    e-mails: jourdain@cermics.enpc.fr, anis.al-gerbi@cermics.enpc.fr - This research benefited
    from the support of the ``Chaire Risques Financiers'', Fondation du
    Risque.}~ and E. Cl\'ement\thanks{Universit\'e Paris-Est, LAMA (UMR 8050), UPEMLV, UPEC, CNRS, F-77454, Marne-la-Vall\'ee, France,
    e-mail: emmanuelle.clement@u-pem.fr.}}

\begin{document}
\maketitle
In this paper, we are interested in the strong convergence properties of the Ninomiya-Victoir scheme which is known to exhibit weak convergence with order 2. We prove strong convergence with order $1/2$. This study is aimed at analysing the use of this scheme either at each level or only at the finest level of a multilevel Monte Carlo estimator: indeed, the variance of a multilevel Monte Carlo estimator is related to the strong error between the two schemes used on the coarse and fine grids at each level.  
Recently, Giles and Szpruch proposed in \cite{GS} a scheme permitting to construct a multilevel Monte Carlo estimator achieving the optimal complexity $O\left(\epsilon^{-2}\right)$  for the precision $\epsilon$. 
In the same spirit, we propose a modified Ninomiya-Victoir scheme, which may be strongly coupled with order $1$ to the Giles-Szpruch scheme at the finest level of a multilevel Monte Carlo estimator. Numerical experiments show that this choice improves the efficiency, since the order $2$ of weak convergence of the Ninomiya-Victoir scheme permits to reduce the number of discretization levels.
\section{Introduction}
This paper is dedicated to the computation of $Y=\mathbb{E}\left[f\left(X_T\right) \right]$, where $f : \mathbb{R}^n \longrightarrow \mathbb{R}$ is a payoff function and $X_T$ is the solution, at time $T \in \mathbb{R}_+^*$, to a multi-dimensional stochastic differential equation of the form
\begin{equation}
\left\{
    \begin{array}{ll}
dX_t = b(X_t) dt + \sum \limits_{j=1}^d \sigma^j(X_t)dW_t^j, ~	 t \in [0,T]\\
X_0 = x.
\end{array}
\right.
\label{EDS_ITO}
\end{equation}
Here, $x \in \mathbb{R}^n$ is the initial condition, $W = \left(W^1,\ldots,W^d \right)$ is a $d-$dimensional standard Brownian motion, $b: \mathbb{R}^n \longrightarrow \mathbb{R}^n$ is the drift coefficient and $\sigma^j: \mathbb{R}^n \longrightarrow \mathbb{R}^n, j \in \left\{1,\ldots,d\right\}$, are the diffusion coefficients.\\

The standard Monte Carlo method consists in estimating $\mathbb{E}\left[f \left(X_T \right)\right]$ by discretizing the stochastic differential equation with $N \in \mathbb{N}^*$ steps and approximating the expectation using $M \in \mathbb{N}^*$ independent path simulations. To be clear, the crude Monte Carlo estimator is given by
\begin{equation*}
\hat{Y}_{CMC} = \frac{1}{M} \sum \limits_{k=1}^M  f \left( X^{N,k}_T\right) 
\end{equation*}
where $X^{N,k}$ are independent copies of a numerical scheme $X^{N}$ with time step $T/N$. Under some regularity assumptions on the coefficients of the SDE and for a smooth payoff, it is well known that to ensure a root mean-square-error $\epsilon$, the computational cost of this method is  $O\left(\epsilon^{-\left(2+\frac1\alpha\right)}\right)$, where $\alpha$ is the order of weak  convergence of the numerical scheme (see theorem 1 in \cite{DG}).
In \cite{NV}, Ninomiya and Victoir proposed a numerical scheme, achieving $\alpha = 2$, which reduces the computational complexity compared to the Euler scheme, for which $\alpha = 1$.
In the optimal complexity $O\left(\epsilon^{-\left(2+\frac1\alpha\right)}\right)$, the term $1/\alpha$ is due to the bias $\mathbb{E}\left[f \left(X_T \right)\right] - \mathbb{E}\left[f \left(X^N_T \right)\right]$.\\

To remove this term, Giles introduced in \cite{Giles} a multilevel Monte Carlo estimator permitting telescopic cancellation of the bias. The multilevel Monte Carlo estimator is built as follows
\begin{equation*}
\hat{Y} = \frac{1}{M_0} \sum \limits_{k=1}^{M_0}  f \left( X^{1,0,k}_T\right)+ \sum \limits_ {l=1}^L \frac{1}{M_l} \sum \limits_{k=1}^{M_l}  \left( f \left( X^{2^l,l,k}_T\right) - f\left(X^{2^{l-1},l,k}_T\right) \right)
\end{equation*}
where $L \in \mathbb{N}^*$ is the last and finest level of discretization with time-step $T/2^L$, $ \left(M_l\right)_{0\leq l \leq L} \in \left(\mathbb{N}^*\right)^{L+1}$ is the vector of sample sizes at each level. Moreover, for all $l \in \left\{1,\ldots, L\right\}$, the two numerical schemes $X^{2^l,l}_T$ and $X^{2^{l-1},l}_T$ are simulated with the same Brownian motion. For each discretization level $l \in \left\{0,\ldots,L\right\}$, $M_l$ independent and identically distributed path simulations independent from the other levels are used. The optimal complexity of this method is driven by the order $\beta$ of convergence to zero of the variance $\mathbb{V}\left(f \left( X^{2^l,l}_T\right) - f\left(X^{2^{l-1},l}_T\right)\right)$, which is related to the strong convergence order $\gamma$ of the scheme. For a Lipschitz payoff $f$, using the strong convergence properties of the scheme in the estimation of the variance, one gets $\beta \ge 2 \gamma$. For $\beta >  1$, the optimal complexity is $O\left(\epsilon^{-2} \right)$. This complexity is the same as in a simple Monte Carlo method with independent and identically distributed unbiased random variables. The condition $\beta>1$ is satisfied by the Milstein scheme for which $\gamma = 1$. Unfortunately, to simulate the Milstein scheme, one needs, in general, to simulate L\'evy areas for which there is no known efficient method when the dimension of the Brownian motion $d$ is larger than 2. Unless the diffusion coefficients $\sigma^j, j \in \left\{1,\ldots,d\right\}$, are constant, the strong order of the Euler scheme is $\gamma = 1/2$, which leads to $\beta = 1$ and to the optimal complexity $O\left(\epsilon^{-2} \left(\log\left( \frac{1}{\epsilon} \right) \right)^2\right)$. \\

Recently, two approaches have been developed to improve the case $\gamma = 1/2$. In \cite{GS}, Giles and Szpruch introduced a modified Milstein scheme, with the L\'evy areas set to zero, and its antithetic version based on the swapping of each successive pair of Brownian increments in the scheme. Regarding the multilevel Monte Carlo estimator, at each discretization level $l \in \left\{1,\ldots,L\right\}$, on the finest grid, instead of using a simple scheme, Giles and Szpruch employed the arithmetic average of the scheme and its antithetic version as follows   
\begin{equation*}
\hat{Y} = \frac{1}{M_0} \sum \limits_{k=1}^{M_0}  f \left( X^{1,0,k}_T\right)+ \sum \limits_ {l=1}^L \frac{1}{M_l} \sum \limits_{k=1}^{M_l} \left( \frac12 \left( f \left( \tilde{X}^{2^l,l,k}_T\right) + f \left( X^{2^l,l,k}_T\right) \right) - f\left(X^{2^{l-1},l,k}_T\right) \right)
\end{equation*}
where $\tilde{X}^{2^l}$ denotes the antithetic version of the Giles-Szpruch scheme with time-step $T/2^l$. Under some regularity assumptions on the coefficients of the SDE and for a smooth payoff, Giles and Szpruch showed that  despites $\gamma$ is equal to $1/2$,  $\beta$ is equal to $ 2$ which leads to an optimal complexity $O\left(\epsilon^{-2} \right)$. The second approach called multilevel Richardson-Romberg method and investigated by Lemaire and Pag\`es in \cite{PL}, fully takes advantage of the existence of a weak error expansion while keeping the multilevel Monte Carlo estimator properties.  The multilevel Richardson-Romberg estimator is a weighted version of the multilevel Monte Carlo method which integrates the multi-step Richardson-Romberg extrapolation developed by Pag\`es in \cite{Pages}. Lemaire and Pag\`es obtained an optimal complexity $O\left(\epsilon^{-2} \log\left( \frac{1}{\epsilon} \right)\right)$ when $\beta = 1$ which improves the standard multilevel Monte Carlo method. When $\beta > 1$, the optimal complexity $O\left(\epsilon^{-2} \right)$ is preserved.\\

In this paper, we propose to use the Ninomiya-Victoir scheme, which is known to exhibit weak convergence with order 2, on the finest grid at the last level $L$ of a multilevel Monte Carlo estimator. This idea is inspired by Debrabant and R\"ossler \cite{DR} who suggest to use a scheme with high order of weak convergence on the finest grid at the finest level $L$ of the multilevel Monte Carlo method. By this way, Debrabant and R\"ossler reduce the constant in the computational complexity by decreasing the number of discretization levels.

In section 2, to derive the strong convergence order of the Ninomiya-Victoir scheme, we provide a suitable interpolation between time grid points. Then we prove strong convergence with order $\gamma = 1/2$ under some regularity assumptions on the coefficients of the SDE. In section 3, we propose a modified Ninomiya-Victoir scheme, which may be strongly coupled with order $1$ to the Giles-Szpruch scheme. This result allows us to derive an antithetic version of the Ninomiya-Victoir scheme and combine the ideas of Giles-Szpruch and Debrabant-R\"ossler by building the multilevel Monte Carlo estimator with the Giles-Szpruch scheme from level 0 to level $L-1$ and the coupling between the Ninomiya-Victoir scheme and the Giles-Szpruch scheme at the last level $L$.  The efficiency of this estimator is confirmed in section 4, where we present and comment, in details, numerical experiments carried out on the Clark-Cameron SDE and Heston SDE as in \cite{GS}.

\section{Strong convergence of the Ninomiya-Victoir scheme}
We begin this section by introducing some notations which will be used throughout this paper. To discretize \eqref{EDS_ITO} we consider a uniform grid with time step $h = T/N$ where $N \in \mathbb{N}^*$ and we denote:
\begin{itemize}
\item $\left(t_k\right)_{k \in [\![0;N]\!]} = k h$ the subdivision of $[0,T]$ with equal time step h,
\item $\hat{\tau}_s$ the last time discretization before $s \in [0,T] $, ie $\hat{\tau}_s = t_k$ if $s \in \left(t_k,t_{k+1}\right]$, and for $s = t_0 = 0 $, we set $\hat{\tau}_0 = t_{0} = 0$,
\item $\check{\tau}_s$ the first time discretization after $s \in [0,T]$, ie $\check{\tau}_s = t_{k+1}$ if $s \in \left(t_k,t_{k+1}\right]$, and for $s = t_0 = 0 $, we set $\check{\tau}_0 = t_0 = 0$,
\item $\forall j \in \left\{1,\ldots,d\right\}, \forall s \in [0,T],$ such that $ t_k < s \leq t_{k+1}$, $\Delta W^j_s = W^j_s - W^j_{t_k}$,
\item $\forall s \in [0,T],$ such that $ t_k < s \leq t_{k+1}$, $\Delta s = s - t_k$,
\item $\eta = \left(\eta_1,\ldots,\eta_{N}\right)$ a sequence of independent, identically distributed Rademacher random variables independent of $W$,
\item By a slight abuse of notation, we set $\eta_s = \eta_{k+1}$ if $s \in (t_k,t_{k+1}]$,
\item $\forall x \in \mathbb{R}_+$, $\left \lceil x \right \rceil$ denotes the unique $n \in \mathbb{N}^*$ satisfying  $n - 1 < x \leq n$, 
\item $\forall x \in \mathbb{R}_+$, $\left \lfloor x \right\rfloor$ denotes the unique $n \in \mathbb{N}^*$ satisfying  $n \leq x < n+ 1$. 
\end{itemize}
Let $V: \mathbb{R}^n \longrightarrow \mathbb{R}^n$ Lipschitz continuous and consider the ordinary differential equation in $\mathbb{R}^n$:
\begin{equation}
\left\{
    \begin{array}{ll}
 \frac{dx(t)}{dt}  = V\left(x(t)\right) \\
  x(0) = x_0 . 
\end{array}
\right.
\label{ODE}
\end{equation}
The solution of \eqref{ODE}  at time $t$, $t \in \mathbb{R}$ is denoted by
\begin{equation*}
 x(t) = \exp(tV)x_0,
\end{equation*}
and the integral form of \eqref{ODE} is given by
\begin{equation*}
 x(t) = \exp(tV)x_0 = x_0 + \displaystyle\int_0^t V(x(s))ds = x + \displaystyle\int_0^t V\left(\exp(sV)x_0\right)ds.
\label{EDO*}
\end{equation*}
We recall that in \eqref{EDS_ITO}, each coordinate $ i \in \left\{1,\ldots,n\right\}$ evolves according to the following stochastic differential equation
\begin{equation*}
 dX^i_{t} = b^i(X_t) dt + \sum \limits_{j=1}^d \sigma^{ij}(X_t)dW_t^j. 
\end{equation*}
Then, assuming $\mathcal{C}^{1}$ regularity for the diffusion coefficients, one can write \eqref{EDS_ITO} in Stratonovich form
\begin{equation}
\left\{
    \begin{array}{ll}
dX_t = \sigma^0(X_t) dt + \sum \limits_{j=1}^d \sigma^j(X_t)\circ dW_t^j \\
X_0 = x
\end{array}
\right.
\label{EDS_STO}
\end{equation}
where $\sigma^0 = b - \frac{1}{2} \sum \limits_{j=1}^d \partial \sigma^j \sigma^j $ and $\partial \sigma^j$ is the Jacobian matrix of $\sigma^j$ defined as follows
\begin{equation*}
\partial \sigma^j = \left(\left(\partial \sigma^j \right)_{ik}\right)_{i,k \in [\![1;n]\!] } = \left(\partial_{x_k} \sigma^{ij} \right)_{i,k \in [\![1;n]\!] }. 
\end{equation*} 
Now, we present the Ninomiya-Victoir scheme introduced in \cite{NV}.
\begin{itemize}
\item Starting point: $X^{NV,\eta}_{t_0} = x$.
\item For $k \in \left\{0\ldots,N-1\right\}$,
 if $\eta_{k+1} = 1 $:
\begin{equation}
 X^{NV,\eta}_{t_{k+1}} = \exp\left(\frac{h}{2}\sigma^0\right) \exp\left (\Delta W^d_{t_{k+1}}\sigma^d \right) \ldots \exp\left (\Delta W^1_{t_{k+1}}\sigma^1 \right)  \exp\left(\frac{h}{2}\sigma^0\right) X^{NV,\eta}_{t_{k}}, 
\label{case 1 NV}
\end{equation}
and if $\eta_{k+1} = -1 $:
\begin{equation}
X^{NV,\eta}_{t_{k+1}} = \exp\left(\frac{h}{2}\sigma^0\right) \exp\left (\Delta W^1_{t_{k+1}}\sigma^1 \right) \ldots \exp\left (\Delta W^d_{t_{k+1}}\sigma^d \right)  \exp\left(\frac{h}{2}\sigma^0\right)  X^{NV,\eta}_{t_{k}}.
\label{case 2 NV}
\end{equation}
\end{itemize}
The Stratonovich form is preferred when we use the Ninomiya-Victoir scheme since the Stratonovich drift appears in the definition of the scheme. Moreover, using It\^o's formula, one has
\begin{equation}
\forall t,s \in \mathbb{R}_+, s \leq t, ~ \exp\left(\left(W^j_t-W^j_s\right)V\right)y = y + \displaystyle\int_s^t V\left(\exp \left(\left(W^j_u - W^j_s\right)V\right)y\right)\circ dW^j_u.
\label{ITO-EDO}
\end{equation} 
Then, rewriting \eqref{case 1 NV} and \eqref{case 2 NV}, one obtains
\begin{equation}
X^{NV,\eta}_{t_{k+1}} =  X^{NV,\eta}_{t_{k}} +  \sum \limits_{j=1}^d \displaystyle\int_{t_k}^{t_{k+1}} \sigma^j\left(\bar{X}^{j,\eta}_s\right) \circ dW^j_s +  \displaystyle\int_{t_k}^{t_{k+1}} \frac{1}{2} \left( \sigma^0\left(\bar{X}^{0,\eta}_s\right) + \sigma^0\left(\bar{X}^{d+1,\eta}_s\right) \right)  ds,
\label{NV_Intergral}
\end{equation}
where, for $s \in \left(t_k, t_{k+1}\right]$,
\begin{equation}
\bar{X}^{0,\eta}_s = \exp\left(\frac{\Delta s}{2}\sigma^0\right) \left( X^{NV,\eta}_{t_{k}} \mathds{1}_{\left\{\eta_{k+1} = 1\right\}} + \bar{X}^{1,\eta}_{t_{k+1}} \mathds{1}_{\left\{\eta_{k+1} = -1\right\}} \right),
\end{equation}
for $s \in \left(t_k, t_{k+1}\right]$, $j \in \left\{1,\ldots,d\right\}$,
\begin{equation}
\bar{X}^{j,\eta}_s = \exp\left(\Delta W_s^j\sigma^j\right) \left( \bar{X}^{j-1,\eta}_{t_{k+1}} \mathds{1}_{\left\{\eta_{k+1} = 1\right\}} + \bar{X}^{j+1,\eta}_{t_{k+1}} \mathds{1}_{\left\{\eta_{k+1} = -1\right\}} \right),
\label{Bar_j}
\end{equation}
for $s \in \left(t_k, t_{k+1}\right]$,
\begin{equation}
\bar{X}^{d+1,\eta}_s = \exp\left(\frac{\Delta s}{2}\sigma^0\right) \left( \bar{X}^{d,\eta}_{t_{k+1}} \mathds{1}_{\left\{\eta_{k+1} = 1\right\}} +  X^{NV,\eta}_{t_{k}} \mathds{1}_{\left\{\eta_{k+1} = -1\right\}}\right).
\end{equation}
Denoting $\bar{X}^{-1,\eta}_{t_{k+1}} = \bar{X}^{d+2,\eta}_{t_{k+1}} = X^{NV,\eta}_{t_{k}}$, one gets an expression similar to \eqref{Bar_j} for $j \in \left\{0,d+1\right\}$ and $s \in \left(t_k, t_{k+1}\right]$
\begin{equation}
\bar{X}^{j,\eta}_s = \exp\left(\frac{\Delta s}{2}\sigma^0\right) \left( \bar{X}^{j-1,\eta}_{t_{k+1}} \mathds{1}_{\left\{\eta_{k+1} = 1\right\}} + \bar{X}^{j+1,\eta}_{t_{k+1}} \mathds{1}_{\left\{\eta_{k+1} = -1\right\}} \right).
\label{exp_compact}
\end{equation}
Then, one can observe that the Ninomiya-Victoir scheme is obtained by replacing the exact solution $X$ by one of the intermediate processes $\bar{X}^{j,\eta}$ in the Stratonovich formulation \eqref{EDS_STO} of the SDE \eqref{EDS_ITO}.   
\begin{arem}
The stochastic processes $\left(\bar{X}^{j,\eta}_{t}\right)_{t\in[0,T]}$, $j \in \left\{1,\ldots d+1\right\}$, are not adapted to the natural filtration $\mathcal{F}_t = \sigma\left(W_s, s\leq t\right)$ of the Brownian motion. To get around this problem, we work with the following filtration $\tilde{\mathcal{F}}^j_t =\sigma\left( W^j_s, s\leq t   \right)\underset{k \neq j}{\bigvee} \sigma\left( W^k_s, s\leq T   \right), \forall j \in \left\{1,\ldots,d\right\}$. Then, for $j \in \left\{1,\ldots,d\right\}$, by independence, $W^j$ is a $\tilde{\mathcal{F}}^j$ Brownian motion, and $\bar{X}^{j,\eta}$ is adapted to the filtration $\tilde{\mathcal{F}}^j$. This ensures that each stochastic integral is well defined.
\end{arem}

In order to study the strong convergence, we have to build an interpolated scheme. Let $\left(X^{NV,\eta}_t\right)_{t \in [0,T]}$ be the following  It\^o process
\begin{equation}
\left\{
    \begin{array}{ll}
dX^{NV,\eta}_t = \sum \limits_{j=1}^d \sigma^j(\bar{X}^{j,\eta}_t) \circ  dW_t^j + \frac{1}{2} \left(\sigma^0\left(\bar{X}^{0,\eta}_t\right) + \sigma^0\left(\bar{X}^{d+1,\eta}_t\right)\right) dt \\
X^{NV,\eta}_0 = x.
\end{array}
\right.
\label{NV-Interpol_STO}
\end{equation}
Using \eqref{NV_Intergral} and forward induction, one can show that $\left(X^{NV,\eta}_t\right)_{0\leq t \leq T}$ is an interpolation of the Ninomiya-Victoir scheme $\left(X^{NV,\eta}_{t_k}\right)_{ k \in [\![0;N]\!]}$.
The It\^o decomposition of  $\left( X^{NV,\eta}_{t}\right)_{t \in [0,T]}$ is given by\\
\begin{equation}
\left\{
    \begin{array}{ll}
dX^{NV,\eta}_t = \sum \limits_{j=1}^d \sigma^j(\bar{X}^{j,\eta}_t)  dW_t^j + \frac{1}{2} \sum \limits_{j=1}^d \partial \sigma^j \sigma^j\left(\bar{X}^{j,\eta}_t\right) dt + \frac{1}{2} \left(\sigma^0\left(\bar{X}^{0,\eta}_t\right) + \sigma^0\left(\bar{X}^{d+1,\eta}_t\right)\right) dt   \\
X^{NV,\eta}_0 = x.
\end{array}
\right.
\label{NV-Interpol_ITO}
\end{equation}
\begin{arem}
A natural and adapted interpolation for this scheme could be
\begin{equation}
 X^{NV,\eta}_{t} = h_{\eta_t}\left(\frac{\Delta t}{2},\Delta W_t, \frac{\Delta t}{2}; X^{NV,\eta}_{\hat{\tau}_t} \right) 
\end{equation} 
where  
\begin{equation}
 h_{-1}\left(t_0,\ldots,t_{d+1};x\right) = \exp\left(t_0 \sigma^0\right) \exp\left(t_1 \sigma^1\right) \ldots \exp\left(t_d \sigma^d\right)  \exp\left(t_{d+1} \sigma^0\right)  x  
\end{equation} 
and 
\begin{equation}
 h_{1}\left(t_0,\ldots,t_{d+1};x\right) = \exp\left(t_0 \sigma^0\right) \exp\left(t_d \sigma^d\right) \ldots \exp\left(t_1 \sigma^1\right)  \exp\left(t_{d+1} \sigma^0\right)  x.  
\end{equation} 
In both cases $\Delta W_t = \left(\Delta W_t^1, \ldots , \Delta W_t^d \right)$. In order to obtain the It\^o decomposition of $X^{NV,\eta}$, we have to apply the It\^o formula. To do so, we have to compute the derivatives of $h_{\eta}$. In the general case, the computation of derivatives of this function is quite complicated. That is why we will not focus on this interpolation.   
\end{arem}
\subsection{Strong convergence}
We recall that  $\sigma^j \in \mathcal{C}^{1}\left(\mathbb{R}^n,\mathbb{R}^n\right)$, $\forall j \in \left\{1,\ldots,d\right\}$, and we assume that the vector fields, $\sigma^j, \forall j \in \left\{0,\ldots,d\right\},$ and $\partial \sigma^j \sigma^j, \forall j \in \left\{1,\ldots,d\right\}$, are Lipschitz continuous functions. Obviously, $b$ is also Lipschitz continuous, since $b = \sigma^0 + \frac{1}{2}\sum \limits_{j=1}^d \partial \sigma^j \sigma^j$. Let $L \in \mathbb{R}_+^*$ denote their common Lipschitz constant:
\begin{equation*}
\begin{split}
\forall j \in \left\{0,\ldots,d\right\}, \forall x, y \in \mathbb{R}^n,  & \left\| \sigma^j(x) - \sigma^j(y)\right\| \leq L \left\| x - y\right\|, \\
\forall j \in \left\{1,\ldots,d\right\}, \forall x, y \in \mathbb{R}^n,  & \left\|\partial \sigma^j \sigma^j(x) - \partial \sigma^j \sigma^j(y)\right\| \leq L \left\| x - y\right\| 
\end{split}
\end{equation*}
where the Euclidean norm is denoted by $\left\| . \right\|$.
\begin{thm}
\label{SC_NV}
Let $p \in [1,+\infty)$. Under the previous Lipschitz assumption, there exists a deterministic constant $C_{NV} \in \mathbb{R}_+^*$ such that
\begin{equation*}
\forall N \in \mathbb{N}^*, ~ \mathbb{E}\left[ \underset{t\leq T}{\sup}\left\|X_t - X^{NV,\eta}_{t}\right\|^{2p} \Bigg| \eta  \right] \leq C_{NV} \left(1+ \left\| x\right\|^{2p}\right) h^p.
\end{equation*}
Of course, this result implies that
\begin{equation*}
\forall N \in \mathbb{N}^*, ~ \mathbb{E}\left[ \underset{t\leq T}{\sup}\left\|X_t - X^{NV,\eta}_{t}\right\|^{2p} \right] \leq C_{NV} \left(1+ \left\| x\right\|^{2p}\right) h^p.
\end{equation*}
\end{thm} 
Obviously, $\left(X^{NV,\eta}_t\right)_{0 \leq t \leq T}$ and $h$ depend on $N$, but in order to keep the notations simple, the dependence on $N$ is not made explicit.
The following proposition will be used to prove the theorem.
\begin{aprop}
\label{Prop1}
Let $p \ge 1$, $Y= \left( Y_t \right)_{0\leq t \leq h}$ be the solution of the following $n$-dimensional SDE, driven by a $d$-dimensional Brownian motion, until $t = h$
\begin{equation*}
\left\{
    \begin{array}{ll}
 dY_s  = \alpha(Y_s) ds + \beta(Y_s) dW_s \\
  Y_0  \text{ independent of } \left(W_t\right)_{t \in [0,h]} \text{such that~} \mathbb{E}\left[\left\| Y_0 \right\|^{2p} \right] <+ \infty .
\end{array}
\right.
\end{equation*}
Assume that $\alpha$ and $\beta$ are Lipschitz continuous functions, then: $\exists C_0 \in \mathbb{R}^*_+, \forall t,s \in [0,h], s \leq t,$
\begin{enumerate}[(i)]
\item \begin{equation}
 \mathbb{E}\left[ 1 + \left\| Z_t \right\|^{2p} \right] \leq \mathbb{E}\left[ 1 + \left\| Z_0 \right\|^{2p} \right] \exp\left(C_0h\right). 
\label{Res1}
\end{equation}
\item \begin{equation}
\label{Res4}
 \mathbb{E}\left[ \left\| Z_t - Z_s \right\|^{2p} \right] \leq C_0 \left(1+ \mathbb{E}\left[\left\| Z_0 \right\|^{2p}\right]\right) \left(t - s\right)^{p}.
\end{equation}
If $ \beta = 0$, we have a better result:
\begin{equation}
\label{Res5}
\mathbb{E}\left[ \left\| Z_t - Z_s \right\|^{2p} \right] \leq C_0 \left(1+ \mathbb{E}\left[\left\| Z_0 \right\|^{2p}\right]\right) \left(t - s\right)^{2p}.
\end{equation}
\end{enumerate}
The constant $C_0$  only depends on $ \left\| \alpha(0) \right\|,  \left\| \beta(0) \right\|$, $T$, $p$, and the Lipschitz constants of the functions $\alpha$ and $ \beta $.
\end{aprop}
All these results are well known (see \cite{RY} for example).
\subsection{Intermediate results}
By using the previous proposition, one can show that the scheme has uniformly bounded moments.
\begin{alem}
\label{Lemme0}
$\forall p \ge 1,  \exists C_1 \in \mathbb{R}^*_+, \forall t \in [0,T], \forall N \in \mathbb{N}^*, \forall j \in \left\{0,\ldots,d+1\right\}$,
\begin{equation*}
\mathbb{E}\left[1 + \left\| \bar{X}^{j,\eta}_t \right\|^{2p} \Bigg| \eta \right] \leq \exp(C_1 \check{\tau}_t ) \left(1+ \left\| x\right\|^{2p}\right). 
\end{equation*}
\end{alem}
\begin{adem}
Let $p \ge 1$ and $ t \in [0,T]$, then  $\exists k \in  \left\{0,\ldots,N-1\right\}$ such that $t_k < t \leq  t_{k+1}. $ For $ j= 0$, $\left(\bar{X}^{0,\eta}_s\right)_{t_k < s \leq t_{k+1}}$ is the solution of the following ODE
\begin{equation*}
\left\{
    \begin{array}{ll}
 dZ_s  = \frac{1}{2} \sigma^0(Z_s) ds \\\\
  Z_{t_k} = X^{NV,\eta}_{t_{k}} \mathds{1}_{\left\{\eta_{k+1} = 1\right\}} + \bar{X}^{1,\eta}_{t_{k+1}} \mathds{1}_{\left\{\eta_{k+1} = -1\right\}}.
\end{array}
\right.
\end{equation*}
The independence between $\eta$ and $W$ combined with \eqref{Res1} ensures that:
\begin{equation}
\begin{split}
 \mathbb{E}\left[ 1+ \left\| \bar{X}^{0,\eta}_t \right\|^{2p}  \Bigg |  \eta \right] & \leq  \mathbb{E}\left[1+ \left\| X^{NV,\eta}_{t_{k}} \mathds{1}_{\left\{\eta_{k+1} = 1\right\}} + \bar{X}^{1,\eta}_{t_{k+1}} \mathds{1}_{\left\{\eta_{k+1} = -1\right\}} \right\|^{2p} \Bigg | \eta \right] \exp \left(\frac{1}{2}C_0 h \right)\\
& =   \left( \mathds{1}_{\left\{\eta_{k+1} = 1\right\}} \mathbb{E}\left[1+ \left\| X^{NV,\eta}_{t_{k}} \right\|^{2p} \Bigg | \eta \right] + \mathds{1}_{\left\{\eta_{k+1} = -1\right\}} \mathbb{E}\left[1+ \left\| \bar{X}^{1,\eta}_{t_{k+1}} \right\|^{2p} \Bigg | \eta \right] \right) \exp \left(\frac{1}{2}C_0 h \right).
\label{b1} 
\end{split}
\end{equation}
Similarly, for $1 \leq j \leq d $: $\left(\bar{X}^{j,\eta}_s\right)_{t_k < s \leq t_{k+1}}$ is the solution of the following SDE:
\begin{equation*}
\left\{
    \begin{array}{ll}
 dZ_s  = \frac{1}{2}\partial \sigma^j \sigma^j\left(Z_s\right) ds + \sigma^j(Z_s) dW^j_s \\
  Z_{t_k} = \bar{X}^{j-1,\eta}_{t_{k+1}}.
\end{array}
\right.
\end{equation*}
Using the same argument, one gets: 
\begin{equation}
\begin{split}
 \mathbb{E}\left[ 1+ \left\| \bar{X}^{j,\eta}_t \right\|^{2p}  \Bigg |  \eta \right]  \leq   \left( \mathds{1}_{\left\{\eta_{k+1} = 1\right\}} \mathbb{E}\left[1+ \left\| \bar{X}^{j-1,\eta}_{t_{k+1}} \right\|^{2p} \Bigg | \eta \right] + \mathds{1}_{\left\{\eta_{k+1} = -1\right\}} \mathbb{E}\left[1+ \left\| \bar{X}^{j+ 1,\eta}_{t_{k+1}} \right\|^{2p} \Bigg | \eta \right] \right) \exp \left(C_0 h \right).
\label{b2} 
\end{split}
\end{equation}
Obviously, for $ j= d+1$, one has a similar result:
\begin{equation}
\begin{split}
 \mathbb{E}\left[ 1+ \left\| \bar{X}^{j,\eta}_t \right\|^{2p}  \Bigg |  \eta \right]  \leq  \left( \mathds{1}_{\left\{\eta_{k+1} = 1\right\}} \mathbb{E}\left[1+ \left\| \bar{X}^{d,\eta}_{t_{k+1}} \right\|^{2p} \Bigg | \eta \right] + \mathds{1}_{\left\{\eta_{k+1} = -1\right\}} \mathbb{E}\left[1+ \left\| X^{NV,\eta}_{t_{k}} \right\|^{2p} \Bigg | \eta \right] \right) \exp \left(\frac{1}{2}C_0 h \right).
\label{b3} 
\end{split}
\end{equation}
The global Lipschitz constant $L$ is the same for all vector fields, therefore, the same constant $C_0$ is involved in the three inequalities. In both ODEs, the vector field $\sigma^0$ is multiplied by $1/2$, it is equivalent to integrate the equation until $h/2$ and simply remove the multiplicative factor $1/2$. That is why one gets a factor $1/2$ in both inequalities \eqref{b1} and \eqref{b3}. Since for all $ k \in \left\{0,\ldots,N\right\}$, $X^{NV,\eta}_{t_k} = \mathds{1}_{\left\{\eta_{k} = 1\right\}} \bar{X}^{d+1,\eta}_{t_k} + \mathds{1}_{\left\{\eta_{k} = -1\right\}} \bar{X}^{0,\eta}_{t_k}$, one can use forward induction on $k$ combined with forward induction (respectively backward) on $j \in \left\{0,\ldots,d+1\right\}$ if $\eta_{k+1} = 1$ (respectively $\eta_{k+1} = -1$) to get:
\begin{equation*}
 \mathbb{E}\left[ 1 + \left\| \bar{X}^{j,\eta}_t\right\|^{2p}  \Bigg |  \eta \right] \leq  \exp\left(C_1 t_{k+1} \right) \left( 1 + \left\|x\right\|^{2p} \right) 
\end{equation*}
where $C_1 = \left(d+1\right) C_0$. 
\end{adem}
The following lemma is a direct application of Proposition \ref{Prop1}, together with Lemma \ref{Lemme0}.   
\begin{alem}
\label{Lemme1}
$\forall p \ge 1,  \exists C_2 \in \mathbb{R}^*_+, \forall t \in [0,T], \forall N \in \mathbb{N}^*,$ $ \forall j \in \left\{1,\ldots,d\right\}$,
\begin{equation*}
 \mathbb{E}\left[ \left\| \bar{X}^{j,\eta}_t - \bar{X}^{j-1,\eta}_{\check{\tau}_t} \mathds{1}_{\left\{\eta_{t} = 1\right\}}  - \bar{X}^{j+1,\eta}_{\check{\tau}_t} \mathds{1}_{\left\{\eta_{t} = -1\right\}} \right\|^{2p} \Bigg | \eta \right] \leq C_2 \left(1+ \left\| x\right\|^{2p}\right) h^{p},
\end{equation*}
and for $ j \in \left\{0,d+1\right\}$,
\begin{equation*}
\mathbb{E}\left[ \left\| \bar{X}^{j,\eta}_t - \bar{X}^{j-1,\eta}_{\check{\tau}_t} \mathds{1}_{\left\{\eta_{t} = 1\right\}}  - \bar{X}^{j+1,\eta}_{\check{\tau}_t} \mathds{1}_{\left\{\eta_{t} = -1\right\}}\right\|^{2p} \Bigg | \eta \right] \leq   C_2 \left(1+ \left\| x\right\|^{2p}\right) h^{2p},
\end{equation*}
where by convention $\bar{X}^{-1,\eta}_{\check{\tau}_t} = \bar{X}^{d+2,\eta}_{\check{\tau}_t} = X^{NV,\eta}_{\hat{\tau}_t}$.
\end{alem}
\begin{adem}
Let $p\ge 1$, $t \in  [0,T]$ and $j \in \left\{1,\ldots,d\right\}$. Thanks to \eqref{Res4} in Proposition \ref{Prop1} we have:
\begin{equation*}
\mathbb{E}\left[ \left\| \bar{X}^{j,\eta}_t - \bar{X}^{j-1,\eta}_{\check{\tau}_t} \mathds{1}_{\left\{\eta_{t} = 1\right\}}  - \bar{X}^{j+1,\eta}_{\check{\tau}_t} \mathds{1}_{\left\{\eta_{t} = -1\right\}} \right\|^{2p} \Bigg | \eta \right] \leq C_0 \left(1+ \mathbb{E}\left[\left\|  \bar{X}^{j-1,\eta}_{\check{\tau}_t} \mathds{1}_{\left\{\eta_{t} = 1\right\}}  - \bar{X}^{j+1,\eta}_{\check{\tau}_t} \mathds{1}_{\left\{\eta_{t} = -1\right\}} \right\|^{2p} \Bigg | \eta\right]\right) h^{p}. 
\end{equation*}
Since
\begin{equation*}
1+ \mathbb{E}\left[\left\|  \bar{X}^{j-1,\eta}_{\check{\tau}_t} \mathds{1}_{\left\{\eta_{t} = 1\right\}}  - \bar{X}^{j+1,\eta}_{\check{\tau}_t} \mathds{1}_{\left\{\eta_{t} = -1\right\}} \right\|^{2p} \Bigg | \eta\right] = \mathds{1}_{\left\{\eta_{t} = 1\right\}} \mathbb{E}\left[ 1 + \left\| \bar{X}^{j-1,\eta}_{\check{\tau}_t}\right\|^{2p} \Bigg |  \eta \right] + \mathds{1}_{\left\{\eta_{t} = -1\right\}}\mathbb{E}\left[ 1 + \left\| \bar{X}^{j+1,\eta}_{\check{\tau}_t}\right\|^{2p} \Bigg |  \eta\right]
\end{equation*}
combining this estimation with lemma \ref{Lemme0} we get
\begin{equation*}
\begin{split}
 \mathbb{E}\left[ \left\| \bar{X}^{j,\eta}_t - \bar{X}^{j-1,\eta}_{\check{\tau}_t} \mathds{1}_{\left\{\eta_{t} = 1\right\}}  - \bar{X}^{j+1,\eta}_{\check{\tau}_t} \mathds{1}_{\left\{\eta_{t} = -1\right\}} \right\|^{2p} \Bigg | \eta \right] & \leq C_0 \exp(C_1 \check{\tau}_t ) \left(1+ \left\| x\right\|^{2p}\right)  h^{p}\\
& \leq C_0 \exp(C_1 T ) \left(1+ \left\| x\right\|^{2p}\right)  h^{p}.  
\end{split}
\end{equation*}
Applying a similar argument, using \eqref{Res5} from Proposition \ref{Prop1}, we get the same result for $\bar{X}^{0,\eta}$ and $\bar{X}^{d+1,\eta}$. We conclude by setting $C_2 = C_0 \exp\left(C_1 T\right)$. 
\end{adem}
The following lemma deals with the estimation of the difference between the scheme $X^{NV,\eta} $ and the intermediate process $\bar{X}^{j,\eta}$ for $ j \in \left\{0,\ldots,d+1\right\}$.
\begin{alem}
\label{Lemme2}
$\forall p \ge 1,  \exists C_3 \in \mathbb{R}^*_+, \forall t \in [0,T], \forall N \in \mathbb{N}^*, \forall j \in \left\{0,\ldots,d+1\right\}$,
\begin{equation*}
 \mathbb{E}\left[ \left\| \bar{X}^{j,\eta}_t - X^{NV,\eta}_{\hat{\tau}_t}\right\|^{2p} \Bigg | \eta  \right] \leq C_3 \left(1+ \left\| x\right\|^{2p}\right) h^{p}. 
\end{equation*}
\end{alem}
\begin{adem}
Let $p\ge 1, t \in  [0,T]$ and $j \in \left\{1,\ldots,d+1\right\}$. Using telescopic summation and convexity inequality, we get
\begin{equation*}
\begin{split}
 \left\| \bar{X}^{j,\eta}_t - X^{NV,\eta}_{\hat{\tau}_t}\right\|^{2p} & \leq  \left(d+2\right)^{2p-1} \Bigg( \left\| \bar{X}^{j,\eta}_t - \bar{X}^{j-1,\eta}_{\check{\tau}_t} \mathds{1}_{\left\{\eta_{t} = 1\right\}}  - \bar{X}^{j+1,\eta}_{\check{\tau}_t} \mathds{1}_{\left\{\eta_{t} = -1\right\}} \right\|^{2p}\\
& + \sum \limits_{\eta_t m < \eta_t j } \left\| \bar{X}^{m,\eta}_{\check{\tau}_t} - \bar{X}^{m-1,\eta}_{\check{\tau}_t} \mathds{1}_{\left\{\eta_{t} = 1\right\}}  - \bar{X}^{m+1,\eta}_{\check{\tau}_t} \mathds{1}_{\left\{\eta_{t} = -1\right\}} \right\|^{2p}
\Bigg).
\end{split}
\end{equation*}
Taking the conditional expectation, and using  Lemma \ref{Lemme1}, we obtain
\begin{equation*}
 \mathbb{E}\left[ \left\| \bar{X}^{j,\eta}_t - X^{NV,\eta}_{\hat{\tau}_t}\right\|^{2p} \Bigg | \eta \right] \leq \left(d+2\right)^{2p-1}\left(d+2T^p\right) C_2\left(1+ \left\| x\right\|^{2p}\right) h^{p} = C_3 \left(1+ \left\| x\right\|^{2p}\right) h^{p}  
\end{equation*}
with $C_3 = \left(d+2\right)^{2p-1} \left(d+2T^p\right) C_2$. 
\end{adem}

\subsection{Proof of the strong convergence}
\begin{adem}
Let $p \in [1,+\infty)$, $t \in [0,T]$ and $s\in[0,t]$. Subtracting \eqref{NV-Interpol_ITO} from \eqref{EDS_ITO}, we can evaluate the difference between the exact solution and the scheme
\begin{equation*}
\begin{split}
X_s - X^{NV,\eta}_{s} & = \frac{1}{2} \left( \displaystyle\int_{0}^{s}\left( \sigma^0(X_u) -  \sigma^0\left(\bar{X}^{0,\eta}_u\right) \right) du  +   \displaystyle\int_{0}^{t} \left( \sigma^0(X_u) -  \sigma^0\left(\bar{X}^{d+1,\eta}_u\right) \right) du \right) \\
& + \sum \limits_{j=1}^d  \displaystyle\int_{0}^{s} \left( \sigma^j(X_u) -   \sigma^j(\bar{X}^{j,\eta}_u) \right)  dW_u^j \\
& +   \frac{1}{2} \sum \limits_{j=1}^d \displaystyle\int_{0}^{s} \left( \partial \sigma^j \sigma^j\left(X_u\right) -  \partial \sigma^j \sigma^j \left(\bar{X}^{j,\eta}_u\right) \right)du. 
\end{split}
\end{equation*}
Using a convexity inequality and taking the conditional expectation of the supremum, we get:
\begin{equation}
\label{Proof_SC_Estimation}
\mathbb{E}\left[ \underset{s\leq t}{\sup}\left\|X_s - X^{NV,\eta}_{s}\right\|^{2p} \Bigg | \eta \right] \leq \left(2\left(d+1\right)\right)^{2p-1} \left( \sum \limits_{j=1}^d E_j + \frac{1}{2^{2p}}\sum \limits_{j=0}^{d+1} I_j \right)
\end{equation}
where
\begin{equation*}
I_0 = \mathbb{E}\left[ \underset{s\leq t}{\sup}\left\|\displaystyle\int_{0}^{s}\left( \sigma^0(X_u) -  \sigma^0\left(\bar{X}^{0,\eta}_u\right)\right) du \right\|^{2p} \Bigg | \eta  \right],  
\end{equation*}
\begin{equation*}
I_{d+1} = \mathbb{E}\left[ \underset{s\leq t}{\sup}\left\|\displaystyle\int_{0}^{s}\left( \sigma^0(X_u) -  \sigma^0\left(\bar{X}^{d+1,\eta}_u\right)\right) du  \right\|^{2p} \Bigg | \eta  \right], 
 \end{equation*}
and for $j \in \left\{1,\ldots,d \right\}$
\begin{equation*}
E_j = \mathbb{E}\left[ \underset{s\leq t}{\sup}\left\|\displaystyle\int_{0}^{s}\left( \sigma^j(X_u) -   \sigma^j(\bar{X}^{j,\eta}_u) \right)  dW_u^j \right\|^{2p}\Bigg | \eta   \right],  
\end{equation*}
\begin{equation*}
I_j = \mathbb{E}\left[ \underset{s\leq t}{\sup}\left\|\displaystyle\int_{0}^{s}\left( \partial \sigma^j \sigma^j \left(X_u\right) -  \partial \sigma^j \sigma^j \left(\bar{X}^{j,\eta}_u\right)\right) du  \right\|^{2p} \Bigg | \eta  \right].  
\end{equation*}
Let us focus on $E_j$ and $I_j$, for $j \in \left\{1,\ldots,d \right\}$. The independence between $W$ and $\eta$ permits to apply the Burkholder-Davis-Gundy inequality to obtain
\begin{equation*}
 E_j  \leq K ~ \mathbb{E}\left[ \left(\displaystyle\int_{0}^{t} \left\|  \sigma^j(X_u) -   \sigma^j(\bar{X}^{j,\eta}_u) \right\|^{2} du \right)^p \Bigg | \eta  \right]  \leq KT^{p-1} \displaystyle\int_{0}^{t}\mathbb{E}\left[ \left\|  \sigma^j(X_u) -   \sigma^j(\bar{X}^{j,\eta}_u) \right\|^{2p} \Bigg | \eta \right] du
\end{equation*}
where $K$ is the constant that appears in the Burkholder-Davis-Gundy inequality.
By the Lipschitz assumption
\begin{equation}
\label{Proof_SC_Ej}
 E_j  \leq KT^{p-1} L^{2p}  \displaystyle\int_{0}^{t}\mathbb{E}\left[ \left\|  X_u -  \bar{X}^{j,\eta}_u \right\|^{2p}\Bigg | \eta  \right] du.
\end{equation} 
Applying a convexity inequality, we obtain
\begin{equation*}
I_j  \leq   T^{2p-1}\displaystyle\int_{0}^{t} \mathbb{E}\left[ \left\| \partial \sigma^j \sigma^j \left(X_s\right) -  \partial \sigma^j \sigma^j \left(\bar{X}^{j,\eta}_s\right)  \right\|^{2p} \Bigg | \eta \right] ds.
\end{equation*}
Again, by the Lipschitz assumption, we also get
\begin{equation}
\label{Proof_SC_Ij}
 I_j  \leq T^{2p-1}L^{2p}  \displaystyle\int_{0}^{t}\mathbb{E}\left[ \left\|  X_u -  \bar{X}^{j,\eta}_u \right\|^{2p} \Bigg | \eta \right] du.
\end{equation} 
Using the same approach, we get a similar result for $ I_{0}$ and $I_{d+1}$. Combining \eqref{Proof_SC_Ej} and \eqref{Proof_SC_Ej}, together with \eqref{Proof_SC_Estimation}, we obtain
\begin{equation}
\mathbb{E}\left[ \underset{s\leq t}{\sup}\left\|X_u - X^{NV,\eta}_{u}\right\|^{2p} \Bigg | \eta \right] \leq \alpha \sum \limits_{j=0}^{d+1} \displaystyle\int_{0}^{t}\mathbb{E}\left[ \left\|  X_u -  \bar{X}^{j,\eta}_u \right\|^{2p} \Bigg | \eta \right] du
\label{Proof_Strong_NV}
\end{equation}
where $\alpha = \left(2\left(d+1\right)\right)^{2p-1} L^{2p} \left(KT^{p-1} + \frac{T^{2p-1}}{2^{2p}}\right)  $. Now, we look at $\left\|  X_u -  \bar{X}^{j,\eta}_u \right\|$. Let $j \in \left\{0,\ldots,d+1\right\}$ and $ u \in [0,t]$. Introducing the solution $X$ and the Ninomiya-Victoir scheme
$X^{NV,\eta}$ at time $\hat{\tau}_u$, and using a convexity inequality we get
\begin{equation*}
 \mathbb{E}\left[ \left\|  X_u -  \bar{X}^{j,\eta}_u \right\|^{2p} \Bigg | \eta \right] \leq 3^{2p-1}  ~ \mathbb{E}\left[ \left\|  X_u -  X_{\hat{\tau}_u} \right\|^{2p} + \left\|  X_{\hat{\tau}_u} -  X^{NV,\eta}_{\hat{\tau}_u} \right\|^{2p} + \left\| X^{NV,\eta}_{\hat{\tau}_u} - \bar{X}^{j,\eta}_u \right\|^{2p} \Bigg | \eta \right].
 \end{equation*}
Then, using the estimation \eqref{Res4} from Proposition \ref{Prop1}
\begin{equation*}
 \mathbb{E}\left[ \left\|  X_u -  X_{\hat{\tau}_u} \right\|^{2p} \Bigg | \eta \right] \leq C_0 \left(1+ \left\| x\right\|^{2p}\right) \left(u - \hat{\tau}_u\right)^p \leq C_0 \left(1+ \left\| x\right\|^{2p}\right)  h^p 
\end{equation*}
and from Lemma \ref{Lemme2}
\begin{equation*}
 \mathbb{E}\left[ \left\| \bar{X}^{j,\eta}_u - X^{NV,\eta}_{\hat{\tau}_u} \right\|^{2p} \Bigg | \eta \right] \leq C_3 \left(1+ \left\| x\right\|^{2p}\right) h^{p}. 
\end{equation*}
Moreover
\begin{equation*}
 \mathbb{E}\left[ \left\|  X_{\hat{\tau}_u} -  X^{NV,\eta}_{\hat{\tau}_u} \right\|^{2p}  \Bigg | \eta \right] \leq \mathbb{E}\left[ \underset{v\leq u}{\sup}\left\|X_v - X^{NV,\eta}_{v}\right\|^{2p} \Bigg | \eta\right]. 
\end{equation*}
We finally get
\begin{equation}
\mathbb{E}\left[ \underset{s\leq t}{\sup}\left\|X_s -  X^{NV,\eta}_{s}\right\|^{2p} \Bigg | \eta \right] \leq \beta \displaystyle\int_{0}^{t}\mathbb{E}\left[ \underset{v\leq u}{\sup}\left\|X_v -  X^{NV,\eta}_{v}\right\|^{2p} \Bigg | \eta \right] du + \gamma \left(1+ \left\| x\right\|^{2p}\right) h^p,
\end{equation}
where
\begin{equation*}
\beta = 3^{2p-1} \left(d+2\right) \alpha
\end{equation*}
and
\begin{equation*}
\gamma = \beta T \left(C_0 + C_3\right). 
\end{equation*} 
Before applying Gronwall's lemma, let us remark, by \eqref{Proof_Strong_NV},\eqref{Res1} and Lemma \ref{Lemme0}, that $\mathbb{E}\left[ \underset{s\leq t}{\sup}\left\|X_s -  X^{NV,\eta}_{s}\right\|^{2p} \right]$ is finite.  \\
We conclude thanks to Gronwall's lemma
\begin{equation*}
\mathbb{E}\left[ \underset{t\leq T}{\sup}\left\|X_t -  X^{NV,\eta}_{t}\right\|^{2p} \Bigg | \eta \right] \leq \exp\left(\beta T\right) \gamma \left(1+ \left\| x\right\|^{2p}\right) h^p.
\end{equation*}
\end{adem}
We conclude this section with a lemma which will be useful for the next section.
\begin{alem}
Let $F \in \mathcal{C}^2\left(\mathbb{R}^n,\mathbb{R}^n\right)$ and assume that its first and second order derivatives have a polynomial growth.
Under the assumptions of the Theorem \ref{SC_NV} we have the following result: $\forall p\in [1,+\infty),\exists C_4 \in \mathbb{R}_+^*,\forall j \in \left\{0,\ldots,d+1\right\},\forall N \in \mathbb{N}^*,$
\begin{equation*}
 \mathbb{E}\left[ \underset{t\leq T}{\sup}\left\| \displaystyle\int_{0}^{t} F\left( \bar{X}^{j,\eta}_s\right) - F\left( X^{NV,\eta}_{\hat{\tau}_s}\right) ds  \right\|^{2p} \Bigg | \eta \right]\leq C_4 h^{2p}. 
\end{equation*}
\label{Lemme_Utile}
\end{alem}
\begin{adem}
Let $j \in \left\{0,\ldots,d+1\right\}, i \in \left\{1,\ldots,n\right\}$, and $ t \in [0,T]$. Using the integration by parts formula 
\begin{equation*} 
\begin{split}
\displaystyle\int_{0}^{t}\left(  F^{i}\left( \bar{X}^{j,\eta}_s\right) - F^{i}\left( X^{NV,\eta}_{\hat{\tau}_s}\right) \right) du & = \displaystyle\int_{0}^{t}\left( t \wedge\check{\tau}_s - s\right)   d\left(F^{i} \left(\bar{X}^{j,\eta}_s\right)\right) +  \displaystyle\int_{0}^{\check{\tau}_t}\sum \limits_{\eta_s m < \eta_s j} \left(\check{\tau}_s - s\right)   d\left(F^{i} \left(\bar{X}^{m,\eta}_s\right)\right).
\end{split}
\end{equation*}
Then, using the chain rule for $m \in \left\{0,d+1\right\}$, we get
\begin{equation*}
d\left(F^{i} \left(\bar{X}^{m,\eta}_s\right)\right) =  \frac12 \sigma^0\left(\bar{X}^{m,\eta}_s\right).~\nabla F^{i} \left(\bar{X}^{m,\eta}_s\right) ds.
\end{equation*}
Applying It\^o's formula for $m \in \left\{1,\ldots,d\right\}$, we obtain
\begin{equation*}
\begin{split}
d\left(F^{i} \left(\bar{X}^{m,\eta}_s\right)\right) &= \left( \frac{1}{2} \partial \sigma^m \sigma^m\left(\bar{X}^{m,\eta}_s\right).~\nabla F^{i} \left(\bar{X}^{m,\eta}_s\right) + \frac{1}{2} tr\left(\sigma^m \left(\sigma^m\right)^*\left(\bar{X}^{m,\eta}_s\right) \nabla^2 F^{i} \left(\bar{X}^{m,\eta}_s\right)\right) \right) ds\\
& + \sigma^m\left(\bar{X}^{m,\eta}_s\right).~\nabla F^{i} \left(\bar{X}^{m,\eta}_s\right) dW^m_s.
\end{split}
\end{equation*}
In both cases, combining a convexity inequality, the Burkholder-Davis-Gundy inequality, the Holder inequality, the Lipschitz assumption on $\sigma^m, \partial \sigma^m  \sigma^m $ for $ m \in \left\{0,\ldots,d\right\}$, the polynomial growth assumption for the first and second order derivatives of $F$, and $t \wedge \check{\tau}_s - s \leq h, ~ \forall s \in [0,\check{\tau}_t]$, we get two constants $\gamma \in \mathbb{R}_+^*$ and $q \in \mathbb{N}^*$, independent of $N$, such that
\begin{equation*}
\mathbb{E}\left[\underset{t\leq T}{\sup} \left| \displaystyle\int_{0}^{t} F^i\left( \bar{X}^{j,\eta}_s\right) -F^i\left( X^{NV,\eta}_{\hat{\tau}_s}\right)  ds \right |^{2p} \Bigg | \eta \right] \leq \gamma  h^{2p} \sum \limits_{m=0}^{d+1} \displaystyle\int_{0}^{T}\mathbb{E}\left[ 1+ \left\| \bar{X}^{m,\eta}_s  \right\|^{2q} \Bigg | \eta \right] ds. 
\end{equation*}
We conclude by using Lemma \ref{Lemme0} and taking the Euclidean norm. 
\end{adem}

\section{Coupling with Giles-Szpruch scheme}
In \cite{GS}, Giles and Szpruch proposed a modified Milstein scheme defined as follows
\begin{equation}
\left\{
    \begin{split}
X_{t_{k+1}}^{GS} & = X_{t_{k}}^{GS} + b\left(X_{t_{k}}^{GS} \right) \left(t_{k+1} - t_k\right) + \sum \limits_{j=1}^d \sigma^j\left(X_{t_{k}}^{GS} \right) \Delta W^j_{t_{k+1}} \\
& +  \frac{1}{2} \sum \limits_{j,m=1}^d \partial \sigma^j \sigma^m \left(X_{t_k}^{GS} \right)\left(\Delta W^j_{t_{k+1}} \Delta W^m_{t_{k+1}} - \mathds{1}_{\left\{j=m\right\}} h \right) \\
X^{GS}_{t_0} & = x.
\end{split}
\right.
\label{GS-Scheme}
\end{equation}
In comparison with the Milstein scheme, the terms involving the L\'evy areas $ \displaystyle\int_{t_k}^{t_{k+1}} \Delta W^j_s dW^m_s - \displaystyle\int_{t_k}^{t_{k+1}} \Delta W^m_s dW^j_s$  have been removed. According to Lemma 4.2 in \cite{GS}, the strong order of convergence is $\gamma = 1/2$. 
\begin{alem}
\label{GS_Order}
Assume that $ b, \sigma^j \in \mathcal{C}^2\left(\mathbb{R}^n,\mathbb{R}^n\right), \forall j \in \left\{1,\ldots,d\right\}$, with bounded first and second order derivatives, and that $\partial \sigma^j \sigma^m$ ,$\forall j,m \in \left\{1,\ldots,d\right\}$, has bounded first order derivatives. Then:
\begin{equation}
\exists C_{GS} \in \mathbb{R}_+^*, \forall N \in \mathbb{R}_+^*, ~ \mathbb{E}\left[ \underset{k \in \left\{0,\ldots,N\right\}}{\max}\left\|X_{t_k} - X^{GS}_{t_k}\right\|^{2p} \right] \leq C_{GS} h^p.
\end{equation}
\end{alem}
Giles and Szpruch also proposed an antithetic version of their scheme based on the swapping of each successive pair of the Brownian increments in the scheme. With regards to the multilevel Monte Carlo estimator, Giles and Szpruch use the arithmetic average of the scheme \eqref{GS-Scheme} and its antithetic version on the fine grids, at each level $l  \in \left\{1,\ldots,L\right\}$ as follows
\begin{equation*}
\hat{Y} = \frac{1}{M_0} \sum \limits_{k=1}^{M_0}  f \left( X^{1,0,k}_T\right)+ \sum \limits_ {l=1}^L \frac{1}{M_l} \sum \limits_{k=1}^{M_l}  \left( \frac12 \left( f \left( \tilde{X}^{2^l,l,k}_T\right) + f \left( X^{2^l,l,k}_T\right) \right) - f\left(X^{2^{l-1},l,k}_T\right) \right).
\end{equation*}
The swapping of each successive pair of  Brownian increments provides a strong convergence of order 1 between the schemes used on the coarse and fine grids, and so Giles and Szpruch obtained the convergence rate $\beta = 2$ of the variance $\mathbb{V}\left(\frac12 \left( f \left( \tilde{X}^{2^l,l,k}_T\right) + f \left( X^{2^l,l,k}_T\right) \right) - f\left(X^{2^{l-1},l,k}_T\right) \right)$, when the payoff $f$ is smooth. 
In this way, using this multilevel Monte Carlo estimator leads to the computational complexity $O \left(\epsilon^{-2}\right)$ for the mean-square-root error $\epsilon$.
To use the Ninomiya-Victoir scheme either at each level or only at the finest level of a multilevel Monte Carlo estimator, we study in this section the coupling between the Ninomiya-Victoir and Giles-Szpruch schemes. 
To keep $\beta = 2$, we suggest to compare the Giles-Szpruch scheme with the following modified Ninomiya-Victoir scheme
\begin{equation}
\bar{X}^{NV,\eta} =  \frac{1}{2} \left( X^{NV,\eta} + X^{NV,-\eta}  \right).
\end{equation}
To be consistent with the interpolation of the Ninomiya-Victoir scheme, we define the interpolation of the scheme between the grid points as follows
\begin{equation}
\begin{split}
\label{Iterpol_Mod_NV}
X_{t}^{GS} &= x + \displaystyle\int_{0}^{s} b\left(X_{\hat{\tau}_u}^{GS} \right) du  + \sum \limits_{j=1}^d  \displaystyle\int_{0}^{s} \sigma^j\left(X_{\hat{\tau}_u}^{GS} \right) dW^j_u+  \sum \limits_{j=1}^d  \displaystyle\int_{0}^{s} \partial \sigma^j \sigma^j\left(X_{\hat{\tau}_u}^{GS} \right) \Delta W^j_u dW^j_u 
\\
& + \frac{1}{2} \sum \limits_{\underset{m \neq j}{j,m=1}}^d  \displaystyle\int_{0}^{s} \partial \sigma^j \sigma^m \left(X_{\hat{\tau}_u}^{GS} \right)  \Delta W^m_{\check{\tau}_u} dW^j_u. 
\end{split}
\end{equation}
\begin{thm}
\label{Coupling}
We assume that $b \in  \mathcal{C}^2 \left(\mathbb{R}^n;\mathbb{R}^n\right)$ with bounded first and second order derivatives, and $ \sigma^j  \in \mathcal{C}^3 \left(\mathbb{R}^n;\mathbb{R}^n\right), \forall j \in \left\{1, \ldots , d\right\},$ with bounded first and second order derivatives and with polynomially growing third order derivatives, and that $\partial \sigma^j \sigma^m$, $\forall j,m \in \left\{1,\ldots,d\right\}$, has bounded first order derivatives. Then:
\begin{equation*}
\exists C \in \mathbb{R}_+^*, \forall N \in \mathbb{N}^*,~ \mathbb{E}\left[ \underset{t\leq T}{\sup}\left\|\bar{X}^{NV,\eta}_{t} - X^{GS}_t\right\|^{2p} \Bigg | \eta \right] \leq C h^{2p}.
\end{equation*}
\end{thm}

\begin{adem}
We denote by $L$ the common Lipschitz constant of $b, \sigma^j$ and $\partial \sigma^j \sigma^m, \forall j,m \in \left\{1, \ldots , d\right\}$. We also denote by $M$ the global bound on first and second derivatives of $b$ and $\sigma^j, \forall j \in \left\{1, \ldots , d\right\}$.
Let $t \in [0,T]$ and $s \in [0,t]$. Writing $\bar{X}^{NV,\eta}$ in integral form, we get
\begin{equation*}
\begin{split}
\bar{X}^{NV,\eta}_s & =  x + \sum \limits_{j=1}^d  \displaystyle\int_{0}^{s} \frac{1}{2} \left( \sigma^j\left( \bar{X}^{j,\eta}_u\right) + \sigma^j\left( \bar{X}^{j,-\eta}_s\right)  \right) dW^j_u  + \sum \limits_{j=1}^d  \displaystyle\int_{0}^{s} \frac{1}{4} \left( \partial \sigma^j  \sigma^j\left( \bar{X}^{j,\eta}_u\right) + \partial \sigma^j  \sigma^j\left( \bar{X}^{j,-\eta}_u\right)  \right) du \\
&+ \displaystyle\int_{0}^{s} \frac{1}{4} \left( \sigma^0\left( \bar{X}^{0,\eta}_u\right) + \sigma^0\left( \bar{X}^{0,-\eta}_u\right)  \right) du + \displaystyle\int_{0}^{s} \frac{1}{4} \left( \sigma^0\left( \bar{X}^{d+1,\eta}_u\right) + \sigma^0\left( \bar{X}^{d+1,-\eta}_u\right)  \right) du .
\end{split}
\end{equation*}
Then using $\frac{1}{2} b  - \frac{1}{4} \sum\limits_{j=1}^d \partial \sigma^j \sigma^j - \frac{1}{2} \sigma^0 = 0$, we get
\begin{equation*}
\begin{split}
\bar{X}^{NV,\eta}_s & =  x + \sum \limits_{j=1}^d  \displaystyle\int_{0}^{s} \frac{1}{2} \left( \sigma^j\left( X^{NV,\eta}_{\hat{\tau}_u}\right) + \sigma^j\left( X^{NV,-\eta}_{\hat{\tau}_u}\right)  \right) dW^j_u  +  \displaystyle\int_{0}^{s} \frac{1}{2} \left( b\left( X^{NV,\eta}_{\hat{\tau}_u}\right) + b\left( X^{NV,-\eta}_{\hat{\tau}_u}\right)  \right) du \\
& + \sum \limits_{j=1}^d  \displaystyle\int_{0}^{s} \frac{1}{2} \left( \sigma^j\left( \bar{X}^{j,\eta}_u\right) - \sigma^j\left( X^{NV,\eta}_{\hat{\tau}_u}\right)  + \sigma^j\left( \bar{X}^{j,-\eta}_u\right) - \sigma^j\left( X^{NV,-\eta}_{\hat{\tau}_u}\right)  \right) dW^j_u  \\ 
&+ \sum \limits_{j=1}^d  \displaystyle\int_{0}^{s} \frac{1}{4} \left( \partial \sigma^j  \sigma^j\left( \bar{X}^{j,\eta}_u\right) - \partial \sigma^j  \sigma^j\left( X^{NV,\eta}_{\hat{\tau}_u}\right) + \partial \sigma^j  \sigma^j\left( \bar{X}^{j,-\eta}_u\right) - \partial \sigma^j  \sigma^j\left( X^{NV,-\eta}_{\hat{\tau}_u}\right) \right) du \\
& + \displaystyle\int_{0}^{s} \frac{1}{4} \left( \sigma^0\left( \bar{X}^{0,\eta}_u\right) - \sigma^0\left( X^{NV,\eta}_{\hat{\tau}_u}\right)  + \sigma^0\left( \bar{X}^{0,-\eta}_u\right) - \sigma^0\left( X^{NV,-\eta}_{\hat{\tau}_u}\right)  \right) du  \\
&  + \displaystyle\int_{0}^{s} \frac{1}{4} \left( \sigma^0\left( \bar{X}^{d+1,\eta}_u\right) - \sigma^0\left( X^{NV,\eta}_{\hat{\tau}_u}\right) + \sigma^0\left( \bar{X}^{d+1,-\eta}_u\right) - \sigma^0\left( X^{NV,-\eta}_{\hat{\tau}_u}\right)  \right) du .
\end{split}
\end{equation*}
Subtracting \eqref{Iterpol_Mod_NV} and using a convexity inequality, we obtain
\begin{equation}
\label{Estimation_Coupling}
\mathbb{E}\left[ \underset{s\leq t}{\sup}\left\|\bar{X}^{NV,\eta}_s - X_{s}^{GS}\right\|^{2p} \Bigg | \eta \right] \leq  3^{2p-1} \left(d+1\right)^{2p-1} \left(\sum \limits_{j=1}^d I_j + \sum \limits_{j=0}^d E_j + \sum \limits_{j=0}^{d+1} R_j\right) 
\end{equation}
where
\begin{equation*}
E_0 = \mathbb{E}\left[ \underset{s\leq t}{\sup}\left\| \displaystyle\int_{0}^{s} \left( \frac{1}{2} \left( b\left( X^{NV,\eta}_{\hat{\tau}_u}\right) + b\left( X^{NV,-\eta}_{\hat{\tau}_u}\right)  \right) - b\left(X_{\hat{\tau}_u}^{GS}\right) \right) du  \right\|^{2p} \Bigg | \eta \right],
\end{equation*}
\begin{equation*}
R_0 = \mathbb{E}\left[ \underset{s\leq t}{\sup}\left\| \displaystyle\int_{0}^{s} \frac{1}{4} \left( \sigma^0\left( \bar{X}^{0,\eta}_u\right) - \sigma^0\left( X^{NV,\eta}_{\hat{\tau}_u}\right)  + \sigma^0\left( \bar{X}^{0,-\eta}_u\right) - \sigma^0\left( X^{NV,-\eta}_{\hat{\tau}_u}\right)  \right) du  \right\|^{2p} \Bigg | \eta \right],
\end{equation*}
\begin{equation*}
R_{d+1} = \mathbb{E}\left[ \underset{s\leq t}{\sup}\left\| \displaystyle\int_{0}^{s} \frac{1}{4} \left( \sigma^0\left( \bar{X}^{d+1,\eta}_u\right) - \sigma^0\left( X^{NV,\eta}_{\hat{\tau}_u}\right) + \sigma^0\left( \bar{X}^{d+1,-\eta}_u\right) - \sigma^0\left( X^{NV,-\eta}_{\hat{\tau}_u}\right)  \right) du	 \right\|^{2p} \Bigg | \eta \right],
\end{equation*}
and for $j \in \left\{1,\ldots,d\right\}$,
\begin{equation*}
\begin{split}
I_j =  \mathbb{E}\Bigg[ \underset{s\leq t}{\sup}\Bigg\| &  \displaystyle\int_{0}^{s} \frac{1}{2} \left( \sigma^j\left( \bar{X}^{j,\eta}_u\right) - \sigma^j\left( X^{NV,\eta}_{\hat{\tau}_u}\right)  + \sigma^j\left( \bar{X}^{j,-\eta}_u\right) - \sigma^j\left( X^{NV,-\eta}_{\hat{\tau}_u}\right)  \right)  dW^j_u \\
& - \displaystyle\int_{0}^{s}  \Bigg(\partial \sigma^j \sigma^j\left(X_{\hat{\tau}_u}^{GS} \right) \Delta W^j_u +  \frac{1}{2} \sum \limits_{\underset{m \neq j}{m=1}}^d   \partial \sigma^j \sigma^m \left(X_{\hat{\tau}_u}^{GS} \right) \Delta W^m_{\check{\tau}_u}   \Bigg)  dW^j_u  \Bigg\|^{2p} \Bigg | \eta  \Bigg], 
\end{split}
\end{equation*}
\begin{equation*}
  E_j = \mathbb{E}\left[ \underset{s\leq t}{\sup}\left\|\displaystyle\int_{0}^{s}\left( \frac{1}{2} \left( \sigma^j\left( X^{NV,\eta}_{\hat{\tau}_u}\right) + \sigma^j\left( X^{NV,-\eta}_{\hat{\tau}_u}\right)  \right) -  \sigma^j\left(X_{\hat{\tau}_u}^{GS} \right) \right) dW^j_u\right\|^{2p} \Bigg | \eta \right], 
\end{equation*}
\begin{equation*}
R_j =  \mathbb{E}\left[ \underset{s\leq t}{\sup}\left\| \displaystyle\int_{0}^{s} \frac{1}{4} \left( \partial \sigma^j  \sigma^j\left( \bar{X}^{j,\eta}_u\right) - \partial \sigma^j  \sigma^j\left( X^{NV,\eta}_{\hat{\tau}_u}\right) + \partial \sigma^j  \sigma^j\left( \bar{X}^{j,-\eta}_u\right) - \partial \sigma^j  \sigma^j\left( X^{NV,-\eta}_{\hat{\tau}_u}\right) \right) du  \right\|^{2p} \Bigg | \eta \right].
\end{equation*}

\textbf{Step 1: estimation of $E_j$, for $j \in \left\{0,\ldots,d\right\}$}.\\\\
Let us start with the estimation of $E_j$, for $j \in \left\{0,\ldots,d\right\}$. We set for $F^0 = b$ and $F^j = \sigma^j$ for $j \in \left\{1,\ldots,d\right\}$. 
Combining the Burkholder-Davis-Gundy inequality and a convexity inequality, we get
\begin{equation*}
E_j \leq \max \left\{T^{2p-1},KT^{p-1}\right\}   \displaystyle\int_{0}^{t} \mathbb{E}\left[ \left\| \frac{1}{2} \left( F^j\left( X^{NV,\eta}_{\hat{\tau}_u}\right) + F^j\left( X^{NV,-\eta}_{\hat{\tau}_u}\right)  \right) - F^j\left(X_{\hat{\tau}_u}^{GS}\right) \right\|^{2p}    \Bigg | \eta \right] du
\end{equation*}
where $K$ is the constant that appears in the Burkholder-Davis-Gundy inequality. For $i \in \left\{1,\ldots,n\right\}$, denoting $Y_u = \frac{1}{2} \left(F^{ij}\left( X^{NV,\eta}_{\hat{\tau}_u}\right) + F^{ij}\left( X^{NV,-\eta}_{\hat{\tau}_u}\right)  \right) - F^{ij}\left(X_{\hat{\tau}_u}^{GS}  \right)$ and  performing a second order Taylor series expansion, we obtain
\begin{equation*}
\begin{split}
Y_u & =  F^{ij}\left(\bar{X}^{NV,\eta}_{\hat{\tau}_u}  \right) - F^{ij}\left(X_{\hat{\tau}_u}^{GS}  \right) 
 +  \frac{1}{16} \left(X^{NV,\eta}_{\hat{\tau}_u} - X^{NV,-\eta}_{\hat{\tau}_u} \right)^* \left(\nabla^2F^{ij}\left(\xi^1_{\hat{\tau}_u}\right) + \nabla^2F^{ij}\left(\xi^2_{\hat{\tau}_u}\right) \right) \left(X^{NV,\eta}_{\hat{\tau}_u} - X^{NV,-\eta}_{\hat{\tau}_u} \right) 
\end{split}
\end{equation*}
where $\xi^1_{\hat{\tau}_u}$ and $\xi^2_{\hat{\tau}_u}$ are points between $X^{NV,\eta}_{\hat{\tau}_u}$ and $X^{NV,-\eta}_{\hat{\tau}_u}$. Then, we easily get
\begin{equation*}
\left\| Y_u  \right\|^{2p}   \leq \alpha_1 \left(\left\|\bar{X}^{NV,\eta}_{\hat{\tau}_u} -X_{\hat{\tau}_u}^{GS} \right\|^{2p}  
+  \left\| X^{NV,\eta}_{\hat{\tau}_u} - X^{NV,-\eta}_{\hat{\tau}_u}  \right\|^{4p} \right)
 \end{equation*}
where $\alpha_1 = 2^{2p-1}\left(L^{2p} + \left(\frac{M}{8}\right)^{2p}\right)$. Thus
\begin{equation*}
E_j \leq \alpha_1 \max \left\{T^{2p-1},K T^{p-1}\right\} \left( \displaystyle\int_{0}^{t} \mathbb{E}\left[ \underset{v\leq u}{\sup}\left\|\bar{X}^{NV,\eta}_v - X_{v}^{GS}\right\|^{2p} \Bigg | \eta \right] du + \displaystyle\int_{0}^{t} \mathbb{E}\left[ \left\| X^{NV,\eta}_{\hat{\tau}_u} - X^{NV,-\eta}_{\hat{\tau}_u} \right\|^{4p}  \Bigg | \eta \right] du \right).
\end{equation*}
Introducing the solution $X$ at time $\hat{\tau}_u$ and using a convexity inequality, we obtain
\begin{equation*}
\mathbb{E}\left[ \left\| X^{NV,\eta}_{\hat{\tau}_u} - X^{NV,-\eta}_{\hat{\tau}_u} \right\|^{4p}  \Bigg | \eta \right]  \leq 2^{4p-1}\left( \mathbb{E}\left[ \left\| X^{NV,\eta}_{\hat{\tau}_u} - X_{\hat{\tau}_u} \right\|^{4p} \Bigg | \eta \right]  + \mathbb{E}\left[ \left\| X_{\hat{\tau}_u} - X^{NV,-\eta}_{\hat{\tau}_u} \right\|^{4p}  \Bigg | \eta \right]  \right).
\end{equation*}
Thanks to Theorem \ref{SC_NV}, we deduce that
\begin{equation*}
\mathbb{E}\left[ \left\| X^{NV,\eta}_{\hat{\tau}_u} - X^{NV,-\eta}_{\hat{\tau}_u} \right\|^{4p}  \Bigg | \eta \right] \leq 2^{4p} C_{NV} \left(1 + \left\|x \right\|^{4p}\right) h^{2p}. 
\end{equation*}
It follows that
\begin{equation}
\label{Step1}
E_j \leq\beta_1 \left( \displaystyle\int_{0}^{t} \mathbb{E}\left[ \underset{v\leq u}{\sup}\left\|\bar{X}^{NV,\eta}_v - X_{v}^{GS}\right\|^{2p} \Bigg | \eta \right] du +  h^{2p}\right)
\end{equation}
where $\beta_1 = \alpha_1 \max\left\{T^{2p-1},KT^{p-1}\right\}   \max\left\{1, 2^{4p} C_{NV} \left(1 + \left\|x \right\|^{4p}\right)\right\}$.

\textbf{Step 2: estimation of $R_j$, for $j \in \left\{0,\ldots,d\right\}$}.\\\\
Turning to the estimation of $R_j$, for $j \in \left\{0,\ldots,d\right\}$, from Lemma \ref{Lemme_Utile} we get a constant $\beta_2 \in \mathbb{R}_+^*$, such that
\begin{equation}
\label{Step2}
 R_j \leq \beta_2 h^{2p}.
\end{equation}

\textbf{Step 3: estimation of $I_j$, for $j \in \left\{1,\ldots,d\right\}$}.\\\\
It remains to estimate $I_j$, for $j \in \left\{1,\ldots,d\right\}$. Using the Burkholder-Davis-Gundy and convexity inequalities, we get
\begin{equation*}
\begin{split}
I_j  \leq \frac{1}{2^{2p}} KT^{p-1}  \displaystyle\int_{0}^{t} & \mathbb{E}\Bigg[ \Big\| \sigma^j\left( \bar{X}^{j,\eta}_u\right) - \sigma^j\left( X^{NV,\eta}_{\hat{\tau}_u}\right)  + \sigma^j\left( \bar{X}^{j,-\eta}_u\right) - \sigma^j\left( X^{NV,-\eta}_{\hat{\tau}_u}\right)  \\  & -2 \partial \sigma^j \sigma^j\left(X_{\hat{\tau}_u}^{GS} \right)  \Delta W^j_{u}  -  \sum \limits_{\underset{m \neq j}{m=1}}^d   \partial \sigma^j \sigma^m \left(X_{\hat{\tau}_u}^{GS} \right) \Delta W^m_{\check{\tau}_u}   \Big\|^{2p}  \Bigg | \eta \Bigg] ds.  
\end{split}
\end{equation*}
Introducing
\begin{equation*}
 \Psi^{j}_u = \Psi^{j,\eta}_u + \Psi^{j,-\eta}_u 
 \end{equation*}
where
\begin{equation}
\label{PSI}
\Psi^{j,\eta}_u = \sigma^j\left( \bar{X}^{j,\eta}_u\right) - \sigma^j\left( X^{NV,\eta}_{\hat{\tau}_u}\right)   - \partial \sigma^j \sigma^j\left(X_{\hat{\tau}_u}^{NV,\eta} \right)  \Delta W^j_{u}  -  \sum \limits_{\eta_u m < \eta_u j}   \partial \sigma^j \sigma^m \left(X_{\hat{\tau}_u}^{NV,\eta} \right) \Delta W^m_{\check{\tau}_u} 
\end{equation}
and
\begin{equation}
\begin{split}
\Phi^j_u & =   \partial \sigma^j \sigma^j\left(X_{\hat{\tau}_u}^{NV,\eta} \right)  \Delta W^j_{u}  +\partial \sigma^j \sigma^j\left(X_{\hat{\tau}_u}^{NV,-\eta} \right)  \Delta W^j_{u} - 2 \partial \sigma^j \sigma^j\left(X_{\hat{\tau}_u}^{GS} \right)  \Delta W^j_{u} + \sum \limits_{\eta_u m < \eta_u j}   \partial \sigma^j \sigma^m \left(X_{\hat{\tau}_u}^{NV,\eta} \right) \Delta W^m_{\check{\tau}_u}  \\
 & +  \sum \limits_{\eta_{u} m > \eta_u j}   \partial \sigma^j \sigma^m \left(X_{\hat{\tau}_u}^{NV,-\eta} \right) \Delta W^m_{\check{\tau}_u} -  \sum \limits_{\underset{m \neq j}{m=1}}^d   \partial \sigma^j \sigma^m \left(X_{\hat{\tau}_u}^{GS} \right) \Delta W^m_{\check{\tau}_u}
\end{split}
\end{equation}
we obtain
\begin{equation*}
I_j \leq \frac{1}{2} KT^{p-1} \left(  \displaystyle \int_0^t \mathbb{E}\left[ \left\| \Psi^j_u \right\|^{2p} +  \left\| \Phi^j_u \right\|^{2p}\Bigg | \eta \right]  du   \right).
\end{equation*}
\textbf{Step 3.1: estimation of $\mathbb{E}\left[ \left\| \Psi^j_u \right\|^{2p} \Bigg | \eta  \right]$, for $j \in \left\{1,\ldots,d\right\}$}.\\\\
Applying It\^o's formula in \eqref{PSI} to compute $ \sigma^j\left( \bar{X}^{j,\eta}_u\right) - \sigma^j\left( X^{NV,\eta}_{\hat{\tau}_u}\right)$ , we get
\begin{equation*}
\begin{split}
\Psi^{j,\eta}_u & = \displaystyle \int_{\hat{\tau}_u}^{u} \left( \partial \sigma^j \sigma^j\left(\bar{X}_v^{j,\eta}\right) - \partial \sigma^j \sigma^j\left(X_{\hat{\tau}_v}^{NV,\eta} \right)\right) dW^j_v ~ + \sum \limits_{\eta_u m < \eta_u j} \displaystyle \int_{\hat{\tau}_u}^{\check{\tau}_u}  \left(\partial \sigma^j \sigma^m\left(\bar{X}_v^{m,\eta}\right) -  \partial \sigma^j \sigma^m \left(X_{\hat{\tau}_v}^{NV,\eta} \right)  \right) dW^m_v \\
& +  \frac{1}{2} \displaystyle \int_{\hat{\tau}_u}^{u} \partial F^{j,j} \sigma^j\left(\bar{X}_v^{j,\eta}\right) ~ dv ~ + \frac12 \sum \limits_{\eta_u m < \eta_u j} \displaystyle \int_{\hat{\tau}_u}^{\check{\tau}_u}  \partial F^{j,m} \sigma^m\left(\bar{X}_v^{m,\eta}\right)  dv
\end{split}
\end{equation*}
where  $F^{j,m} = \partial  \sigma^j \sigma^m$ for $j,m \in \left\{1,\ldots,d\right\}$.  Note that the term $\frac12 \sum \limits_{\eta_u m < \eta_u j} \displaystyle \int_{\hat{\tau}_u}^{\check{\tau}_u}  \partial F^{j,m} \sigma^m\left(\bar{X}_v^{m,\eta}\right)  dv $  is equal to the sum of the drift contribution and the It\^o correction due to the dynamics of $\bar{X}^{m,\eta}$. The assumptions on $\sigma^j$ and Lemma \ref{Lemme0}, ensure that
\begin{itemize}
\item $\forall j,m \in \left\{1,\ldots,d\right\}, \partial \sigma^j\sigma^m$ is Lipschitz continuous.
\item $\forall j,m \in \left\{1,\ldots,d\right\}, \partial F^{j,m} \sigma^m \left( \bar{X}_v^{m,\eta}\right)$ has uniformly bounded moments.
\end{itemize}  
Using Lemma \ref{Lemme2}, the Burkholder-Davis-Gundy and a convexity inequalities, we obtain a constant $\gamma_3 \in \mathbb{R}_+^*$ such that
\begin{equation*}
 \mathbb{E}\left[  \left\|   \Psi^{j,\eta}_u \right\|^{2p}  \Bigg | \eta \right]\leq \gamma_3 h^{2p}.
\end{equation*}
Obviously, we have the same inequality for $\Psi^{j,-\eta}$.\\\\
\textbf{Step 3.2: estimation of $\mathbb{E}\left[ \left\| \Phi^j_u \right\|^{2p}\Bigg | \eta  \right]$, for $j \in \left\{1,\ldots,d\right\}$}.\\\\
By the Lipschitz assumption
\begin{equation}
\begin{split}
 \left\| \Phi^j_u  \right\|^{2p} & \leq \left(d+1\right)^{2p-1} L^{2p} \Bigg( \left\| X_{\hat{\tau}_u}^{NV,\eta} - X_{\hat{\tau}_u}^{GS} \right\|^{2p} \left|\Delta W^j_{u}  \right|^{2p}+\left\| X_{\hat{\tau}_u}^{NV,-\eta} - X_{\hat{\tau}_u}^{GS} \right\|^{2p} \left|\Delta W^j_{u}  \right|^{2p}\\
& +  \sum \limits_{\eta_u m < \eta_u j} \left\| X_{\hat{\tau}_u}^{NV,\eta} - X_{\hat{\tau}_u}^{GS} \right\|^{2p} \left|\ \Delta W^m_{\check{\tau}_u}\right|^{2p} +  \sum \limits_{\eta_u m > \eta_u j} \left\| X_{\hat{\tau}_u}^{NV,-\eta} - X_{\hat{\tau}_u}^{GS} \right\|^{2p} \left|\ \Delta W^m_{\check{\tau}_u}\right|^{2p} \Bigg) .
\label{Phi_}
\end{split}
\end{equation}
By independence
\begin{equation*}
\begin{split}
  \mathbb{E}\left[ \left\| X_{\hat{\tau}_u}^{NV,\eta} - X_{\hat{\tau}_u}^{GS} \right\|^{2p} \left|\ \Delta W^m_{\check{\tau}_u}\right|^{2p} \Bigg | \eta \right] & =  \mathbb{E}\left[ \left\| X_{\hat{\tau}_u}^{NV,\eta} - X_{\hat{\tau}_u}^{GS} \right\|^{2p} \Bigg | \eta \right]  \mathbb{E}\left[  \left|\ \Delta W^m_{\check{\tau}_u}\right|^{2p} \right]
\\ &  \leq  2^{2p-1} \mathbb{E}\left[  \left|\ \Delta W^m_{\check{\tau}_u}\right|^{2p} \right]   \Bigg( \mathbb{E}\left[ \left\| X_{\hat{\tau}_u}^{NV,\eta} - X_{\hat{\tau}_u} \right\|^{2p} \Bigg | \eta \right]  + \mathbb{E}\left[ \left\| X_{\hat{\tau}_u} - X_{\hat{\tau}_u}^{GS} \right\|^{2p} \Bigg | \eta \right]  \Bigg). 
\end{split}
\end{equation*}
Then, using Theorem \ref{SC_NV} and Lemma \ref{GS_Order}, we get
\begin{equation*}
  \mathbb{E}\left[ \left\| X_{\hat{\tau}_u}^{NV,\eta} - X_{\hat{\tau}_u}^{GS} \right\|^{2p} \left|\ \Delta W^m_{\check{\tau}_u}\right|^{2p} \Bigg | \eta \right] \leq 2^{2p} ~\mathbb{E}\left[ \left| G\right|^{2p} \right] \left(C_{NV}\left(1+\left\|x\right\|^{2p}\right) + C_{GS} \right) h^{2p}
\end{equation*}
where $G$ is a normal random variable. Using the same approach, we get the same result for the other terms in the in the right-hand side of \eqref{Phi_}. Thus, we deduce that there exists a constant $\alpha_3 \in \mathbb{R}_+^*$ such that:
\begin{equation*}
 \mathbb{E}\left[  \left\| \Phi^j_u \right\|^{2p} \Bigg | \eta \right] \leq \alpha_3 h^{2p}.
\end{equation*}
Combining our different inequalities, we obtain
\begin{equation}
\label{Step3}
I_j \leq \beta_3 h^{2p} 
\end{equation}
where $\beta_3 = \frac{1}{2} KT^p \left(\alpha_3 + 2^{2p} \gamma_3\right)$.\\\\
\textbf{Step 4: conclusion}\\\\
Finally, by combining \eqref{Step1}, \eqref{Step2}, \eqref{Step3}, together with \eqref{Estimation_Coupling},  we complete the proof using Gronwall's lemma
\begin{equation*}
 \mathbb{E}\left[ \underset{t\leq T}{\sup}\left\| \bar{X}^{NV}_t - X^{GS}_{t} \right\|^{2p} \Bigg | \eta \right] \leq Ch^{2p} 
\end{equation*}
where $C=  3^{2p-1} \left(d+1\right)^{2p-1} \left(d \beta_1 + (d+1)\beta_2 + \left(d+2\right) \beta_3\right) \exp\left(3^{2p-1} \left(d+1\right)^{2p-1} d \beta_1 T \right)$.
\end{adem}

\section{Multilevel methods for SDEs} 
In this section, we are interested in the computation, by Monte Carlo methods, of the expectation $Y= \mathbb{E}\left[ f\left(X_T\right)\right]$, where $X = \left(X_t\right)_{t\in[0,T]}$ is the solution of the stochastic differential equation \eqref{EDS_ITO} and $f : \mathbb{R}^n \mapsto \mathbb{R}$ a given function such that $\mathbb{E}\left[ f\left(X_T\right)^2\right]$ is finite. We will focus on minimizing the computational complexity subject to a given target error $\epsilon$. 
To measure the accuracy of an estimator $\hat{Y}$, we will consider the root mean square error
\begin{equation*}
RMSE\left(\hat{Y},Y\right) = \mathbb{E}^{\frac{1}{2}}\left[ \left| Y - \hat{Y} \right|^2\right].
\end{equation*}
\subsection{Multilevel Monte Carlo}
The multilevel Monte Carlo method, introduced by Giles in \cite{Giles}, consists in combining multiple levels of discretization, using a geometric sequence of time steps $h_l = T/2^l$ for example. Denoting by $X^N$ a numerical scheme, with time step $T/N$, the main idea of this technique is to use the following telescopic summation to control the bias 
\begin{equation*}
\mathbb{E}\left[f\left(X^{2^L}_T\right) \right] = \mathbb{E}\left[f\left(X^{1}_T\right) \right] + \sum \limits_{l=1}^L \mathbb{E}\left[f\left(X^{2^l}_T\right) - f\left(X^{2^{l-1}}_T\right)\right].
\end{equation*}
Then, a generalized multilevel Monte Carlo estimator is built as follows
\begin{equation}
\hat{Y}_{MLMC} = \sum \limits_{l=0}^L \frac{1}{M_l} \sum \limits_{k=1}^{M_l} Z^l_k
\label{MLMC}
\end{equation}
where $\left(Z^l_k\right)_{0\leq l\leq L, 1\leq k \leq M_l}$ are independent random variables such that for, a given discretization level $l \in \left\{0,\ldots,L\right\}$, the sequence $\left(Z^l_k\right)_{1\leq k \leq M_l}$ is identically distributed and satisfies
\begin{equation}
\label{RQ1}
\mathbb{E}\left[ Z^0 \right] = \mathbb{E}\left[f\left(X^1_T\right) \right]
\end{equation}
and
\begin{equation}
\label{RQ2}
\forall l \in \left\{1,\ldots,L\right\},  \mathbb{E}\left[ Z^l \right] =\mathbb{E}\left[f\left(X^{2^l}_T\right) - f\left(X^{2^{l-1}}_T\right)\right].
\end{equation}
Assume that, for a given discretization level $l \in \left\{0,\ldots,L\right\}$, the computational cost of simulating one sample $Z^l$ is $ C \lambda_l 2^l$, where $C \in \mathbb{R}_+$ is a constant, depending only on the discretization scheme and  $\lambda_l \in \mathbb{Q}^*_+$ is a weight, depending only on $l$. The computational complexity of $\hat{Y}_{MLMC}$, denoted by $\mathcal{C}_{MLMC}$, is given by
\begin{equation}
\mathcal{C}_{MLMC} = C \sum \limits_{l=0}^L M_l \lambda_l 2^l.
\end{equation}
The natural choice for $Z^l, l \in \left\{0,\ldots,L\right\}$ considered in \cite{Giles} is
\begin{equation}
Z^0 = f\left(X^{1}_T\right)
\end{equation}  	
\begin{equation}
Z^l = f\left(X^{2^l}_T\right) - f\left(X^{2^{l-1}}_T\right), \forall l \in \left\{1,\ldots,L\right\}.
\end{equation}  	
For this canonical choice,  it is natural to take $\lambda_0 = 1$ and $\lambda_l =3/2$, $\forall l \in \left\{1,\ldots,L\right\}$. According to Theorem 3.1 in \cite{Giles} the optimal complexity $\mathcal{C}^*_{MLMC}$, depends on the order $\alpha$ of weak convergence of the scheme and the order $\beta$ of convergence to $0$ of the variance of $Z^l$. Here, we recall this complexity theorem.
\begin{thm}
\label{Complexity}
Assume that 
\begin{equation}
\mathbb{E}\left[ f\left(X^{2^l}_T\right) \right] - Y  = \frac{c_1}{2^{\alpha l}} + o\left(\frac{1}{2^{\alpha l}} \right)
\label{Biais}
\end{equation}
and
\begin{equation}
\mathbb{V}\left( Z^l \right)  = \frac{c_2}{2^{\beta l}} + o\left(\frac{1}{2^{\beta l}} \right)
\label{Var}
\end{equation}
for some constants $c_1 \in \mathbb{R}^* $ and $c_2 \in \mathbb{R}_+^*$ independent of $l$. 
Then, by choosing:
\begin{equation}
\label{LMax}
L^* = \left \lceil \frac{\log_2\left(\frac{\sqrt{2}\left|c_1 \right| }{\epsilon} \right)}{\alpha} \right \rceil 
\end{equation}
and  
\begin{equation}
\label{MLMC_Sample}
\forall l \in \left\{0,\ldots,L^*\right\}, M^*_l = \left \lceil \frac{2}{\epsilon^2} \sqrt{\frac{\mathbb{V}\left( Z^l \right)}{\lambda_l 2^l}} \sum \limits_{j=0}^{L^*} \sqrt{\lambda_j 2^j \mathbb{V}\left( Z^j \right)}  \right \rceil 
\end{equation}
we get an optimal computational complexity:
\begin{equation}
  \left\{
      \begin{aligned}
        \mathcal{C}^*_{MLMC} &=  O\left( \epsilon^{-2}\right) \text{   if } \beta > 1\\
        \mathcal{C}^*_{MLMC} &=  O\left( \epsilon^{-2} \left(\log\left(\frac{1}{\epsilon}\right)\right)^2\right) \text{   if } \beta = 1\\
        \mathcal{C}^*_{MLMC} &= O\left( \epsilon^{-2+ \frac{\beta -1}{\alpha}}\right) \text{   if } \beta < 1
      \end{aligned}
    \right.
\end{equation}
with $RMSE\left(\hat{Y}_{MLMC},Y\right)$ bounded by $\epsilon$. 
\end{thm}
To obtain the estimation \eqref{Var}, the key point is that the simulation of $X^{2^l}$ and $X^{2^{l-1}}$ comes from the same Brownian path. We easily bound the variance convergence rate from below using the strong convergence rate $\gamma$ of the numerical scheme, since in general, $\beta \ge 2 \gamma$  for a smooth payoff.    
To attain $\gamma = 1$, one has, in general, to simulate iterated Brownian integrals involving L\'evy areas, for which there is no known efficient method. To get around this difficulty, Giles and Szpruch introduced a Milstein scheme without  L\'evy areas and its antithetic version by swapping the Brownian increments. In a multilevel Monte Carlo method, using the arithmetic average of the modified Milstein scheme and its antithetic version in the finest grid, and the modified Milstein scheme in the coarsest grid leads to $\beta = 2$.  
By this way, Giles and Szpruch managed to improve the variance convergence rate without simulating the L\'evy areas. To be precise, they choose $Z^l$ as follows
\begin{equation}
\label{ZGS0}
Z_{GS}^0 = \ f\left(X^{GS,1}_T\right)
\end{equation}  	
\begin{equation}
\label{ZGS}
Z_{GS}^l = \frac{1}{2} \left( f\left(\tilde{X}^{GS,2^l}_T\right) + f\left(X^{GS,2^l}_T\right) \right) - f\left(X^{GS,2^{l-1}}_T\right), \forall l \in \left\{1,\ldots,L\right\}.
\end{equation}  	
Here, $X^{GS,2^l}$ is the Giles and Szpruch scheme defined by \eqref{GS-Scheme} using a grid with time step $h_l = T/2^l$ and $\tilde{X}^{GS,2^l}$ is an antithetic discretization defined by swapping each successive pair of Brownian increments in the scheme.
To be more precise, we define two grids, a coarse grid with time step $h_{l-1}$ and a fine grid with time step $h_{l}$. The discretization times $\left(t_k\right)_{0\leq k \leq 2^{l-1}}$ and $\left(t_{k+\frac12}\right)_{0\leq k \leq 2^{l-1}-1}$ are defined by $t_k = k h_{l-1}, \forall k \in \left\{0,\ldots,2^{l-1}\right\},$ and $t_{k+\frac12} = \left(k+\frac12\right) h_{l-1}, \forall k \in \left\{0,\ldots,2^{l-1}-1\right\}$. 
Then, on the coarsest grid, $\left(X_{t_{k+1}}^{GS,2^{l-1}}\right)_{k \in \left\{0,\ldots,2^{l-1}\right\}}$ is defined inductively by $X_{t_{0}}^{GS,2^{l-1}} =  x $ and
\begin{equation}
\begin{split}
X_{t_{k+1}}^{GS,2^{l-1}} &= X_{t_{k}}^{GS,2^{l-1}} + b\left(X_{t_{k}}^{GS,2^{l-1}} \right) h_{l-1} + \sum \limits_{j=1}^d \sigma^j\left(X_{t_{k}}^{GS,2^{l-1}} \right) \Delta W^{j,c}_{t_{k+1}}\\
& + \frac{1}{2} \sum \limits_{j,m=1}^d \partial \sigma^j \sigma^m \left(X_{t_k}^{GS,2^{l-1}} \right)\left(\Delta W^{j,c}_{t_{k+1}} \Delta W^{m,c}_{t_{k+1}} - \mathds{1}_{\left\{m=j\right\}} h_{l-1} \right) 
\end{split}
\label{Giles-S-C}
\end{equation}
where $\Delta W^c_{t_{k+1}} = W_{t_{k+1}} - W_{t_{k}}$. Similarly, on the finest grid, $\left(X_{t_{k+1}}^{GS,2^{l}}\right)_{k \in \left\{0,\ldots,2^{l-1}\right\}}$ is defined inductively by $X_{t_{0}}^{GS,2^{l}} =  x $ and
\begin{equation}
\left\{
    \begin{array}{ll}
X_{t_{k+\frac12}}^{GS,2^{l}} &= X_{t_{k}}^{GS,2^{l}} + b\left(X_{t_{k}}^{GS,2^{l}} \right) h_{l} + \sum \limits_{j=1}^d \sigma^j\left(X_{t_{k}}^{GS,2^{l}} \right) \Delta W^{j,f}_{t_{k+\frac12}}\\
& + \frac{1}{2} \sum \limits_{j,m=1}^d \partial \sigma^j \sigma^m \left(X_{t_k}^{GS,2^{l}} \right)\left(\Delta W^{j,f}_{t_{k+\frac12}} \Delta W^{m,f}_{t_{k+\frac12}} - \mathds{1}_{\left\{m=j\right\}} h_{l} \right) \\
X_{t_{k+1}}^{GS,2^{l}} &= X_{t_{k+\frac12}}^{GS,2^{l}} + b\left(X_{t_{k+\frac12}}^{GS,2^{l}} \right) h_{l} + \sum \limits_{j=1}^d \sigma^j\left(X_{t_{k+\frac12}}^{GS,2^{l}} \right)\Delta W^{j,f}_{t_{k+1}}\\
& + \frac{1}{2} \sum \limits_{j,m=1}^d \partial \sigma^j \sigma^m \left(X_{t_{k+\frac12}}^{GS,2^{l}} \right)\left(\Delta W^{j,f}_{t_{k+1}} \Delta W^{m,f}_{t_{k+1}} - \mathds{1}_{\left\{m=j\right\}} h_{l} \right) 
\end{array}
\right.
\label{Giles-S-F}
\end{equation}
where $\Delta W^{f}_{t_{k+\frac12}} = W_{t_{k+\frac12}}  - W_{t_{k}}$, $\Delta W^{f}_{t_{k+1}} =  W^{f}_{t_{k+1}} -  W^{f}_{t_{k+\frac12}}$. The antithetic scheme is defined by the same iterative equations, except that the Brownian increment $\Delta W^{f}_{t_{k+\frac12}}$ and $\Delta W^{f}_{t_{k+1}}$ are swapped
\begin{equation}
\left\{
    \begin{array}{ll}
\tilde{X}_{t_{k+\frac12}}^{GS,2^{l}} &= \tilde{X}_{t_{k}}^{GS,2^{l}} + b\left(\tilde{X}_{t_{k}}^{GS,2^{l}} \right) h_{l} + \sum \limits_{j=1}^d \sigma^j\left(\tilde{X}_{t_{k}}^{GS,2^{l}} \right)\Delta W^{j,f}_{t_{k+1}}\\
& + \frac{1}{2} \sum \limits_{j,m=1}^d \partial \sigma^j \sigma^m \left(\tilde{X}_{t_k}^{GS,2^{l}} \right)\left(\Delta W^{j,f}_{t_{k+1}} \Delta W^{m,f}_{t_{k+1}} - \mathds{1}_{\left\{m=j\right\}} h_{l} \right) \\
\tilde{X}_{t_{k+1}}^{GS,2^{l}} &= \tilde{X}_{t_{k+\frac12}}^{GS,2^{l}} + b\left(\tilde{X}_{t_{k+\frac12}}^{GS,2^{l}} \right) h_{l} + \sum \limits_{j=1}^d \sigma^j\left(\tilde{X}_{t_{k+\frac12}}^{GS,2^{l}} \right)\Delta W^{j,f}_{t_{k+\frac12}}\\
& + \frac{1}{2} \sum \limits_{j,m=1}^d \partial \sigma^j \sigma^m \left(\tilde{X}_{t_{k+\frac12}}^{GS,2^{l}} \right)\left(\Delta W^{j,f}_{t_{k+\frac12}} \Delta W^{m,f}_{t_{k+\frac12}} - \mathds{1}_{\left\{m=j\right\}} h_{l} \right). 
\end{array}
\right.
\label{Giles-S-A}
\end{equation}
Theorem 4.10, Lemma 2.2 and Lemma 4.6 in \cite{GS} ensure that $\beta = 2$ under some regularity assumptions on $f$ and the coefficients of the SDE.  
\begin{thm}
\label{AGS_thm}
Assume that $f \in \mathcal{C}^{2}\left(\mathbb{R}^n,\mathbb{R}\right) $ with bounded first and second order derivatives, $b,\sigma^j \in \mathcal{C}^{2}\left(\mathbb{R}^n,\mathbb{R}^n\right), \forall j \in \left\{1,\ldots,d\right\},$ with bounded first and second order derivatives, and that $\partial \sigma^j \sigma^m$, $\forall j,m \in \left\{1,\ldots,d\right\}$, has bounded first order derivatives. Then:
\begin{equation*}
\forall p \ge 1, \exists c \in \mathbb{R}_+^*, \forall l \in \mathbb{N}^*, ~ \mathbb{E}\left[ \left| Z_{GS}^l \right|^{2p} \right] \leq \frac{c}{2^{2pl}}
\end{equation*}
where $Z^l_{GS}$ is defined by \eqref{ZGS}.
\end{thm}
To account for the use of three schemes in the levels $l \in \left\{1,\ldots,L^*\right\}$ instead of one in level $0$ we choose $\lambda_0 = 1$ and $\lambda_l = 5/2, \forall l \in \left\{1,\ldots, L^*\right\}$ . 
Then, the multilevel Monte Carlo estimator $\hat{Y}_{MLMC}^{GS} = \sum \limits_{l=0}^{L^*} \frac{1}{M_l^*} Z^{l}_{GS}$, where $L^*$ and $M_l^*$ are given by \eqref{LMax} and \eqref{MLMC_Sample}, respectively, achieves a complexity $O\left( \epsilon^{-2}\right)$. In \cite{DR}, Debrabant R\"ossler improved the multilevel Monte Carlo method by using, in the last level $L$, a scheme with high order of weak convergence. Although this modified method attains the same complexity, it reduces the computation time by reducing the bias. We can follow this idea using the Ninomiya-Victoir scheme at the last level $L$, thereby taking advantage of its order 2 of weak convergence.  More precisely, we propose to choose
\begin{equation}
Z_{GS}^0 = f\left(X^{GS,1}_T\right)
\end{equation}  	
\begin{equation}
Z_{GS}^l = \frac{1}{2} \left( f\left(\tilde{X}^{GS,2^l}_T\right) + f\left(X^{GS,2^l}_T\right) \right) - f\left(X^{GS,2^{l-1}}_T\right), \forall l \in \left\{1,\ldots,L-1\right\}
\end{equation}  	
\begin{equation}
\label{ZGSNV}
Z_{GS-NV}^L = \frac{1}{4} \left( f\left(\tilde{X}^{NV,2^L,\eta}_T\right)  + f\left(\tilde{X}^{NV,2^L,-\eta}_T\right) + f\left(X^{NV,2^L,\eta}_T\right) + f\left(X^{NV,2^L,-\eta}_T\right) \right) -  f\left(X^{GS,2^{L-1}}_T\right).
\end{equation}  	
Here, $\tilde{X}^{NV,2^L,\eta}$ (respectively $\tilde{X}^{NV,2^L,-\eta}$) is the antithetic discretization of the Ninomiya-Victoir scheme $X^{NV,2^L,\eta}$ (respectively $X^{NV,2^L,-\eta}$), obtained by swapping each successive pair of Brownian increments.
Theorem \ref{Coupling} ensures that \eqref{Var} the order of convergence of the variance at the last level $L$ is $2$.
\begin{aprop}
\label{AGS-ANV}
We assume that $f \in \mathcal{C}^{2}\left(\mathbb{R}^n,\mathbb{R}\right) $ with bounded first and second order derivatives, $b \in \mathcal{C}^{2}\left(\mathbb{R}^n,\mathbb{R}^n\right)$ with bounded first and second order derivatives, $\sigma^j \in \mathcal{C}^3\left(\mathbb{R}^n,\mathbb{R}^n\right)$, $ \forall j \in \left\{1,\ldots,d\right\}$, with bounded first and second order derivatives and with polynomially growing third order derivatives, and that $\partial \sigma^j \sigma^m$, $\forall j,m \in \left\{1,\ldots,d\right\}$, has bounded first order derivatives.  
Then:
\begin{equation*}
\forall p \ge 1, \exists c \in \mathbb{R}_+^*, \forall l \in \mathbb{N}^*, ~ \mathbb{E}\left[ \left| Z_{GS-NV}^l \right|^{2p} \right] \leq \frac{c}{2^{2pl}}
\end{equation*}
where $Z^l_{GS-NV}$ is defined by \eqref{ZGSNV}.
\end{aprop}
\begin{adem}
Let $p \ge 1$, introducing $\frac{1}{2} \left( f\left(\tilde{X}^{GS,2^l}_T\right) + f\left(X^{GS,2^l}_T\right) \right)$ and using a convexity inequality, we get
\begin{equation*}
\begin{split}
\left| Z_{GS-NV}^l \right|^{2p} &\leq \frac{3^{2p-1}}{2^{2p}} \Bigg( \left| \frac{1}{2} \left(f\left(X^{NV,2^l,\eta}_T\right)  + f\left(X^{NV,2^l,-\eta}_T\right)  \right) -  f\left(X^{GS,2^l}_T\right) \right|^{2p}\\
& + \left| \frac{1}{2} \left(f\left(\tilde{X}^{NV,2^l,\eta}_T\right)  + f\left(\tilde{X}^{NV,2^l,-\eta}_T\right)  \right) -  f\left(\tilde{X}^{GS,2^l}_T\right) \right|^{2p} \Bigg)  +  3^{2p-1} \left| Z_{GS}^l \right|^{2p}. 
\end{split}
\end{equation*}
However, $\left(X^{NV,2^l,\eta}_T, X^{NV,2^l,-\eta}_T,X^{GS,2^l}_T \right)$ and $\left(\tilde{X}^{NV,2^l,\eta}_T,\tilde{X}^{NV,2^l,-\eta}_T,\tilde{X}^{GS,2^l}_T\right)$ have exactly the same distribution. Then, by taking the expectation we obtain
\begin{equation*}
\begin{split}
\mathbb{E}\left[\left| Z_{GS-NV}^l \right|^{2p}\right] &\leq \frac{3^{2p-1}}{2^{2p-1}} \mathbb{E}\left[ \left| \frac{1}{2} \left(f\left(X^{NV,2^l,\eta}_T\right)  + f\left(X^{NV,2^l,-\eta}_T\right)  \right) -  f\left(X^{GS,2^l}_T\right) \right|^{2p}\right]
+ 3^{2p-1} \mathbb{E}\left[ \left| Z_{GS}^l \right|^{2p} \right].
\end{split}
\end{equation*}
Denoting $\bar{X}^{NV,2^l,\eta}_T = \frac{1}{2}\left( X^{NV,2^l,\eta}_T +  X^{NV,2^l,-\eta}_T\right) $ and performing a second order Taylor expansion as in Lemma 2.2 in \cite{GS}, we get a constant $C \in \mathbb{R}_+^*$, which only depends on $f$ and $p$, such that
 \begin{equation*}
\begin{split}
\mathbb{E}\left[\left| Z_{GS-NV}^l \right|^{2p}\right] &\leq C \left(\mathbb{E}\left[ \left\| \bar{X}^{NV,2^l,\eta}_T -  X^{GS,2^l}_T \right\|^{2p}\right] + \mathbb{E}\left[ \left\| X^{NV,2^l,\eta}_T - X^{NV,2^l,-\eta}_T \right\|^{4p}\right]
+ \mathbb{E}\left[ \left| Z_{GS}^l \right|^{2p} \right] \right).
\end{split}
\end{equation*}
Introducing the exact solution $X$ at time $T$ in $\mathbb{E}\left[ \left\| X^{NV,2^l,\eta}_T - X^{NV,2^l,-\eta}_T \right\|^{4p}\right]$, we get
 \begin{equation*}
\begin{split}
\mathbb{E}\left[ \left\| X^{NV,2^l,\eta}_T - X^{NV,2^l,-\eta}_T \right\|^{4p}\right] &\leq 2^{4p-1}\left( \mathbb{E}\left[ \left\| X^{NV,2^l,\eta}_T - X_T \right\|^{4p}\right] + \mathbb{E}\left[ \left\| X_T  - X^{NV,2^l,-\eta}_T \right\|^{4p}\right]\right). 
\end{split}
\end{equation*}
Since $\left(X^{NV,2^l,\eta}_T, X_T \right)$ and $\left(X^{NV,2^l,-\eta}_T,X_T\right)$ have the same distribution, we deduce that 
 \begin{equation*}
\begin{split}
\mathbb{E}\left[ \left\| X^{NV,2^l,\eta}_T - X^{NV,2^l,-\eta}_T \right\|^{4p}\right] & \leq  2^{4p} \mathbb{E}\left[ \left\| X^{NV,2^l,\eta}_T - X_T \right\|^{4p}\right]. 
\end{split}
\end{equation*}
Hence:
\begin{equation*}
\begin{split}
\mathbb{E}\left[\left| Z_{GS-NV}^l \right|^{2p}\right] &\leq 2^{4p} C \left(\mathbb{E}\left[ \left\| \bar{X}^{NV,2^l,\eta}_T -  X^{GS,2^l}_T \right\|^{2p}\right] + \mathbb{E}\left[ \left\| X^{NV,2^l,\eta}_T - X_T \right\|^{4p}\right]
+ \mathbb{E}\left[ \left| Z_{GS}^l \right|^{2p} \right] \right).
\end{split}
\end{equation*}
Then we conclude using Theorems \ref{SC_NV}, \ref{Coupling} and \ref{AGS_thm}. 
\end{adem}
Exploiting the telescoping summation, one can change the constraint \eqref{RQ2} on the last level $L$ and assume:
\begin{equation}
\label{RQ2_DR}
\mathbb{E}\left[ Z^L \right] =\mathbb{E}\left[f\left(\hat{X}^{2^L}_T\right) - f\left(X^{2^{L-1}}_T\right)\right].
\end{equation}
Here $\hat{X}$ is an other scheme, and to be consistent, \eqref{Biais} becomes
\begin{equation}
\mathbb{E}\left[ f\left(\hat{X}^{2^l}_T\right) \right] - Y  = \frac{c_1}{2^{\alpha l}} + o\left(\frac{1}{2^{\alpha l}} \right).
\end{equation}
Then we propose to use the estimator $\hat{Y}_{MLMC}^{GS-NV} = \sum \limits_{l=0}^{L-1} \frac{1}{M_l} \sum \limits_{k=1}^{M_l} Z_{GS}^{l,k} + \frac{1}{M_L} \sum \limits_{k=1}^{M_L} Z_{GS-NV}^{L,k}$ . Of course, the bias of this estimator is given by the bias of the Ninomiya-Victoir scheme. Thanks to its weak order 2, we hope to decrease the value of $L$, and so to reduce the computation time.
We can also use the Ninomiya-Victoir scheme at each level and choose $\left(Z^l_{NV}\right)_{0\leq l \leq L}$ as follows
\begin{equation}
\label{ZNV0}
Z_{NV}^0 =  f\left(X^{NV,1,\eta}_T\right) 
\end{equation}
or  	
\begin{equation}
\label{ZNV00}
Z_{NV}^0 = \frac{1}{2} \left( f\left(X^{NV,1,\eta}_T\right)  + f\left(X^{NV,1,-\eta}_T\right) \right)
\end{equation}  	
and
\begin{equation}
\begin{split}
Z_{NV}^l &= \frac{1}{4} \left( f\left(\tilde{X}^{NV,2^l,\eta}_T\right)  + f\left(\tilde{X}^{NV,2^l,-\eta}_T\right) + f\left(X^{NV,2^l,\eta}_T\right) + f\left(X^{NV,2^l,-\eta}_T\right) \right) \\
& -  \frac{1}{2} \left(f\left(X^{NV,2^{l-1},\eta}_T\right) + f\left(X^{NV,2^{l-1},-\eta}_T\right) \right),\forall l \in \left\{1,\ldots,L\right\}.
\end{split}
\label{ZNV}
\end{equation}  	
Actually, there is an abuse of notation in \eqref{ZNV}, we use the same notation $\eta$ for the $2^l$-dimensional vector $\left(\eta_1,\ldots,\eta_{2^l}\right)$ of the independent and identically distributed Rademacher random variables needed to generate the Ninomiya-Victoir scheme on the fine grid with $2^l$ steps and for the $2^{l-1}$-dimensional subvector $\left(\eta_1,\eta_3 \ldots,\eta_{2^{l}-1}\right)$ used to generate the Ninomiya-Victoir scheme on the coarse grid with $2^{l-1}$ steps. The extraction of the $2^{l-1}$-dimensional vector from the $2^{l}$-dimensional one is aimed at reducing the variance.
As previously, we obtain the same rates $\alpha$ and $\beta$, but the main drawback is the simulation of six schemes at each level $l \in \left\{1,\ldots,L-1\right\}$ instead of three. Reasoning like in the proof of Proposition \ref{AGS-ANV}, since
\begin{equation*}
Z^l_{NV} = Z^l_{GS-NV}  + f\left( \bar{X}^{NV,2^{l-1},\eta}_T\right)  -  \frac{1}{2} \left(f\left(X^{NV,2^{l-1},\eta}_T\right)  + f\left(X^{NV,2^{l-1},-\eta}_T\right)  \right) +  f\left(X^{GS,2^{l-1}}_T\right) - f\left( \bar{X}^{NV,2^{l-1},\eta}_T\right) 
\end{equation*}
one obtains:
\begin{aprop}
\label{ANV}
Assume that $f \in \mathcal{C}^{2}\left(\mathbb{R}^n,\mathbb{R}\right) $ with bounded first and second order derivatives,  $b \in \mathcal{C}^{2}\left(\mathbb{R}^n,\mathbb{R}^n\right)$ with bounded first and second order derivatives, $\sigma^j \in \mathcal{C}^3\left(\mathbb{R}^n,\mathbb{R}^n\right)$ ,$\forall j \in \left\{1,\ldots,d\right\}$, with bounded first and second order derivatives and with polynomially growing third order derivatives, and that $\partial \sigma^j \sigma^m$,$\forall j,m \in \left\{1,\ldots,d\right\}$, has bounded first order derivatives.   
Then:
\begin{equation*}
\forall p \ge 1, \exists c \in \mathbb{R}_+^*, \forall l \in \mathbb{N}^*, ~ \mathbb{E}\left[ \left| Z_{NV}^l \right|^{2p} \right] \leq \frac{c}{2^{2pl}}
\end{equation*}
where $Z^l_{NV}$ is defined by \eqref{ZNV}.
\end{aprop}
\subsection{Multilevel Richardson-Romberg extrapolation}
Recently, in \cite{PL} Lemaire and Pag\`es developed a new method called multilevel Richardson-Romberg extrapolation (ML2R). This method combines the ideas behind the multilevel Monte Carlo approach and the multi-step Richardson-Romberg extrapolation introduced in \cite{Pages}. Actually, the multilevel Richardson-Romberg extrapolation can be seen as a weighted version of the multilevel Monte Carlo estimator.    
Adapting the notation of Lemaire and Pag\`es \cite{PL}, the multilevel Richardson-Romberg extrapolation estimator is built as follows
\begin{equation}
\label{ML2R}
\hat{Y}_{ML2R} = \sum \limits_{l=0}^L \frac{W_l}{M_l} \sum \limits_{k=0}^{M_l} Z^l_{k}
\end{equation}
where $\left(Z^l_{k}\right)_{0\leq l\leq L, 1\leq k \leq M_l}$ are independent random variables satisfying \eqref{RQ1}, \eqref{RQ2} and a bias error expansion
\begin{equation}
\label{Bias}
\exists \alpha \in \mathbb{R}_+^*, \exists R \in \mathbb{N}^*, \exists c^{\prime}_1,\ldots,c^{\prime}_R \in \mathbb{R}, \forall l \in \mathbb{N},  \mathbb{E}\left[ f\left(X^{2^l}_T\right)\right] - Y = \sum \limits_{j=1}^{R} c^{\prime}_j h_l^{\alpha j} + O\left(h_l^{\alpha \left(R+1\right)}\right) 
\end{equation}
where $h_l = T/2^l$ is the time step. As previously, $\alpha$ is the order of weak convergence of the discretization scheme. By introducing the weights $\left( W_l \right)_{0\leq l \leq L}$, one can get a smaller bias\footnote{See \cite{PL} and \cite{Pages} for more details.} by canceling the successive bias terms in the expansion \eqref{Bias}.  
Following \cite{PL}, the computational complexity of $\hat{Y}_{ML2R}$, denoted by $\mathcal{C}_{ML2R}$ is defined as $\mathcal{C}_{MLMC}$, except that we do not take into account the weights $\left(\lambda_l\right)_{0\leq l \leq L}$.  
Under some assumptions (see \cite{PL} for further information), the optimal complexity $\mathcal{C}^*_{ML2R}$ is given by Theorem 3.11 in \cite{PL}, which states that $\mathcal{C}^*_{ML2R}$ depends on $\alpha$, and the variance convergence rate \footnote{In \cite{PL}, Lemaire and G. Pag\`es assume that: $\forall l \in \mathbb{N}^*, \exists V_1 \in \mathbb{R}_+, \mathbb{E}\left[\left\| f\left(X^{2^l}_T\right) - f\left(X_T\right)\right\|^2 \right] \leq V_1 h_l^{\beta} $. One can easily adapt the proof with assumption \eqref{BetaNew}.} of $Z^l$, denoted as previously by $\beta$
\begin{equation}
\exists c_2 \in \mathbb{R}_+ , \forall l \in \mathbb{N}^*,   \mathbb{V}\left(Z^l \right) \leq \frac{c_2}{2^{\beta l}}. 
\label{BetaNew}
\end{equation}
\begin{itemize}
\item $\mathcal{C}^*_{ML2R} =  O\left( \epsilon^{-2}\right) \text{   if } \beta > 1$, 
 \item $  \mathcal{C}^*_{ML2R} =  O\left( \epsilon^{-2} \log\left(\frac{1}{\epsilon}\right)\right) \text{   if } \beta = 1$,
 \item  $  \mathcal{C}^*_{ML2R} = O\left( \epsilon^{-2 } \exp\left(- \frac{\beta -1}{\sqrt{\alpha}}\sqrt{2\log\left(2\right)  \log\left(\frac{1}{\epsilon} \right)}\right)\right) \text{   if } \beta < 1.$
\end{itemize}
Similarly to the multilevel Monte Carlo method, the best complexity, obtained when $\beta > 1$, is the same as in a simple Monte Carlo method with independent and identically distributed unbiased random variables. With a view to achieving this complexity by applying Theorem \ref{AGS_thm} or Proposition \ref{ANV} we will choose $\left(Z_{GS}^l\right)_{0\leq l \leq L}$ and $\left(Z_{NV}^l\right)_{0\leq l \leq L}$ with $Z_{NV}^0 =  f\left(X^{NV,1,\eta}_T\right)$.
Here, we recall the asymptotic\footnote{When $\epsilon$ goes to $0$.} optimal parameters for the multilevel Richardson-Romberg extrapolation estimator:
\begin{equation}
\label{LMax_PL}
L^* =  \left \lfloor \sqrt{\left(\frac12 + \log_2\left(T\right)\right)^2 + \frac{2}{\alpha}\log_2\left(\frac{\sqrt{1+4\alpha}}{\epsilon}\right) } + \log_2\left(T\right) - \frac12 \right \rfloor ,
\end{equation}
\begin{equation}
M_l^* = \left \lceil q^*_l N^* \right \rceil,
\label{ML2R-ML}
\end{equation}
\begin{equation}
W_l = \sum \limits_{j=l}^{L^*} w_j,
\end{equation}
where:
\begin{equation}
w_j = (-1)^{L^*-j} ~ \frac{2^{-\frac{\alpha}{2}\left(L^*-j\right)\left(L^*-j+1\right)}}{\prod \limits_{k=1}^{j} \left(1 - 2^{-k\alpha} \right)\prod \limits_{k=1}^{L^*-j}\left(1 - 2^{-k\alpha} \right) },  \\\end{equation}
\begin{equation}
\left\{
    \begin{array}{ll}
q^*_0 \propto  \left(1 + \theta\right) \\
q^*_l \propto \theta \left| W_l \right| \frac{2^{-\frac{\beta}{2}l} + 2^{-\frac{\beta}{2}\left(l-1\right)}}{\sqrt{2^l + 2^{l-1}}}, ~ \forall l \in \left\{1,\ldots,L^*\right\} \\
\sum \limits_{l=0}^{L^*} q^*_l = 1,
\label{ML2R-QL}
\end{array}
\right.
\end{equation}
\begin{equation}
N^* = \left(1+ \frac{1}{2\alpha \left(L^*+1\right)} \right) \frac{\mathbb{V}\left(f\left(X_T\right)\right) \left(1+\theta  \left(1 + \sum \limits_{l=1}^{L^*} \left| W_l \right| \left(2^{-\frac{\beta}{2}l} + 2^{-\frac{\beta}{2}\left(l-1\right)}\right) \sqrt{2^l + 2^{l-1}} \right)\right)^2}{\epsilon^2 \left( q^*_0 + \sum \limits_{l=1}^{L^*} q^*_l \left(2^{l} + 2^{l-1} \right) \right)},
\label{ML2R-N}
\end{equation}
and
\begin{equation}
\theta = T^{-\frac{\beta}{2}} \sqrt{\frac{c_2}{\mathbb{V}\left(f\left(X_T\right)\right)}}.
\label{ML2R_end}
\end{equation}
\subsection{Numerical experiments}
In this section we present numerical tests in which we compare the multilevel Monte Carlo and the multilevel Richardson-Romberg estimators. Although we have not proved a theoretical expansion of the bias like \eqref{Bias} for the Ninomiya-Victoir and the Giles-Szpruch schemes, we will use these schemes in the multilevel Richardson-Romberg estimators (see \cite{Fujiwara} and \cite{OTV} for extrapolation methods based on the Ninomiya-Victoir scheme).
More precisely, we compare the following estimators:
\begin{itemize}
\item The multilevel Monte Carlo estimator with the Giles-Szpruch scheme
\begin{equation*}
\hat{Y}_{MLMC}^{GS} = \sum \limits_{l=0}^{L^*} \frac{1}{M^*_l} \sum \limits_{k=1}^{M^*_l} Z^{l,k}_{GS}
\end{equation*}
where $Z_{GS}^0$  and $Z_{GS}^l$ are respectively given by \eqref{ZGS0} and \eqref{ZGS}. 
\item The multilevel Monte Carlo estimator with the Ninomiya-Victoir scheme
\begin{equation*}
\hat{Y}_{MLMC}^{NV} = \sum \limits_{l=0}^{L^*} \frac{1}{M^*_l} \sum \limits_{k=1}^{M^*_l} Z^{l,k}_{NV}
\end{equation*}
where $Z_{NV}^0$  and $Z_{NV}^l$ are respectively given by \eqref{ZNV0} or\footnote{The choice of level 0 will be discussed later.} \eqref{ZNV00}  and \eqref{ZNV}.   
\item The multilevel Monte Carlo estimator with the Giles-Szpruch scheme from level 0 to level $L^*-1$, and the coupling  between the Ninomiya-Victoir and the Giles-Szpruch scheme at the last level $L^*$
\begin{equation*}
\hat{Y}_{MLMC}^{GS-NV} = \sum \limits_{l=0}^{L^*-1} \frac{1}{M^*_l} \sum \limits_{k=1}^{M^*_l} Z^{l,k}_{GS} +  \frac{1}{M^*_{L^*}} \sum \limits_{k=1}^{M^*_{L^*}} Z^{L^*,k}_{GS-NV}
\end{equation*}
where $Z_{GS-NV}^{L^*}$ is given by \eqref{ZGSNV}.
\item The multilevel Richardson-Romberg estimator with the Giles-Szpruch scheme
\begin{equation*}
\hat{Y}_{ML2R}^{GS} = \sum \limits_{l=0}^{L^*} \frac{W_l}{M^*_l} \sum \limits_{k=1}^{M^*_l} Z^{l,k}_{GS}.
\end{equation*}
\item The multilevel Richardson-Romberg estimator with the Ninomiya-Victoir scheme
\begin{equation*}
\hat{Y}_{ML2R}^{NV} = \sum \limits_{l=0}^{L^*} \frac{W_l}{M^*_l} \sum \limits_{k=1}^{M^*_l} Z^{l,k}_{NV}.
\end{equation*}
Here, $Z_{NV}^{0}$ is given by \eqref{ZNV0}.
\end{itemize}  
\subsubsection{Clark-Cameron SDE}
\begin{figure}[!ht]
 \begin{center} 
 \includegraphics[width=12.5cm, height=8.5cm]{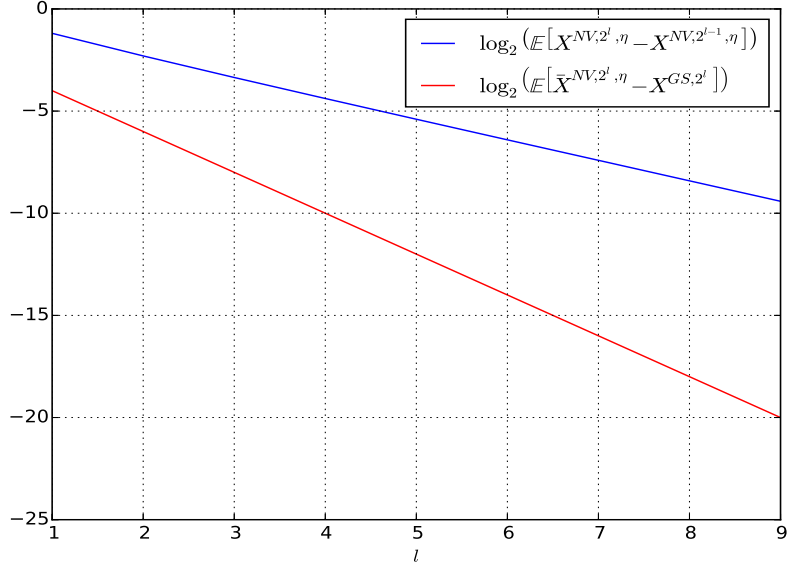}
 \caption{Strong convergence order. Strong error ($y$-axis $\log_2$ scale) as a function of $l$ ($x$-axis).}
 \label{Fig_SC}
 \end{center}
 \end{figure}

For our first numerical test, we consider the Clark-Cameron SDE with drift which is defined as follows
\begin{equation}
  \left\{
      \begin{aligned}
        dU_t &=  S_t dW_t^1\\
       dS_t &= \mu dt +  dW_t^2 
      \end{aligned}
  \right.
\end{equation}
where $\mu \in \mathbb{R}$. In this $2-$dimensional stochastic differential equation, the diffusion coefficients are given by
 $\sigma^1\begin{pmatrix} 
 u \\ 
 s 
 \end{pmatrix} = \begin{pmatrix} 
 s \\ 
 0 
 \end{pmatrix}$, $\sigma^2\begin{pmatrix} 
 u \\ 
 s 
 \end{pmatrix} = \begin{pmatrix} 
 0 \\ 
 1 
 \end{pmatrix}$ and the drift coefficient is $b \begin{pmatrix} 
 u \\ 
 s 
 \end{pmatrix} = \begin{pmatrix} 
 0 \\ 
 \mu 
 \end{pmatrix}$. The Stratonovich drift is given by 
\begin{equation*}
\sigma^0\begin{pmatrix} 
 u \\ 
 s 
 \end{pmatrix} = \left\{b - \frac12\left(\partial \sigma^1 \sigma^1 + \partial \sigma^2 \sigma^2 \right)\right\} \begin{pmatrix} 
 u \\ 
 s 
 \end{pmatrix} = \begin{pmatrix} 
 0 \\ 
 \mu 
 \end{pmatrix} - \frac{1}{2} \left(\begin{pmatrix} 
 0 & 1\\ 
 0 & 0
 \end{pmatrix} \begin{pmatrix} 
 s\\ 
 0 
 \end{pmatrix} + \begin{pmatrix} 
 0 & 0\\ 
 0 & 0
 \end{pmatrix} \begin{pmatrix} 
 0 \\ 
 1 
 \end{pmatrix} \right) = \begin{pmatrix} 
 0 \\ 
 \mu 
 \end{pmatrix}
\end{equation*} These functions are smooth and satisfy the assumptions of Theorems \ref{SC_NV}, \ref{Coupling}, Propositions \ref{AGS-ANV} and \ref{ANV}.   
By a straightforward calculation, the Giles-Szpruch scheme is given by
\begin{equation}
  \left\{
      \begin{aligned}
        U^{GS}_{t_{k+1}} &= U^{GS}_{t_{k}} + S^{GS}_{t_{k}} \left(W^1_{t_{k+1}} - W^1_{t_{k}}\right) +\frac12 \left(W^1_{t_{k+1}} - W^1_{t_{k}}\right)\left(W^2_{t_{k+1}} - W^2_{t_{k}}\right)\\
        S^{GS}_{t_{k+1}} &= S^{GS}_{t_{k}} +  \mu \left(t_{k+1} - t_{k}\right) +  \left(W^2_{t_{k+1}} - W^2_{t_{k}}\right)         
      \end{aligned}
  \right.
\end{equation}
and the Ninomiya-Victoir scheme is given by
\begin{equation}
  \left\{
      \begin{aligned}
        U^{NV,\eta}_{t_{k+1}} &= U^{NV,\eta}_{t_{k}} + S^{NV,\eta}_{t_{k}}  \left(W^1_{t_{k+1}} - W_{t_k}^1\right) + \frac12 \mu \left(t_{k+1} - t_k\right) \left(W^1_{t_{k+1}} - W^1_{t_k}\right) \\
        &+  \mathds{1}_{\left\{\eta_{{k+1}} = 1\right\}}\left(W^1_{t_{k+1}} - W_{t_k}^1\right) \left(W^2_{t_{k+1}} - W^2_{t_k}\right)\\
         S^{NV,\eta}_{t_{k+1}} &= S^{NV,\eta}_{t_{k}} + \mu \left(t_{k+1} - t_k\right) + \left(W^2_{t_{k+1}} - W^2_{t_k}\right).
      \end{aligned}
  \right.
\end{equation}
Before comparing these estimators, we will illustrate Theorems \ref{SC_NV}, \ref{Coupling}, Propositions \ref{AGS-ANV} and \ref{ANV}. In order to check the strong convergence rate of the Ninomiya-Victoir scheme, we will look at the expectation of the square $L^2$-norm of the difference, at time $T$, between the schemes with steps $h_{l}$,  and $h_{l-1}$, simulated with the same Brownian path. 
Denoting by $X^{NV,2^l,\eta}_T = \left(U^{NV,2^l,\eta}_T,S^{NV,2^l,\eta}_T \right)$ and  $X^{NV,2^{l-1},\eta}_T = \left(U^{NV,2^{l-1},\eta}_T,S^{NV,2^{l-1},\eta}_T\right)$, it follows, from Theorem \ref{SC_NV} that
\begin{equation}
\mathbb{E}\left[\left\| X^{NV,2^l,\eta}_T - X^{NV,2^{l-1},\eta}_T \right\|^2 \right] \leq \frac{c}{2^l}.
\end{equation} 
For the simulations, we choose the initial conditions $U_0 = V_0 = 0$, the final time $T = 1$ and the parameter $\mu = 1$. \\
In figure \ref{Fig_SC}, the blue line shows the behavior of $\log_2\left(\mathbb{E}\left[ \left\| X^{NV,2^l,\eta}_T - X^{NV,2^{l-1},\eta}_T\right\|^2 \right]\right)$ and the red line shows the behavior of $\log_2\left(\mathbb{E}\left[ \left\| \bar{X}^{NV,2^l,\eta}_T - X^{GS,2^{l},\eta}_T\right\|^2 \right]\right)$ as a function of the discretization level $l$.  
These expectations are estimated with a standard Monte Carlo method with $M_l = 10^6 $ samples for all $l$. This choice ensures that the confidence intervals are very tight, that is why they are not represented in our plot.  
The blue line illustrates the strong convergence order of the Ninomiya-Victoir scheme. As expected, we obtain a line with slope -1. The red line illustrates the strong convergence order of the coupling between the Ninomiya-Victoir and the Giles-Szpruch scheme. It follows, from Theorem \ref{Coupling} that
\begin{equation}
\mathbb{E}\left[\left\| \bar{X}^{NV,2^l,\eta}_T   - X^{GS,2^{l}}_T \right\|^2 \right] \leq \frac{c}{2^{2l}}.
\end{equation} 
Again, as expected, we obtain a line with slope -2. These numerical results are consistent with Theorems \ref{SC_NV} and \ref{Coupling} stated and proved in this paper.
 
\begin{figure}[!ht]
\begin{center}
\includegraphics[width=12.5cm, height=8.5cm]{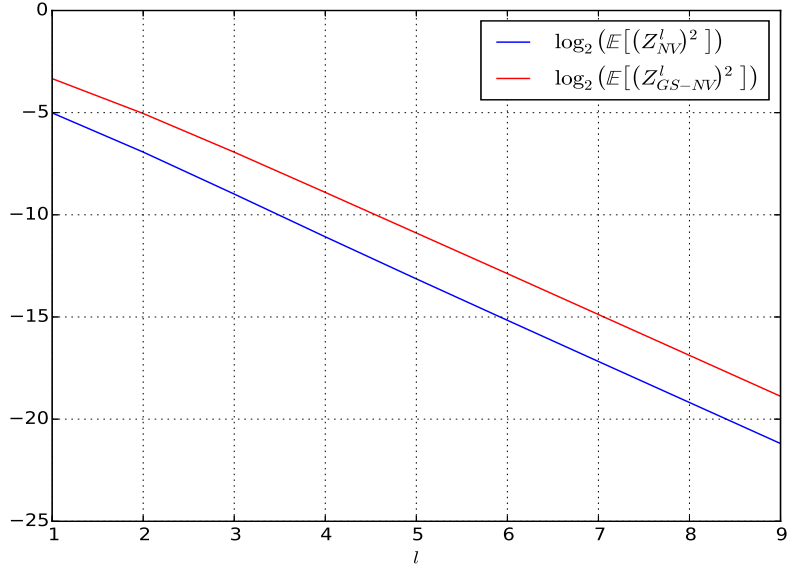}
\caption{Variance convergence order with $f(u,s) = cos(u)$. Second order moment ($y$-axis $\log_2$ scale) as a function of $l$ ($x$-axis).}
\label{Fig_SCV}
\end{center}
\end{figure}

To illustrate Propositions \ref{AGS-ANV} and \ref{ANV}, we choose a smooth payoff function, satisfying the assumptions of Propositions \ref{AGS-ANV} and \ref{ANV}: $f\left(u,s\right) = \cos(u)$. In figure \ref{Fig_SCV}, the top plot shows the behavior of $\log_2\left(\mathbb{E}\left[ \left(Z^l_{GS-NV}\right)^2 \right]\right)$ defined by \eqref{ZGSNV} whereas the bottom plot shows the behavior of $\log_2\left(\mathbb{E}\left[ \left(Z^l_{NV}\right)^2 \right]\right)$ defined by \eqref{ZNV}. Both lines have slope -2.

\begin{figure}[!ht]
\begin{center}
\includegraphics[width=12.5cm, height=8.5cm]{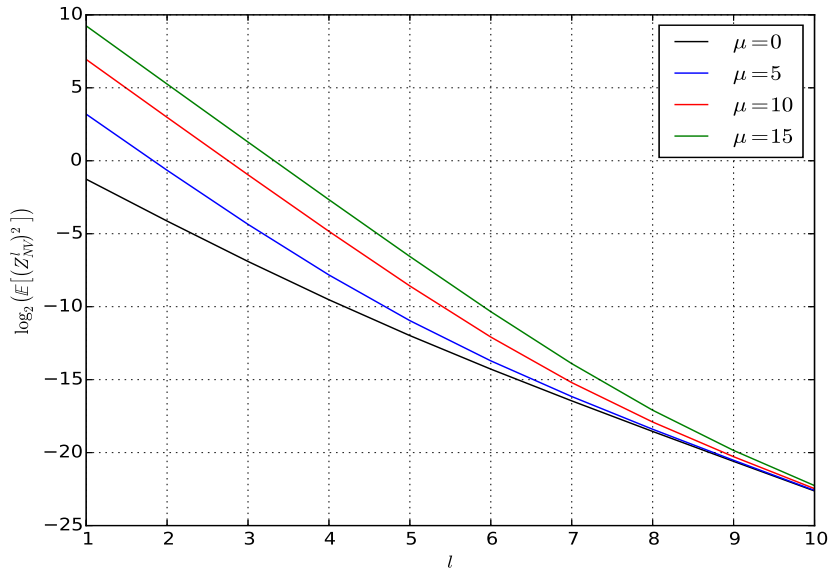}
\caption{Variance convergence order with $f(u,s) = u^2$. Second order moment ($y$-axis $\log_2$ scale) as a function of $l$ ($x$-axis).}
\label{square}
\end{center}
\end{figure}

By increasing the value of $\mu$, we noticed that the theoretical rate of convergence is reached for larger and larger values of $l$. For small values of $l$, the variance decreases faster than the theoretical rate. Figure \ref{square} shows this phenomenon for $Z_{NV}^l$, with the payoff $f(u,s) = u^2$. 
Actually, by choosing this payoff we can check that
\begin{equation}
\begin{split}
\mathbb{E}\left[\left(Z^l_{NV}\right)^2\right]  &= 2^{-4 l } \left( \frac{3}{16} \mu^4 T^6 + \frac{9}{16} \mu^2 T^5\right) + 2^{-3l} \left(\frac{11}{64} \mu^2 T^5 + \frac{1545}{512}T^4 \right) + 2^{-2l} \left(\frac{163}{1024}T^4 \right). 
\end{split}
\label{P_Carre}
\end{equation}
The details of this tedious calculation are postponed to the Appendix.
The previous formula \eqref{P_Carre} contains higher order terms which overshadow the theoretical behavior of the variance. The following plot shows the behavior of $\log_2\left(\mathbb{E}\left[\left( Z^l_{NV}\right)^2 \right]\right)$ as a function of $l$.
For large values of $\mu$ and for small values of $l$, the ratio $\mathbb{E}\left[\left(Z^{l+1}_{NV}\right)^2\right]\Big / ~ \mathbb{E}\left[\left(Z^{l}_{NV}\right)^2\right]$ is close to 16, which shows that the leading term is $2^{-4 l }$.  Asymptotically, the slopes of the curves are 2. From a numerical point of view and given the structure of multilevel methods, this is an important point to emphasize. In particular, the choice \eqref{ML2R-ML} of parameters $\left(M_l^*\right)_{0\leq l \leq L^*}$ in the multilevel Richardson-Romberg estimator is based on asymptotic properties and will not be optimal when this asymptotic behavior fails for the first levels. 

Now we present the practical procedure used to implement the multilevel estimators.
Putting together the elements already discussed, the algorithm that we use for the multilevel Monte Carlo with the Ninomiya-Victoir scheme or the Giles-Szpruch scheme  is as follows. We begin by estimating the weak error constant $c_1$ in \eqref{Biais}, the constant $c_2$ which comes from the variance estimation \eqref{Var} and checking the orders of weak and strong convergence. When the asymptotic behavior \eqref{Biais} of the bias of the scheme is satisfied, one has
\begin{equation}
\mathbb{E}\left[ Z^l \right]  \sim \frac{c_1 \left(1 - 2^{\alpha}\right)}{2^{\alpha l}}.
\end{equation} 
Using a regression with few values of $\left(l,\left| \mathbb{E}\left[ Z^l \right] \right|\right)$, we estimate $c_1$ and check the order $\alpha$ of weak convergence.
In the same way, we estimate $c_2$ and check the strong order $\beta$ of variance convergence to $0$, using a regression in \eqref{Var}. Then we estimate $\mathbb{V}\left( Z^0 \right)$ using a standard Monte Carlo estimator $\hat{V}_0$. After that, for a given $\epsilon$  we define $L^*$ using \eqref{LMax} then we set
\begin{equation}
\label{MLMC_0}
M^*_0 = \left \lceil \frac{2}{\epsilon^2} \sqrt{\frac{\hat{V}^0}{\lambda_0}}  \left( \sqrt{\lambda_0 \hat{V}^0} + \sum \limits_{j=1}^{L^*} \sqrt{c_2\lambda_j 2^{j\left(1-\beta\right)} } \right) \right \rceil 
\end{equation}
and  
\begin{equation}
\label{MLMC_l}
\forall l \in \left\{1,\ldots,L^*\right\}, M^*_l = \left \lceil \frac{2}{\epsilon^2} \sqrt{\frac{c_2}{\lambda_l 2^{l\left(\beta +1 \right)}}}  \left( \sqrt{\lambda_0 \hat{V}^0} + \sum \limits_{j=1}^{L^*} \sqrt{c_2\lambda_j 2^{j\left(1-\beta\right)} } \right) \right \rceil. 
\end{equation}
When we use the Ninomiya-Victoir scheme we have the choice between $Z_{NV}^0 =  f\left(X^{NV,1,\eta}_T\right)$ and $Z_{NV}^0 = \frac{1}{2} \left( f\left(X^{NV,1,\eta}_T\right)  + f\left(X^{NV,1,-\eta}_T\right) \right)$. The second choice reduces the variance of level 0 if $X^{NV,1,\eta}_T$ effectively depends on $\eta$. So, in general, using $Z_{NV}^0 = \frac{1}{2} \left( f\left(X^{NV,1,\eta}_T\right)  + f\left(X^{NV,1,-\eta}_T\right) \right)$ reduces the sample size of the multilevel Monte Carlo estimator. Thus, although we use two schemes in the level 0, the method is slightly faster with this choice in practice.
As already mentioned, for the Giles-Szpruch scheme we choose $\lambda_0 = 1$ and $\lambda_l = 5/2, \forall l \in \left\{1,\ldots, L^*\right\}$, to balance the lower cost of level $l=0$.  
Following this idea, for the Ninomiya-Victoir scheme we choose the same sequence if $Z_{NV}^0 = \frac{1}{2} \left( f\left(X^{NV,1,\eta}_T\right)  + f\left(X^{NV,1,-\eta}_T\right) \right)$, and we propose to choose $\lambda_0 = 1$ and $\lambda_l = 5, \forall l \in \left\{1,\ldots, L^*\right\}$, if $Z_{NV}^0 =  f\left(X^{NV,1,\eta}_T\right)$.

Let us discuss the implementation of the multilevel Monte Carlo estimator with the Giles-Szpruch scheme from level 0 to level $L^*-1$ and the coupling between the Ninomiya-Victoir and the Giles-Szpruch scheme at the last level $L^*$. The practical procedure is slightly different. As already discussed, in the case of $\hat{Y}_{MLMC}^{GS-NV}$ the bias is given by the bias of the Ninomiya-Victoir scheme, so we begin with the estimation of the weak error constant $c_1$ using the Ninomiya-Victoir scheme. The next step is to estimate the constant $c_2$ using the Giles-Szpruch scheme. Then, we estimate $\mathbb{V}\left( Z^0_{GS} \right)$ (respectively $\mathbb{V}\left(  Z_{GS-NV}^{L^*} \right)$) using a standard Monte Carlo estimator $\hat{V}^0_{GS}$ (respectively $ \hat{V}_{GS-NV}^{L^*}$). Finally, we define $L^*$ using \eqref{LMax} and set
\begin{equation}
\label{MLMC_C0}
M^*_0 = \left \lceil \frac{2}{\epsilon^2} \sqrt{\frac{\hat{V}_{GS}^{0}}{\lambda_0}}  \left( \sqrt{\lambda_0 \hat{V}_{GS}^{0}} + \sum \limits_{j=1}^{L^*-1} \sqrt{c_2\lambda_j 2^{j\left(1-\beta\right)} }  + \sqrt{\lambda_L^* 2^{L^*\left(1-\beta\right)} \hat{V}_{GS-NV}^{L^*}}\right) \right \rceil, 
\end{equation}
\begin{equation}
\begin{split}
\label{MLMC_Cl}
\forall l \in \left\{1,\ldots,L^*-1\right\}, M^*_l = \left \lceil \frac{2}{\epsilon^2} \sqrt{\frac{c_2}{\lambda_l 2^{l\left(\beta +1 \right)}}}  \left( \sqrt{\lambda_0 \hat{V}_{GS}^{0}} + \sum \limits_{j=1}^{L^*-1} \sqrt{c_2\lambda_j 2^{l\left(1-\beta\right)} }  + \sqrt{\lambda_L^* 2^{L^*\left(1-\beta\right)} \hat{V}_{GS-NV}^{L^*}}\right) \right \rceil, 
\end{split}
\end{equation}
\begin{equation}
\label{MLMC_CL}
M^*_{L^*} = \left \lceil \frac{2}{\epsilon^2} \sqrt{\frac{\hat{V}_{GS-NV}^{L^*}}{\lambda_{L^*} 2^{L^*} }}  \left( \sqrt{\lambda_0 \hat{V}_{GS}^{0}} + \sum \limits_{j=1}^{L^*-1} \sqrt{c_2\lambda_j 2^{j\left(1-\beta\right)} }  + \sqrt{\lambda_L^* 2^{L^*\left(1-\beta\right)} \hat{V}_{GS-NV}^{L^*}}\right) \right \rceil. 
\end{equation}
We suggest to choose $\lambda_0 = 1$, $\lambda_l = 5/2, \forall l \in \left\{1,\ldots, L^*-1\right\},$ and $\lambda_{L^*} = 9/2$ to balance the higher cost of level $L^*$. 

Since all parameters are explicit, implementing the Multilevel Richardson-Romberg estimator is quite simple. As noted in \cite{PL}, we only need to estimate $\mathbb{V}\left(f\left(X_T\right)\right)$ and the constant $c_2$ in \eqref{Var} which comes from the variance estimation. The variance $\mathbb{V}\left(f\left(X_T\right)\right)$ is estimated using a crude Monte Carlo method. 

\begin{figure}[!ht]
\begin{center}
\includegraphics[width=12.5cm, height=9.5cm]{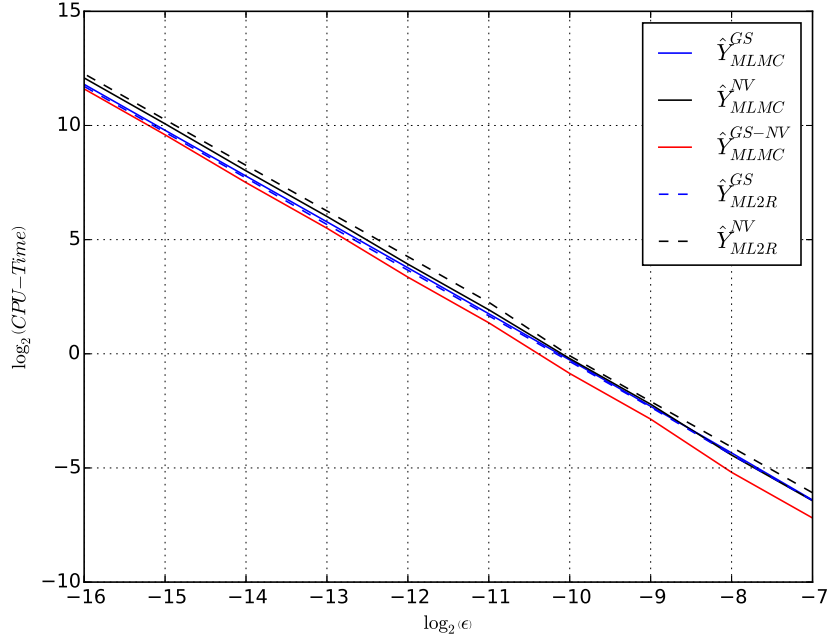}
\caption{Clarl-Cameron SDE with $f(u,s) = cos(u)$, CPU-time in second ($y$-axis $\log_2$ scale) as a function of $\epsilon$ ($x$-axis $\log_2$ scale).}
\label{CCD_Cos}
\end{center}
\end{figure}

Now we present our numerical tests in which we compare the computing time of each estimator as a function of the upper bound, denoted by $\epsilon$, on the root mean squared error. For our first test we choose a smooth payoff $f(u,s) = \cos(u)$. We estimate the two constants $c_1$ and $c_2$ using the above-mentioned procedure. To compute our regression, we estimate $\mathbb{E}\left[Z^l\right]$ and $\mathbb{V}\left[Z^l\right]$ for $l \in \left\{1,\ldots,4\right\}$, using a standard Monte Carlo method. The sample size used must be adjusted to get a rather good estimate, but without spending too much time during this step. In our numerical experiment, we choose a sample size $M = 10^4$. Using this approach, we estimate the theoretical values of the orders of weak and variance convergences. More precisely we get $\alpha = 1$, $\beta = 2$ for the Giles-Szpruch scheme and $\alpha = 2$, $\beta = 2$ for the Ninomiya-Victoir scheme. In figure \ref{CCD_Cos} is depicted the CPU-time in seconds (in $\log_2$ scale) of each multilevel method as a function of $\epsilon$ (in $\log_2$ scale). It provides a direct comparison of the performance of the different estimators. The red line is for $\hat{Y}_{MLMC}^{GS-NV}$. This line is below the other lines, which indicates clearly that, for this experiment, $\hat{Y}_{MLMC}^{GS-NV}$ is faster than the other estimators. Moreover, we observe a rather close behavior of the Multilevel Richardson-Romberg estimator and the Multilevel Monte Carlo estimator. Indeed the black line, representing $\hat{Y}_{MLMC}^{NV}$ is close to the black dashed line representing $\hat{Y}_{ML2R}^{NV}$. Similarly the blue line, representing $\hat{Y}_{MLMC}^{GS}$ is close to the blue dashed line representing $\hat{Y}_{ML2R}^{GS}$. Finally, one can notice that all slopes are equal to $-2$, which indicates that all these estimators achieve a $O\left(\epsilon^{-2}\right)$ complexity. 

\begin{figure}[!ht]
\begin{center}
\includegraphics[width=12.5cm, height=9.5cm]{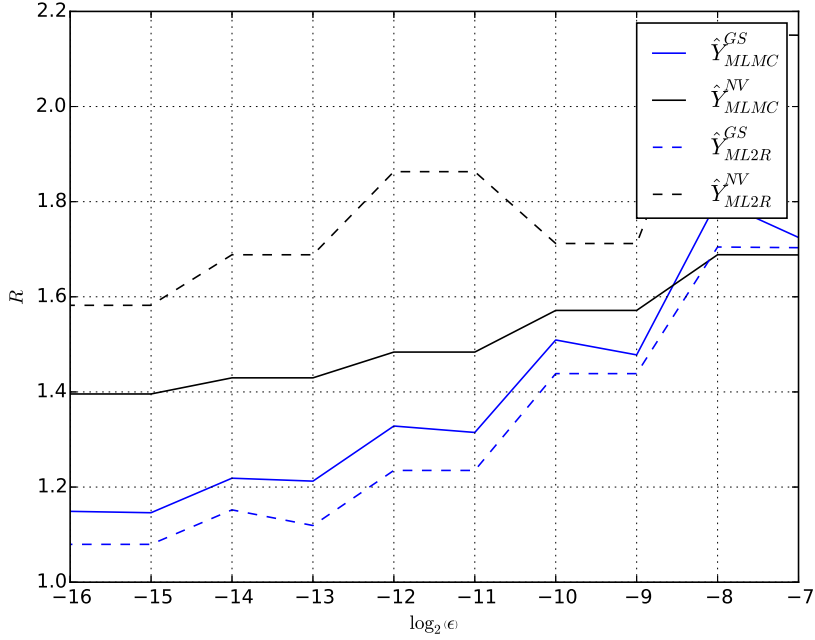}
\caption{Clarl-Cameron SDE with $f(u,s) = cos(u)$, CPU-time ratios ($y$-axis) as a function of $\epsilon$ ($x$-axis $\log_2$ scale).}
\label{Rate_CCD_Cos}
\end{center}
\end{figure}

To measure the efficiency of $\hat{Y}_{MLMC}^{GS-NV}$ with respect to other estimators, we plot in figure \ref{Rate_CCD_Cos} the following CPU-time ratios: 
\begin{equation}
R = \frac{CPU-time\left( \hat{Y}\right)}{CPU-time\left( \hat{Y}_{MLMC}^{GS-NV}\right)}.
\end{equation}

The estimator $\hat{Y}_{MLMC}^{GS-NV}$  is about $1.1$ to $1.6$ faster than $\hat{Y}_{MLMC}^{GS}$ or $\hat{Y}_{ML2R}^{GS}$ when $\epsilon$ goes from $2^{-16}$ to $2^{-7}$.  
In comparison with $\hat{Y}_{MLMC}^{GS-NV}$, $\hat{Y}_{MLMC}^{NV}$ and $\hat{Y}_{ML2R}^{NV}$ perform poorly. \\

In order to understand what is going on, let us provide a theoretical calculation of the CPU-time for the multilevel Monte Carlo estimator.
Denoting by $\tau_l$ the theoretical computing time of level $l \in \left\{0,\ldots,L^*\right\}$, one has
\begin{equation}
\tau^l  \propto M^*_l 2^l.
\end{equation}

Replacing\footnote{Obviously, the constant $C \left(\epsilon \right)$ depend on the estimator. For $\hat{Y}_{MLMC}^{GS}$ and $\hat{Y}_{MLMC}^{NV}$, the constants are given by formulas \eqref{MLMC_0} and \eqref{MLMC_l}: $C_0\left(\epsilon\right) = \frac{2}{\epsilon^2} \sqrt{\frac{\hat{V}^0}{\lambda_0}}  \left( \sqrt{\lambda_0 \hat{V}^0} + \sum \limits_{j=1}^{L^*} \sqrt{c_2\lambda_j 2^{j\left(1-\beta\right)} } \right)$ and $C_l\left(\epsilon\right) = \frac{2}{\epsilon^2} \sqrt{\frac{c_2}{\lambda_l}}  \left( \sqrt{\lambda_0 \hat{V}^0} + \sum \limits_{j=1}^{L^*} \sqrt{c_2\lambda_j 2^{j\left(1-\beta\right)} } \right)$, $\forall l \in \left\{1,\ldots, L^*\right\}$. For $\hat{Y}_{MLMC}^{GS-NV}$ the constans are given by formulas \eqref{MLMC_C0}, \eqref{MLMC_Cl} and \eqref{MLMC_CL}.} $M^*_l$, one can write
\begin{equation}
\tau^l  = C_l \left(\epsilon\right) 2^{-l\left(\frac{\beta+1}{2}\right)}2^l = C_l \left(\epsilon\right) 2^{-l\left(\frac{\beta - 1}{2}\right)}
\end{equation}
The theoretical computing time, denoted by $\tau$, is given by
\begin{equation}
\tau \left(\epsilon\right)  =  \sum \limits_{l=0}^{L^* \left(\epsilon\right)} \tau_l = \sum \limits_{l=0}^{L^* \left(\epsilon\right)} C_l\left(\epsilon\right) 2^{-l\left(\frac{\beta - 1}{2}\right)}. 
\end{equation}  
In the multilevel Monte Carlo estimator studied in this paper $C_l = C_1, \forall l \in \left\{1,\ldots,L^*-1\right\}$, then one has
\begin{equation}
\label{Theo}
\tau \left(\epsilon\right) = \begin{cases} C_0 \left(\epsilon\right) +   \frac{C_1\left(\epsilon\right)  }{1 - 2^{-\frac{\beta - 1}{2}}}\left( 2^{-\frac{\beta - 1}{2}} - 2^{-L^*\left(\epsilon\right) \frac{\beta - 1}{2}} \right) + C_{L^*}\left(\epsilon\right) 2^{-L^*\left(\epsilon\right) \frac{\beta - 1}{2}}  & \mbox{if } \beta \neq 1\\  C_0\left(\epsilon\right) +  \left( L^*\left(\epsilon\right) -2 \right)  C_1\left(\epsilon\right) + C_{L^*}\left(\epsilon\right)   &\mbox{if } \beta = 1. \end{cases}  
\end{equation}  
Now, it is easy to understand why $\hat{Y}_{MLMC}^{GS-NV}$ is faster than $\hat{Y}_{MLMC}^{GS}$. As a matter of fact, the two estimators are very close, and in our numerical experiments we observe that $C^{GS}\left(\epsilon\right) \approx C^{GS-NV}\left(\epsilon\right)$. Since using a scheme with second order of weak convergence provides a lower optimal last level $L^*$, in view of \eqref{Theo}, we understand why, in general, we can state that $\tau_{MLMC}^{GS-NV} \leq \tau_{MLMC}^{GS}$. The poor performance of  $\hat{Y}_{MLMC}^{NV}$  or $\hat{Y}_{ML2R}^{NV}$ reflects the use of six schemes in $Z^l_{NV}$.

\begin{figure}[!ht]
\begin{center}
\includegraphics[width=12.5cm, height=9.5cm]{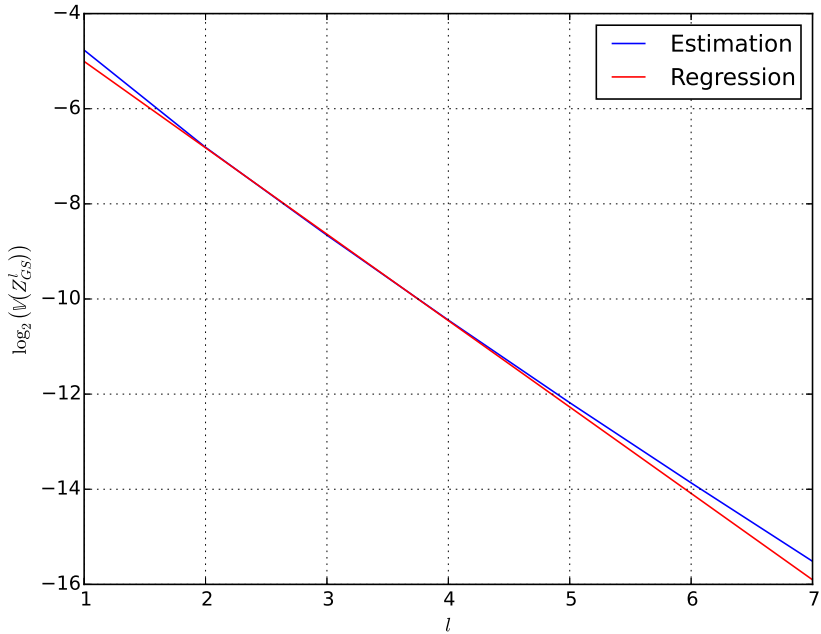}
\caption{Clarl-Cameron SDE with $f(u,s) = u_+$, Variance of the Giles-Szpruch scheme ($y$-axis $\log_2$ scale) as a function of $l$ ($x$-axis).}
\label{Reg}
\end{center}
\end{figure}

For our second experiment, we only change the payoff. We choose the non-smooth payoff $f(u,s) = u_+$.  Theorem 5.2 in \cite{GS} gives the lower bound $\beta = 3/2$ for the Giles-Szpruch scheme. Their proof is, in some ways, generic and it can easily be adapted to the Ninomiya-Victoir scheme. This is enough to keep the $O\left(\epsilon^{-2}\right)$ complexity. To determine the actual values of $\beta$ and $\alpha$, we rely on the numerical results. Using the same automatic process, we get for the Ninomiya-Victoir scheme $\alpha = 3/2$ and $\beta = 3/2$. The non-regularity of the payoff affects both the weak and the variance convergence rates. With regard to the Giles-Szpruch scheme, the regression procedure leads to $\alpha = 1$ and $\beta = 2$, but the situation is quite confusing. Indeed, we noticed that the asymptotic rate $\beta = 3/2$ is reached for $l \ge \bar{l}=5$. Figure \ref{Reg} illustrates this inflection. The blue line is the estimation of $\left(\mathbb{V}\left(Z_{GS}^l\right)\right)_{1\leq l \leq 7}$ whereas the red line is the regression on the first four values. The two lines diverge at level $\bar{l}=5$, which show clearly the inflection.\\

Here, assigning a value for $\left(\beta,c_2\right)$ to implement $\hat{Y}_{MLMC}^{GS}$, $\hat{Y}_{ML2R}^{GS}$ and $\hat{Y}_{MLMC}^{GS-NV}$ by using respectively \eqref{MLMC_0}-\eqref{MLMC_l}, \eqref{MLMC_C0} to \eqref{MLMC_CL}, and \eqref{ML2R-ML} to \eqref{ML2R_end} may not be convenient.  We suggest to apply the numerical procedure described in the following remark to implement the multilevel estimators.

\begin{figure}[!ht]
\begin{center}
\includegraphics[width=12.5cm, height=9.5cm]{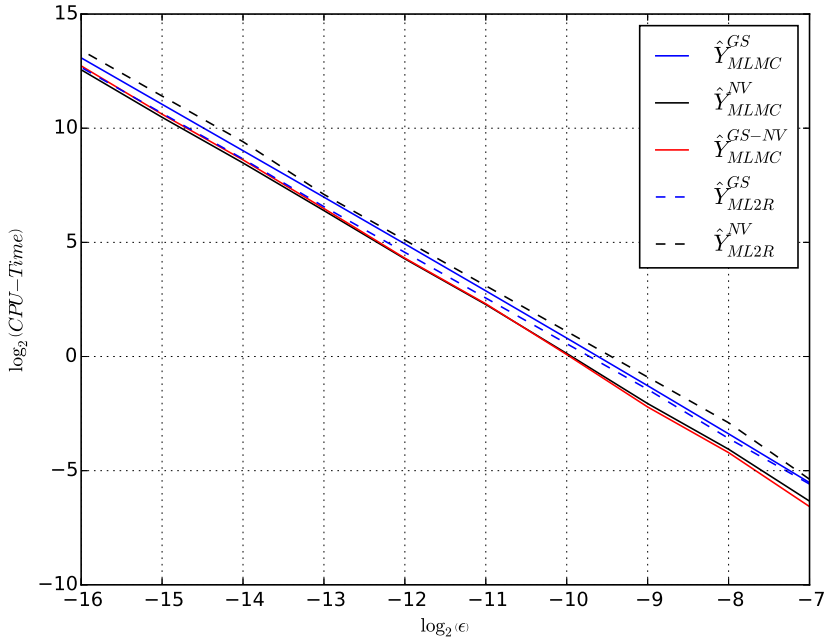}
\caption{Clarl-Cameron SDE with $f(u,s) = u_+$, CPU-time in second ($y$-axis $\log_2$ scale) as a function of $\epsilon$ ($x$-axis $\log_2$ scale).}
\label{CCD_Call}
\end{center}
\end{figure}

\begin{figure}[!ht]
\begin{center}
\includegraphics[width=12.5cm, height=9.5cm]{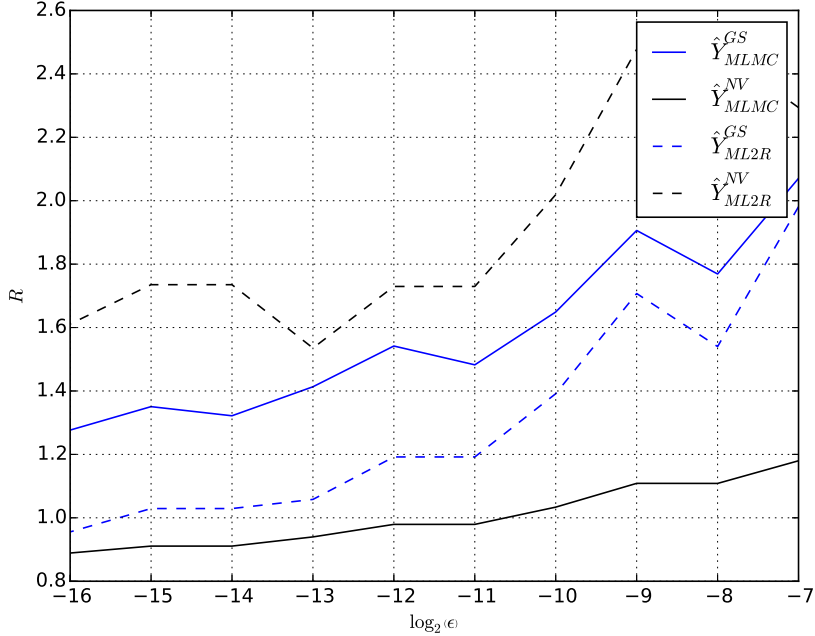}
\caption{Clarl-Cameron SDE with $f(u,s) = u+$, CPU-time ratios ($y$-axis) as a function of $\epsilon$ ($x$-axis $\log_2$ scale).}
\label{Rate_CCD_Call}
\end{center}
\end{figure}

\begin{arem}
\label{Difficile}
In the case of the Clark-Cameron SDE with $\mu = 1,  U_0 = 0$, $S_0 = 0$ and for a smooth payoff, everything is going as expected, but in some cases (see the Heston model or the Clark-Cameron SDE with a large $\mu$) estimating $\left(\beta ,c_2\right)$ may be difficult, especially when the theoretical rate of convergence is reached for a level $\bar{l} \ge 2$ and this may affect the efficiency of the multilevel methods. To get around this problem, a reasonable criterion is to compare $\bar{l}$ and the last level $L^*\left(\epsilon\right)$. If $L^*\left(\epsilon\right) < \bar{l}$, we decide to use the values obtained by the regression and use the usual formulas\footnote{\eqref{MLMC_0}-\eqref{MLMC_l} for $\hat{Y}_{MLMC}^{GS}$ and $\hat{Y}_{MLMC}^{NV}$, \eqref{MLMC_C0} to \eqref{MLMC_CL} for $\hat{Y}_{MLMC}^{GS-NV}$, and \eqref{ML2R-ML} to \eqref{ML2R_end} for $\hat{Y}_{ML2R}$.} to compute $\left(M_l^*\right)_{0\leq l\leq L^*}$ for both methods. If $L^*\left(\epsilon\right) \ge  \bar{l}$, we estimate  $\mathbb{V}\left[Z^l\right]$ for $l \in \left\{0,\ldots,\bar{l}\right\}$ using standard Monte Carlo Method. Then, we approximate, for $l \in \left\{\bar{l}+1 ,\ldots,L^*\right\}$, $\mathbb{V}\left[Z^l\right]$ by $2^{\beta\left(l-\bar{l}\right)} \hat{V}^{\bar{l}}$ where $\hat{V}^{\bar{l}}$ is the estimation of $\mathbb{V}\left[Z^{\bar{l}}\right]$ and $\beta$ is the theoretical order of convergence of the variance. Finally, we compute $M_l^*$ for $l  \in \left\{0,\ldots,L^*\right\}$ using \eqref{MLMC_Sample}. As regards the multilevel Richardson-Romberg estimator, we do not recommend its use in this case.
\end{arem}
In our second experiment, the Giles-Szpruch scheme only appears to be problematic. Indeed the values of $L^*\left(\epsilon\right)$ are given by:
\begin{center}
\begin{tabular}{|c|c|c|c|c|c|c|c|c|c|c|}
  \hline
  $\epsilon$ & $2^{-7}$ & $2^{-8}$ & $2^{-9}$ & $2^{-10}$ & $2^{-11}$ & $2^{-12}$ & $2^{-13}$ & $2^{-14}$ & $2^{-15}$ & $2^{-16}$ \\
  \hline
  $\hat{Y}^{GS}_{MLMC}$ & 6 & 7 & 8 & 9 & 10 & 11 & 12 & 13 & 14 & 15 \\
  $\hat{Y}^{GS-NV}_{MLMC}$ & 3 & 4 & 4 & 5 & 6 & 6 & 7 & 8 & 8 & 9\\
 $\hat{Y}^{GS}_{ML2R}$ & 3 & 3 & 4 & 4 & 4 & 4 & 4 & 5 & 5 & 5 \\
  \hline 
\end{tabular}
~.
\end{center}
If $\epsilon \in \left\{2^{-14}, 2^{-15}, 2^{-16} \right\}$, for $\hat{Y}_{MLMC}^{GS}$, since $\bar{l}$ is exactly equal to $L\left(\epsilon\right)$, we are in a borderline situation. Nevertheless, we keep in the following figures the performance of this estimator for $\epsilon \in \left\{2^{-14}, 2^{-15}, 2^{-16} \right\}$. For the multilevel Monte Carlo estimators with the Giles-Szpruch scheme, we apply the modified procedure of Remark \ref{Difficile} if necessary. Figure \ref{CCD_Call} compares the computing time of the estimator, with the previous graphical conventions.
Unlike the previous experiment, the two fastest estimators are $\hat{Y}_{MLMC}^{NV}$ and $\hat{Y}_{MLMC}^{GS-NV}$. Although we lose the second order of weak convergence, the estimator $\hat{Y}_{MLMC}^{GS-NV}$  is about $1.3$ to $2$ faster than $\hat{Y}_{MLMC}^{GS}$. This is due to the degradation of the variance convergence order $\beta$ from $2$ to $3/2$ in comparison with a smooth payoff. Indeed, thanks to formula \eqref{Theo}, one can see that, in the multilevel Monte Carlo methods, all things being equal, the gain in computing time due to the introduction of a scheme with high order of weak convergence in the last level is all the more significant that $\beta$ is small. This explains why $\hat{Y}_{MLMC}^{NV}$ performs very well. Despite using six schemes in $Z^l_{NV}$, $\hat{Y}_{MLMC}^{NV}$ goes up to 1.1 faster than $\hat{Y}_{MLMC}^{GS-NV}$ (see Figure \ref{Rate_CCD_Call}). In contrast, the use of a scheme with high order of weak convergence like $Z^l_{NV}$ in the multilevel Richardson-Romberg  does not appear to counterbalance its complexity. This difference of behavior is related to the dependence of $L^*\left(\epsilon\right)$ on $\alpha$ . In the multilevel Monte Carlo estimators, the dependence is of the form $1/\alpha$ (see \eqref{LMax}) which provides better results as alpha increases than the multilevel Richardson-Romberg estimator where the dependence on $\alpha$ is given by \eqref{LMax_PL}.  
 \subsubsection{Heston model}
\begin{figure}[!ht]
\begin{center}
\includegraphics[width=12.5cm, height=9.5cm]{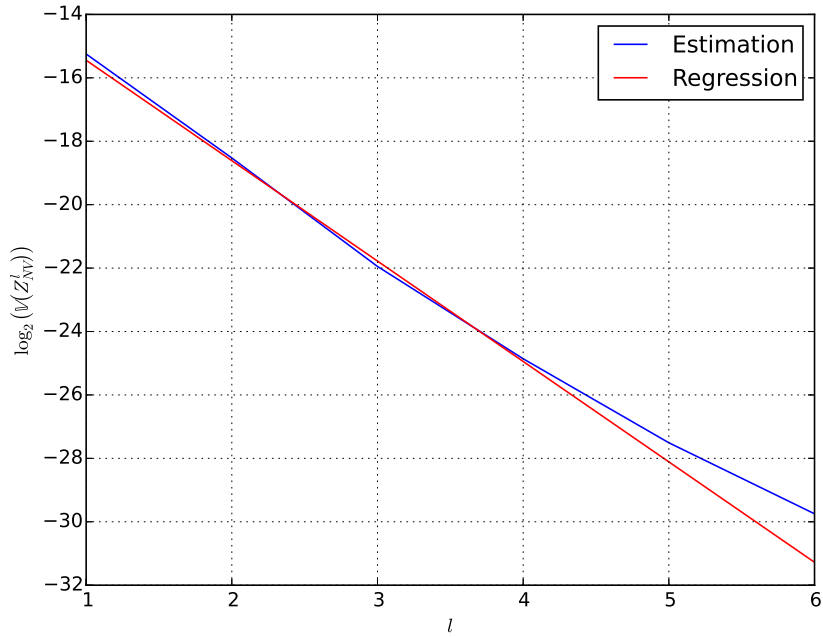}
\caption{Heston SDE with $f(u,v) = \exp\left(-rT\right) \left(\exp\left(u\right) - 1\ \right)_+$, Variance of the Ninomiya-Victoir scheme ($y$-axis $\log_2$ scale) as a function of $l$ ($x$-axis).}
\label{Reg2}
\end{center}
\end{figure}

The Heston model is an asset price model which assumes that volatility, denoted by $V$, evolves according to an autonomous Cox-Ingersoll-Ross SDE:
\begin{equation}
  \left\{
      \begin{aligned}
        dU_t = (r-\frac12V_t)dt + \sqrt{V_t} dW_t^1 \\
        dV_t = \kappa(\theta - V_t)dt + \sigma \sqrt{V_t} dW_t^2.
      \end{aligned}
  \right.
\end{equation}
The asset price $S$ is given by $S_t = \exp(U_t)$.
We assume, for simplicity, no correlation between the Brownian motion driving the asset price and the volatility process.
We  also assume  that $2\kappa \theta \ge \sigma^2$ to ensure that the zero boundary is not attainable for the volatility process. The main difficulty is located in 0, where the square root
is not Lipschitz. 
In this $2-$dimensional model, the diffusion coefficients are given by
 $\sigma^1\begin{pmatrix} 
 u \\ 
 v 
 \end{pmatrix} = \begin{pmatrix} 
 \sqrt{v} \\ 
 0 
 \end{pmatrix}$, $\sigma^2\begin{pmatrix} 
 u \\ 
 v 
 \end{pmatrix} = \begin{pmatrix} 
 0 \\ 
 \sigma \sqrt{v} 
 \end{pmatrix}$ and the drift coefficient is $b \begin{pmatrix} 
 u \\ 
 v 
 \end{pmatrix} = \begin{pmatrix} 
 r - \frac12 v \\ 
 \kappa \left(\theta - v\right) 
 \end{pmatrix}$. The Stratonovich drift is given by
\begin{equation*}
\begin{split}
\sigma^0\begin{pmatrix} 
 u \\ 
 v 
 \end{pmatrix} & = \left\{b - \frac12\left(\partial \sigma^1 \sigma^1 + \partial \sigma^2 \sigma^2 \right)\right\} \begin{pmatrix} 
 u \\ 
 v 
 \end{pmatrix} = \begin{pmatrix} 
 r - \frac12 v \\ 
 \kappa \left(\theta - v\right) 
 \end{pmatrix} - \frac{1}{2} \left(\begin{pmatrix} 
 0 & \frac{1}{2\sqrt{v}}\\ 
 0 & 0
 \end{pmatrix} \begin{pmatrix} 
 \sqrt{v} \\ 
 0 
 \end{pmatrix} + \begin{pmatrix} 
 0 & 0\\ 
 0 & \frac{\sigma}{2\sqrt{v}}
 \end{pmatrix}  \begin{pmatrix} 
 0 \\ 
 \sigma \sqrt{v} 
 \end{pmatrix} \right) \\ & = \begin{pmatrix} 
 r - \frac12 v \\ 
 \kappa \left(\theta - v\right) - \frac{\sigma^2}{4}
 \end{pmatrix}.
\end{split}
\end{equation*}
Then, the Giles-Szpruch scheme is given by
\begin{equation}
  \left\{
      \begin{aligned}
         V^{GS}_{t_{k+1}} &= V^{GS}_{t_{k}} + \kappa\left(\theta - V^{GS}_{t_{k}}\right)h + \sigma \sqrt{V^{GS}_{t_{k}}} \Delta W^2_{t_{k+1}} + \frac14 \sigma^2 \left( \left(\Delta W^2_{t_{k+1}}\right)^2 - h \right)\\
	       U^{GS}_{t_{k+1}} &=  U^{GS}_{t_{k}} + \left(r-\frac12 V^{GS}_{t_{k}}\right)h + \sqrt{V^{GS}_{t_{k}}}\Delta W^1_{t_{k+1}} + \frac14 \sigma \Delta W^1_{t_{k+1}} \Delta W^2_{t_{k+1}}.
     \end{aligned}
  \right.
\end{equation}
Setting $\xi = \theta - \frac{\sigma^2}{4\kappa}$, the Ninomiya-Victoir scheme is given by
\begin{equation}
  \left\{
      \begin{aligned}
        	V^{NV,\eta}_{t_{k+\frac13}}  &= \left({V}^{NV,\eta}_{t_{k}} -\xi\right)\exp\left(-\frac12 \kappa h\right)+\xi \\
	U^{NV,\eta}_{t_{k+\frac13}} &= U^{NV,\eta}_{t_{k}} +  \frac12 \left(r-\frac12 \xi\right)h + \frac1{2\kappa}\left({V}^{NV,\eta}_{t_{k}} - \xi\right)\left(\exp\left(-\frac12 \kappa h\right)-1\right) \\
        	V^{NV,\eta}_{t_{k+\frac23}}  &= \left(\sqrt{V^{\eta,\eta}_{t_{k+\frac13}}} + \frac12 \sigma\Delta W^2_{t_{k+1}}\right)^2 \\ 
U^{NV,\eta}_{t_{k+\frac23}} &= U^{NV,\eta}_{t_{k+\frac13}} + \sqrt{ V^{NV,\eta}_{t_{k+\frac13}}\mathds{1}_{\left\{\eta_{k+1}= 1\right\}} + V^{NV,\eta}_{t_{k+\frac23}} \mathds{1}_{\left\{\eta_{k+1}= -1\right\}}} \Delta W^1_{t_{k+1}}\\
 V^{NV,\eta}_{t_{k+1}}  &= \left({V}^{NV,\eta}_{t_{k+ \frac23}} -\xi\right)\exp\left(-\frac12 \kappa h\right)+\xi \\
	U^{NV,\eta}_{t_{k+1}} &= U^{NV,\eta}_{t_{k + \frac23}} +  \frac12 \left(r-\frac12 \xi\right)h + \frac1{2\kappa}\left({V}^{NV,\eta}_{t_{k+\frac23}} - \xi\right)\left(\exp\left(-\frac12 \kappa h\right)-1\right).
        \end{aligned}
  \right.
\end{equation}
In these formulas close to our implementation of the scheme, the evolution from $\left({U}^{NV,\eta}_{t_{k}},{V}^{NV,\eta}_{t_{k}}\right)$, respectively $\left({U}^{NV,\eta}_{t_{k+\frac23}},{V}^{NV,\eta}_{t_{k+\frac23}}\right)$, to $\left({U}^{NV,\eta}_{t_{k+\frac13}},{V}^{NV,\eta}_{t_{k+\frac13}}\right)$, respectively $\left({U}^{NV,\eta}_{t_{k+1}},{V}^{NV,\eta}_{t_{k+1}}\right)$, corresponds to the integration of the ODE directed by the vector field $\sigma^0$ on half a time step whereas the evolution from $\left({U}^{NV,\eta}_{t_{k+\frac13}},{V}^{NV,\eta}_{t_{k+\frac13}}\right)$ to $\left({U}^{NV,\eta}_{t_{k+\frac23}},{V}^{NV,\eta}_{t_{k+\frac23}}\right)$ corresponds to the integration of the Brownian vector fields. 
The Giles-Szpruch scheme and usual schemes such as the Euler scheme are not well defined  since they can lead to negative values of the volatility process for which the
square root is not defined at the next step. Assuming $\xi \ge 0$, the Ninomiya-Victoir scheme is well defined and the volatility process is always positive (see \cite{Alfonsi}). For $\xi < 0$, in section 3.1 of \cite{Alfonsi}, Alfonsi proposed a modification of the Ninomiya-Victoir scheme preserving the positivity of the volatility and the weak order two. For the simulation studies, we choose, as in \cite{GS}, $S_0 = V_0 = 1,  r = 0.05, T = 1$, $\kappa = 0.5, \theta = 0.9$ and $ \sigma = 0.05$. Then $\xi = 0.89875$, so that the Ninomiya-Victoir scheme is well defined. Using this parameters, we do not observe negative values for the volatility with the Giles-Szpruch scheme. We choose to price the at the money call option. This corresponds to the payoff $f(u,v) = \exp\left(-rT\right) \left(\exp\left(u\right) - 1\ \right)_+$. Estimating the multilevel parameters, we obtain, $\alpha = 1 $ and $\beta = 2$ and not $3/2$ as predicted by the analysis. For the Nynomiya-Victoir scheme the estimation of $\left(\alpha, c_1 \right)$ leads to $\left(2, -3.2\times 10^{-4} \right)$. Since $c_1$ is very small, formula \eqref{LMax} can lead to negative values. If this occurs we set $L^*\left(\epsilon\right)= 1$.  We also observe that $\mathbb{V}\left(Z^l_{NV}\right)$ decreases very quickly and faster than the theoretical rate for the first levels. Actually, an analogy can be drawn between the Ninomiya-Victoir scheme for the Clark-Cameron SDE and the Ninomiya-Victoir scheme for the Heston SDE since we have the same structure. As a matter of fact, when $\sqrt{V^{\eta,\eta}_{t_{k+\frac13}}} + \frac12 \sigma\Delta W^2_{t_{k+1}} \ge 0$,  $V^{NV,\eta}_{t_{k+\frac23}}  = \left(\sqrt{V^{\eta,\eta}_{t_{k+\frac13}}} + \frac12 \sigma\Delta W^2_{t_{k+1}}\right)^2$ rewrites: 
\begin{equation*}
\sqrt{V^{NV,\eta}_{t_{k+\frac23}}} = \sqrt{V^{\eta,\eta}_{t_{k+\frac13}}} + \frac12 \sigma\Delta W^2_{t_{k+1}}
\end{equation*}
This equation, similar to $S^{NV,\eta}_{t_{k+1}} = S^{NV,\eta}_{t_{k}} + \mu \left(t_{k+1} - t_k\right) + \left(W^2_{t_{k+1}} - W^2_{t_k}\right)$ in the Clark-Cameron SDE, is the only place where the Brownian increment $\Delta W^2$ appears in the Ninomiya-Victoir scheme for the Heston model. In the dynamics of the U component the Brownian increment  $\Delta W^1$ is multiplied by $S^{NV,\eta}$ in the Clark-Cameron SDE and $\sqrt{ V^{NV,\eta}_{t_{k+\frac13}}\mathds{1}_{\left\{\eta_{k+1}= 1\right\}} + V^{NV,\eta}_{t_{k+\frac23}} \mathds{1}_{\left\{\eta_{k+1}= -1\right\}}}$ in the Heston SDE. Then, the presence of a non-zero drift probably explains the existence of the higher order terms that disrupt the theoretical behavior like in formula \eqref{P_Carre}.
Figure \ref{Reg2} illustrate this phenomenon. The blue line is the estimation of $\left(\mathbb{V}\left(Z_{NV}^l\right)\right)_{1\leq l \leq 7}$ whereas the red line is the regression on the first four values. To be precise, we estimate $\mathbb{V}\left(Z^l_{NV}\right)$, for $\l \in \left\{1,\ldots,7\right\}$ using $M = 10^7$ samples to get pretty good estimations, but in practice, $M = 10^6$ would be enough to implement the multilevel estimators. The regression leads to $\beta = 3$. As in the Clark-Cameron SDE with $f(u,s) = s_+$, the two lines diverge at level $\bar{l}=5$.   

So to implement the multilevel estimators, we compare $\bar{l}$ and $L^*\left(\epsilon\right)$ as already mentioned in Remark \ref{Difficile}. The values of $L^*\left(\epsilon\right)$ are given by
\begin{center}
\begin{tabular}{|c|c|c|c|c|c|c|c|c|c|c|}
  \hline
  $\epsilon$ & $2^{-7}$ & $2^{-8}$ & $2^{-9}$ & $2^{-10}$ & $2^{-11}$ & $2^{-12}$ & $2^{-13}$ & $2^{-14}$ & $2^{-15}$ & $2^{-16}$ \\
  \hline
  $\hat{Y}^{NV}_{MLMC}$ & 1 & 1 & 1 & 1 & 1 & 1 & 2 & 2 & 3 & 3 \\
   $\hat{Y}^{NV}_{ML2R}$ & 2 & 2 & 2 & 2 & 3 & 3 & 3 & 3 & 3 & 3 \\
  \hline
\end{tabular}
~ .
\end{center}
We notice that even if $\epsilon$ is very small, $L\left(\epsilon \right) < \bar{l}$. So we can implement the multilevel estimators using the standard automatic procedure. With regard to the Ninomiya-Victoir scheme, we remark that: $\mathbb{V}\left(\frac{1}{2} \left( f\left(X^{NV,1,\eta}_T\right)  + f\left(X^{NV,1,-\eta}_T\right) \right)\right) \approx \mathbb{V}\left( f\left(X^{NV,1,\eta}_T\right) \right) $, so we naturally decide to implement the multilevel Monte Carlo with $Z^0_{NV} = f\left(X^{NV,1,\eta}_T\right)$.
In the following plots we compare the five estimators. This time, the fastest estimator is $\hat{Y}_{ML2R}^{NV}$. This is due to the outstanding value of $\beta$ observed for the Ninomiya-Victoir scheme. The poor performance of $\hat{Y}_{MLMC}^{NV}$ is explained by the high variance at the level $0$, while the variance at the higher levels are very small since the numerical value of $\beta $ is $3$.
\begin{figure}[!ht]
\begin{center}
\includegraphics[width=12.5cm, height=9.5cm]{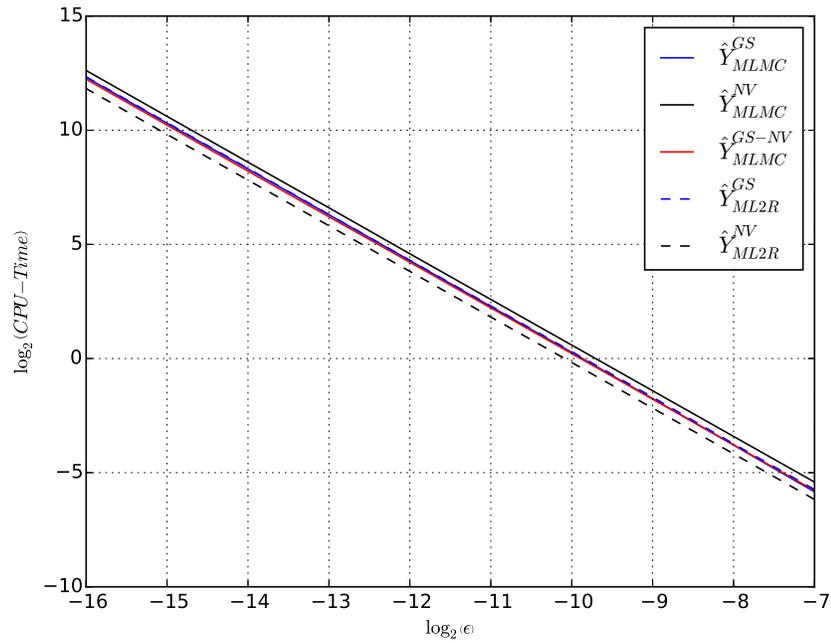}
\caption{Heston SDE with $f(u,v) = \exp\left( -rT \right) \left(\exp\left(u\right) - 1\ \right)_+$, CPU-time in second ($y$-axis $\log_2$ scale) as a function of $\epsilon$ ($x$-axis $\log_2$ scale).}
\label{H_Call}
\end{center}
\end{figure}

\begin{figure}[!ht]
\begin{center}
\includegraphics[width=12.5cm, height=9.5cm]{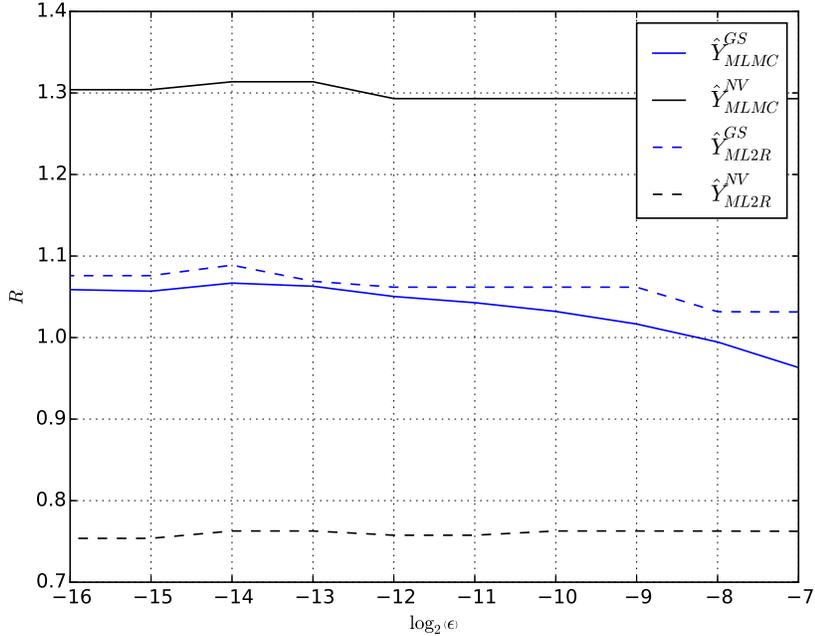}
\caption{Heston SDE with $f(u,v) = \exp\left( -rT \right)\left(\exp\left(u\right) - 1\ \right)_+$, CPU-time ratios ($y$-axis) as a function of $\epsilon$ ($x$-axis $\log_2$ scale).}
\label{Ratio_H_Call}
\end{center}
\end{figure}

Figure \ref{Ratio_H_Call}, which represents our CPU-time ratios defined as previously, emphasizes that $\hat{Y}_{ML2R}^{NV}$ is about 1.75 faster than $\hat{Y}_{MLMC}^{NV}$. $\hat{Y}_{ML2R}^{NV}$ is also 1.3 faster than $\hat{Y}_{MLMC}^{NV-GS}$ since the black dashed curve is always below 1. 
\\\\

\section{Conclusion}
In this paper, we have improved the multilevel Monte Carlo estimator of Giles and Szpruch \cite{GS} using a coupling between the Giles-Szpruch and Ninomiya-Victoir schemes at the last level of the MLMC estimator, which generalize their antithetic method. When the payoff is Lipschitz and piecewise smooth, which is very common in finance for example, the gain is amplified since $\beta = 3/2$. 
We have also highlighted a strange phenomenon: sometimes the numerical rate of convergence of the variance can be better than the theoretical one, at least for the levels used in the multilevel methods. This illustrates the presence of higher order terms which overshadows the theoretical behavior. Therefore, we emphasize that the estimation of the rate $\beta$ and its associated constant $c_2$ should be done cautiously, since the optimal parameters of the multilevel estimators depend on them.
\section{Appendix}

Let $l \in \mathbb{N}^*$ and $N = 2^{l-1}$. 
We recall that the Clark-Cameron SDE is defined as follows
\begin{equation}
  \left\{
      \begin{aligned}
        dU_t &=  S_t dW_t^1\\
       dS_t &= \mu dt +  dW_t^2 
      \end{aligned}
  \right.
\end{equation}
and the Ninomiya-Victoir is given by\\
if $\eta_{t_{k+1}} = -1$
\begin{equation}
  \left\{
      \begin{aligned}
          U^{NV,\eta}_{t_{k+1}} &= U^{NV,\eta}_{t_{k}} + S^{NV,\eta}_{t_{k}}  \left(W^1_{t_{k+1}} - W_{t_k}^1\right) + \frac12 \mu \left(t_{k+1} - t_k\right) \left(W^1_{t_{k+1}} - W^1_{t_k}\right) \\
        & \left(W^1_{t_{k+1}} - W_{t_k}^1\right) \left(W^2_{t_{k+1}} - W^2_{t_k}\right)\\
         S^{NV,\eta}_{t_{k+1}} &= S^{NV,\eta}_{t_{k}} + \mu \left(t_{k+1} - t_k\right) + \left(W^2_{t_{k+1}} - W^2_{t_k}\right)
      \end{aligned}
  \right.
\end{equation}
if $\eta_{t_{k+1}} = 1$
\begin{equation}
  \left\{
      \begin{aligned}
         U^{NV,\eta}_{t_{k+1}} &= U^{NV,\eta}_{t_{k}} + S^{NV,\eta}_{t_{k}}  \left(W^1_{t_{k+1}} - W_{t_k}^1\right) + \frac12 \mu \left(t_{k+1} - t_k\right) \left(W^1_{t_{k+1}} - W^1_{t_k}\right) \\
         S^{NV,\eta}_{t_{k+1}} &= S^{NV,\eta}_{t_{k}} + \mu \left(t_{k+1} - t_k\right) + \left(W^2_{t_{k+1}} - W^2_{t_k}\right).
      \end{aligned}
  \right.
\end{equation}
We define $\bar{U}^{NV,\eta}$ and $\bar{S}^{NV,\eta}$ as $\frac12\left( U^{NV,\eta} + U^{NV,-\eta}\right)$, $\frac12\left( S^{NV,\eta} + S^{NV,-\eta}\right)$ respectively.
\begin{arem}
$S^{NV,\eta}$ does not depend on $\eta$, so $\bar{S}^{NV,\eta} =  S^{NV,\eta}$. 
\end{arem}
The evolution of $\left(\bar{U}^{NV,2^{l-1},\eta},\bar{V}^{NV,2^{l-1},\eta}\right)$ on the coarse grid with time step $T/2^{l-1}$ is given by
\begin{equation}
  \left\{
      \begin{aligned}
        \bar{U}^{NV,2^{l-1},\eta}_{t_{k+1}} &= \bar{U}^{NV,2^{l-1},\eta}_{t_{k}} + \bar{S}^{NV,2^{l-1},\eta}_{t_{k}}  \left(W^1_{t_{k+1}} - W_{t_k}^1\right) + \frac12 \mu \left(t_{k+1} - t_k\right) \left(W^1_{t_{k+1}} - W_{t_k}^1\right) \\
&+ \frac12 \left(W^1_{t_{k+1}} - W_{t_k}^1\right) \left(W^2_{t_{k+1}} - W_{t_k}^2\right)\\
\bar{S}^{NV,2^{l-1},\eta}_{t_{k+1}} &= \bar{S}^{NV,2^{l-1},\eta}_{t_{k}} + \mu \left(t_{k+1} - t_k\right) + \left(W^2_{t_{k+1}} - W_{t_k}^2\right).
      \end{aligned}
  \right.
\end{equation}
Similarly, the evolution of $\left(\bar{U}^{NV,2^{l},\eta},\bar{V}^{NV,2^{l},\eta}\right)$ on the fine grid with time step $T/2^{l}$ is given by
\begin{equation}
  \left\{
      \begin{aligned}
         \bar{U}^{NV,2^l,\eta}_{t_{k+\frac12}} &= \bar{U}^{NV,2^l,\eta}_{t_{k}} + \bar{S}^{NV,2^l,\eta}_{t_{k}}  \left(W^1_{t_{k+\frac12}} - W_{t_k}^1\right) + \frac12 \mu \left(t_{k+\frac12} - t_k\right) \left(W^1_{t_{k+\frac12}} - W_{t_k}^1\right) \\
&+ \frac12 \left(W^1_{t_{k+\frac12}} - W_{t_k}^1\right) \left(W^2_{t_{k+\frac12}} - W_{t_k}^2\right) \\
\bar{S}^{NV,2^l,\eta}_{t_{k+\frac12}} &= \bar{S}^{NV,2^l,\eta}_{t_{k}} + \mu \left(t_{k+\frac12} - t_k\right) + \left(W^2_{t_{k+\frac12}} - W_{t_k}^2\right)
      \end{aligned}
  \right.
\end{equation}
and:
\begin{equation}
  \left\{
      \begin{aligned}
 \bar{U}^{NV,2^l,\eta}_{t_{k+1}} &= \bar{U}^{NV,2^l,\eta}_{t_{k+\frac12}} + \bar{S}^{NV,2^l,\eta}_{t_{k+\frac12}}  \left(W^1_{t_{k+1}} - W_{t_{k+\frac12}}^1\right) + \frac12 \mu \left(t_{k+1} - t_{k+\frac12}\right) \left(W^1_{t_{k+1}} - W_{t_{k+\frac12}}^1\right) \\
&+ \frac12 \left(W^1_{t_{k+1}} - W_{t_{k+\frac12}}^1\right) \left(W^2_{t_{k+1}} - W_{t_{k+\frac12}}^2\right)\\
        \bar{S}^{NV,2^l,\eta}_{t_{k+1}} &= \bar{S}^{NV,2^l,\eta}_{t_{k+\frac12}} + \mu \left(t_{k+1} - t_{k+\frac12}\right) + \left(W^2_{t_{k+1}} - W_{t_{k +\frac12}}^2\right).
      \end{aligned}
  \right.
\end{equation}
By a straightforward calculation, we get
\begin{equation}
  \left\{
      \begin{aligned}
       \bar{U}^{NV,2^l,\eta}_{t_{k+1}} &= \bar{U}^{NV,2^l,\eta}_{t_{k}} + \bar{S}^{NV,2^l,\eta}_{t_{k}}  \left(W^1_{t_{k+1}} - W_{t_k}^1\right)\\
& + \frac12  \mu \left(t_{k+1} - t_k\right) \left(W^1_{t_{k+1}} -  W^1_{t_{k+\frac12}} \right) + \frac14 \mu\left(t_{k+1} - t_k\right) \left(W^1_{t_{k+1}} -  W^1_{t_{k}} \right)  \\
&+  \left(W^1_{t_{k+1}} - W_{t_{k+\frac12}}^1\right)  \left(W^2_{t_{k+\frac12}} - W_{t_k}^2\right)  \\
&+ \frac12  \left(W^1_{t_{k+\frac12}} - W_{t_k}^1\right)\left(W^2_{t_{k+\frac12}} - W_{t_k}^2\right) + \frac12  \left(W^1_{t_{k+1}} - W_{t_{k+\frac12}}^1\right) \left(W^2_{t_{k+1}} - W_{t_{k+\frac12}}^2\right)\\
 \bar{S}^{NV,2^l,\eta}_{t_{k+1}} &= \bar{S}^{NV,2^l,\eta}_{t_{k}} + \mu \left(t_{k+1} - t_k\right) + \left(W^2_{t_{k+1}} - W_{t_k}^2\right).
      \end{aligned}
  \right.
\end{equation}
The antithetic scheme $\left(\tilde{\bar{U}}^{NV,2^l,\eta}, \tilde{\bar{V}}^{NV,2^l,\eta} \right)$ is defined by swapping $W_{t_{k+1}} - W_{t_{k+\frac12}}$ and $W_{t_{k+\frac12}} - W^1_{t_{k}} $:
\begin{equation}
  \left\{
      \begin{aligned}
        \tilde{\bar{U}}^{NV,2^l,\eta}_{t_{k+1}} &= \tilde{\bar{U}}^{NV,2^l,\eta}_{t_{k}} + \tilde{\bar{S}}^{NV,\eta,a}_{t_{k}}  \left(W^1_{t_{k+1}} - W_{t_k}^1\right)\\
& + \frac12 \mu \left(t_{k+1} - t_k\right) \left(W^1_{t_{k+\frac12}} -  W^1_{t_{k}} \right) + \frac14  \mu \left(t_{k+1} - t_k\right) \left(W^1_{t_{k+1}} -  W^1_{t_{k}} \right)  \\
&+   \left(W^1_{t_{k+\frac12}} - W_{t_k}^1\right) \left(W^2_{t_{k+1}} - W_{t_{k+\frac12}}^2\right)\\
&+ \frac12  \left(W^1_{t_{k+\frac12}} - W_{t_k}^1\right) \left(W^2_{t_{k+\frac12}} - W_{t_k}^2\right) + \frac12 \left(W^1_{t_{k+1}} - W_{t_{k+\frac12}}^1\right) \left(W^2_{t_{k+1}} - W_{t_{k+\frac12}}^2\right) \\
 \tilde{\bar{S}}^{NV,2^l,\eta}_{t_{k+1}} &= \tilde{\bar{S}}^{NV,2^l,\eta}_{t_{k}} + \mu \left(t_{k+1} - t_k\right) + \left(W^2_{t_{k+1}} - W_{t_k}^2\right).
      \end{aligned}
  \right.
\end{equation}
 Now we define:
\begin{equation*}
\bar{\bar{U}}^{NV,2^l,\eta}_{t_{k}} := \frac12 \left(\bar{U}^{NV,2^l,\eta}_{t_{k+1}}+ \tilde{\bar{U}}^{NV,2^l,\eta}_{t_{k+1}} \right)
\end{equation*}
and
\begin{equation*}
\bar{\bar{S}}^{NV,2^l,\eta}_{t_{k}} := \frac12 \left(\bar{S}^{NV,2^l,\eta}_{t_{k+1}}+ \tilde{\bar{S}}^{NV,2^l,\eta}_{t_{k+1}} \right).
\end{equation*}
Performing a straightforward calculation, we obtain
\begin{equation}
  \left\{
      \begin{aligned}
       \bar{\bar{U}}^{NV,2^l,\eta}_{t_{k+1}} &= \bar{\bar{U}}^{NV,2^l,\eta}_{t_{k}} + \bar{\bar{S}}^{NV,2^l,\eta}_{t_{k}}  \left(W^1_{t_{k+1}} - W_{t_k}^1\right) + \frac12 \mu \left(t_{k+1} - t_k\right) \left(W^1_{t_{k+1}} - W_{t_k}^1\right) \\
&+ \frac12  \left(W^1_{t_{k+1}} - W_{t_k}^1\right) \left(W^2_{t_{k+1}} - W_{t_k}^2\right) \\
\bar{\bar{S}}^{NV,2^l,\eta}_{t_{k+1}} &= \bar{\bar{S}}^{NV,2^l,\eta}_{t_{k}} + \mu \left(t_{k+1} - t_k\right) + \left(W^2_{t_{k+1}} - W_{t_k}^2\right).
      \end{aligned}
  \right.
\end{equation}
Then, using forward induction on $k$, we easily get that $\forall k \in \left\{0,\ldots,2^{l-1}-1\right\}$
\begin{equation}
\label{FWD}
  \left\{
      \begin{aligned}
       \bar{\bar{U}}^{NV,2^l,\eta}_{t_{k+1}} - \bar{U}^{NV,2^{l-1},\eta}_{t_{k+1}} &= 0  \\
       \bar{\bar{S}}^{NV,2^l,\eta}_{t_{k+1}} - \bar{S}^{NV,2^{l-1},\eta}_{t_{k+1}} &= 0 .
      \end{aligned}
  \right.
\end{equation}
\textbf{We want to calculate}
\begin{equation}
\begin{split}
Y &= \mathbb{E}\left[ \left(Z_{NV}^{l}\right)^2\right]
\end{split}
\end{equation}
where 
\begin{equation}
\begin{split}
Z_{NV}^{l} &= 
\frac14 \left(\left(U^{NV,2^l,\eta}_{T}\right)^2 + \left(\tilde{U}^{NV,2^l,\eta}_{T}\right)^2 + \left(U^{NV,2^l,-\eta}_{T}\right)^2 + \left(\tilde{U}^{NV,2^l,-\eta}_{T}\right)^2\right) \\
& - \frac12 \left(\left(U^{NV,2^{l-1},\eta}_{T}\right)^2 + \left(U^{NV,2^{l-1},-\eta}_{T}\right)^2\right).
\end{split}
\end{equation}
Using
\begin{equation}
\begin{split}
\frac14 \left(x^2 + y^2 + u^2 + v ^2 \right)  -\frac12 \left(z^2 + w^2\right) & = \left(\frac14 \left(x + y + u + v \right)\right)^2  -\left(\frac12 \left(z + w\right)\right)^2\\
& + \frac1{64}  \left(3x - y - u - v\right)^2 + \frac1{64} \left(3y - x - u - v\right)^2 \\ 
& +\frac1{64}  \left(3u - x - y - v\right)^2 + \frac1{64} \left(3v - x - y - u\right)^2   \\
& -  \frac14 \left(z -w\right)^2 
\end{split}
\end{equation}
 and \eqref{FWD}, we get
\begin{equation}
\begin{split}
Y = \frac1{16}\mathbb{E}\left[Z^2\right] 
\end{split}
\end{equation}
where
\begin{equation}
\begin{split}
Z &= \frac1{16}\left(3 U^{NV,2^l,\eta}_T - \tilde{U}^{NV,2^l,\eta}_T - U^{NV,2^l,-\eta}_T  - \tilde{U}^{NV,2^l,-\eta}_T  \right)^2 +  \frac1{16} \left(3 \tilde{U}^{NV,2^l,\eta}_T - U^{NV,2^l,\eta}_T - U^{NV,2^l,-\eta}_T  - \tilde{U}^{NV,2^l,-\eta}_T  \right)^2\\ 
& + \frac1{16} \left(3 U^{NV,2^l,-\eta}_T - U^{NV,2^l,\eta}_T - \tilde{U}^{NV,2^l,\eta}_T  - \tilde{U}^{NV,2^l,-\eta}_T  \right)^2 + \frac1{16} \left(3 \tilde{U}^{NV,2^l,-\eta}_T - U^{NV,2^l,\eta}_T - U^{NV,2^l,-\eta}_T  - \tilde{U}^{NV,2^l,\eta}_T  \right)^2\\ 
& - \left(U^{NV,2^{l-1},\eta}_T - U^{NV,2^{l-1},-\eta}_T \right)^2.
\end{split}
\end{equation}
To lighten the previous expression, we introduce
\begin{equation}
\begin{split}
Z_1 &= \frac14 \left(3 U^{NV,2^l,\eta}_T - \tilde{U}^{NV,2^l,\eta}_T - U^{NV,2^l,-\eta}_T  - \tilde{U}^{NV,2^l,-\eta}_T\right), 
\end{split}
\end{equation}
\begin{equation}
\begin{split}
Z_2 &= \frac14 \left(3 \tilde{U}^{NV,2^l,\eta}_T - U^{NV,2^l,\eta}_T - U^{NV,2^l,-\eta}_T  - \tilde{U}^{NV,2^l,-\eta}_T\right), 
\end{split}
\end{equation}
\begin{equation}
\begin{split}
Z_3 &= \frac14 \left(3 U^{NV,2^l,-\eta}_T - U^{NV,2^l,\eta}_T - \tilde{U}^{NV,2^l,\eta}_T  - \tilde{U}^{NV,2^l,-\eta}_T\right), 
\end{split}
\end{equation}
\begin{equation}
\begin{split}
Z_4 &= \frac14 \left(3 \tilde{U}^{NV,2^l,-\eta}_T - U^{NV,2^l,\eta}_T - U^{NV,2^l,-\eta}_T  - \tilde{U}^{NV,2^l,\eta}_T\right), 
\end{split}
\end{equation}
and
\begin{equation}
Z_0 = U^{NV,2^{l-1},\eta}_T - U^{NV,2^{l-1},-\eta}_T. 
\end{equation}
In order to get an explicit expression for $Z_i, i \in \left\{0,\ldots,4\right\}$, we compute the following differences:
\begin{equation}
     \begin{split}
       U^{NV,2^l,\eta}_{t_{k+1}} - U^{NV,2^l,-\eta}_{t_{k+1}} &= U^{NV,2^l,\eta}_{t_{k}} - U^{NV,2^l,-\eta}_{t_{k}} -\eta_{k+\frac12} \left(W^1_{t_{k+\frac12}} - W^1_{t_{k}}\right) \left(W^2_{t_{k+\frac12}} - W^2_{t_{k}}\right)\\
&  - \eta_{k+1} \left(W^1_{t_{k+1}} - W^1_{t_{k+\frac12}}\right)  \left(W^2_{t_{k+1}} - W^2_{t_{k+\frac12}}\right) 
      \end{split}
\end{equation}
\begin{equation}
     \begin{split}
       \tilde{U}^{NV,2^l,\eta}_{t_{k+1}} - \tilde{U}^{NV,2^l,-\eta}_{t_{k+1}} &= \tilde{U}^{NV,2^l,\eta}_{t_{k}} - \tilde{U}^{NV,2^l,-\eta}_{t_{k}} -\eta_{k+1} \left(W^1_{t_{k+\frac12}} - W^1_{t_{k}}\right) \left(W^2_{t_{k+\frac12}} - W^2_{t_{k}}\right)\\
&  - \eta_{k+\frac12} \left(W^1_{t_{k+1}} - W^1_{t_{k+\frac12}}\right)  \left(W^2_{t_{k+1}} - W^2_{t_{k+\frac12}}\right) 
      \end{split}
\end{equation}
\begin{equation}
     \begin{split}
       U^{NV,2^l,\eta}_{t_{k+1}} - \tilde{U}^{NV,2^l,\eta}_{t_{k+1}} &= U^{NV,2^l,\eta}_{t_{k}} - \tilde{U}^{NV,2^l,\eta}_{t_{k}} +\frac{\mu}2 \frac{T}N \left(W^1_{t_{k+1}} -2W^1_{t_{k+\frac12}} + W^1_{t_{k}}\right) \\
& +   \left(W^1_{t_{k+1}} - W^1_{t_{k+\frac12}}\right) \left(W^2_{t_{k+\frac12}} - W^2_{t_{k}}\right)   - \left(W^1_{t_{k+\frac12}} - W^1_{t_{k}}\right) \left(W^2_{t_{k+1}} - W^2_{t_{k+\frac12}}\right)  \\ 
& +\frac12 \left( \eta_{k+1}  -\eta_{k+\frac12}\right) \left(W^1_{t_{k+\frac12}} - W^1_{t_{k}}\right) \left(W^2_{t_{k+\frac12}} - W^2_{t_{k}}\right)\\
& +\frac12 \left( \eta_{k+\frac12} -\eta_{k+1} \right) \left(W^1_{t_{k+1}} - W^1_{t_{k+\frac12}}\right)  \left(W^2_{t_{k+1}} - W^2_{t_{k+\frac12}}\right)
      \end{split}
\end{equation}
\begin{equation}
     \begin{split}
       U^{NV,2^l,-\eta}_{t_{k+1}} - \tilde{U}^{NV,2^l,-\eta}_{t_{k+1}} &= U^{NV,2^l,-\eta}_{t_{k}} - \tilde{U}^{NV,2^l,-\eta}_{t_{k}} +\frac{\mu}2 \frac{T}N \left(W^1_{t_{k+1}} -2W^1_{t_{k+\frac12}} + W^1_{t_{k}}\right) \\
& +   \left(W^1_{t_{k+1}} - W^1_{t_{k+\frac12}}\right)  \left(W^2_{t_{k+\frac12}} - W^2_{t_{k}}\right)  -  \left(W^1_{t_{k+\frac12}} - W^1_{t_{k}}\right) \left(W^2_{t_{k+1}} - W^2_{t_{k+\frac12}}\right)  \\ 
& +\frac12 \left( \eta_{k+\frac12} - \eta_{k+1}  \right) \left(W^1_{t_{k+\frac12}} - W^1_{t_{k} }\right) \left(W^2_{t_{k+\frac12}} - W^2_{t_{k}}\right)\\
& +\frac12 \left( \eta_{k+1} - \eta_{k+\frac12}  \right)  \left(W^1_{t_{k+1}} - W^1_{t_{k+\frac12}}\right)  \left(W^2_{t_{k+1}} - W^2_{t_{k+\frac12}}\right)
      \end{split}
\end{equation}
\begin{equation}
     \begin{split}
       U^{NV,2^l,\eta}_{t_{k+1}} - \tilde{U}^{NV,2^l,-\eta}_{t_{k+1}} &= U^{NV,2^l,\eta}_{t_{k}} - \tilde{U}^{NV,2^l,-\eta}_{t_{k}} +\frac{\mu}2 \frac{T}N \left(W^1_{t_{k+1}} -2W^1_{t_{k+\frac12}} + W^1_{t_{k}}\right) \\
& +  \left(W^1_{t_{k+1}} - W^1_{t_{k+\frac12}}\right) \left(W^2_{t_{k+\frac12}} - W^2_{t_{k}}\right)    -   \left(W^1_{t_{k+\frac12}} - W^1_{t_{k}}\right)\left(W^2_{t_{k+1}} - W^2_{t_{k+\frac12}}\right) \\ 
      \end{split}
\end{equation}
and
\begin{equation}
     \begin{split}
       U^{NV,2^l,-\eta}_{t_{k+1}} - \tilde{U}^{NV,2^l,\eta}_{t_{k+1}} &= U^{NV,2^l,-\eta}_{t_{k}} - \tilde{U}^{NV,2^l,\eta}_{t_{k}} +\frac{\mu}2 \frac{T}N \left(W^1_{t_{k+1}} -2W^1_{t_{k+\frac12}} + W^1_{t_{k}}\right) \\
& +   \left(W^1_{t_{k+1}} - W^1_{t_{k+\frac12}}\right) \left(W^2_{t_{k+\frac12}} - W^2_{t_{k}}\right)  - \left(W^1_{t_{k+\frac12}} - W^1_{t_{k}}\right) \left(W^2_{t_{k+1}} - W^2_{t_{k+\frac12}}\right)   .
      \end{split}
\end{equation}
Hence, we obtain by summation
\begin{equation}
\begin{split}
Z_i = \sum \limits_{k=0}^{N-1} z_k^i 
\end{split}
\end{equation}
where
\begin{equation}
z_k^0 = \left(W^1_{t_{k+1}} - W^1_{t_k}\right) \left(W^2_{t_{k+1}} - W^2_{t_k}\right),
\end{equation}
\begin{equation}
\begin{split}
z^1_k &=  \left(\frac12 -  \frac18 \left(\eta_{k+\frac12 } - \eta_{k+1} \right) \right) \left(W^1_{t_{k+1}} - W^1_{t_{k+\frac12}}\right) \left(W^2_{t_{k+\frac12}} - W^2_{t_{k}}\right)  \\
& - \left(\frac12 + \frac18 \left(\eta_{k+\frac12 } - \eta_{k+1} \right) \right) \left(W^1_{t_{k+\frac12}} - W^1_{t_{k}}\right) \left(W^2_{t_{k+1}} - W^2_{t_{k+\frac12}}\right)  \\
& + \mu \frac{T}{4N} \left( W^1_{t_{k+1}} - 2 W^1_{t_{k+\frac12}} +  W^1_{t_{k}}\right) -  \frac14  \eta_{k+\frac12 } \left(W^1_{t_{k+\frac12}} - W^1_{t_{k}}\right)  \left(W^2_{t_{k+\frac12}} - W^2_{t_{k}} \right) \\
& -   \frac14 \eta_{k+1}\left( W^1_{t_{k+1}} - W^1_{t_{k+\frac12}}\right)  \left(W^2_{t_{k+1}} -  W^2_{t_{k+\frac12}} \right),
\end{split}
\end{equation}
\begin{equation}
\begin{split}
z^2_k &= - \left(\frac12 + \frac18 \left(\eta_{k+\frac12 } - \eta_{k+1} \right) \right) \left(W^1_{t_{k+1}} - W^1_{t_{k+\frac12}}\right) \left(W^2_{t_{k+\frac12}} - W^2_{t_{k}}\right) \\
& + \left(\frac12 - \frac18 \left(\eta_{k+\frac12 } - \eta_{k+1} \right) \right)\left(W^1_{t_{k+\frac12}} - W^1_{t_{k}}\right) \left(W^2_{t_{k+1}} - W^2_{t_{k+\frac12}}\right) \\
& - \mu \frac{T}{4N} \left( W^1_{t_{k+1}} - 2 W^1_{t_{k+\frac12}} +  W^1_{t_{k}}\right) -    \frac14  \eta_{k+1}\left(W^1_{t_{k+\frac12}} - W^1_{t_{k}}\right)  \left(W^2_{t_{k+\frac12}} - W^2_{t_{k}} \right)  \\
& -\frac14 \eta_{k+\frac12 } \left( W^1_{t_{k+1}} - W^1_{t_{k+\frac12}}\right)  \left(W^2_{t_{k+1}} -  W^2_{t_{k+\frac12}} \right),
\end{split}
\end{equation}
\begin{equation}
\begin{split}
z^3_k &=  \left(\frac12+ \frac18 \left(\eta_{k+\frac12 } - \eta_{k+1} \right) \right) \left(W^1_{t_{k+1}} - W^1_{t_{k+\frac12}}\right) \left(W^2_{t_{k+\frac12}} - W^2_{t_{k}}\right)\\
& - \left(\frac12- \frac18 \left(\eta_{k+\frac12 } - \eta_{k+1} \right) \right) \left(W^1_{t_{k+\frac12}} - W^1_{t_{k}}\right) \left(W^2_{t_{k+1}} - W^2_{t_{k+\frac12}}\right)  \\
& + \mu \frac{T}{4N}\left( W^1_{t_{k+1}} - 2 W^1_{t_{k+\frac12}} +  W^1_{t_{k}}\right) +  \frac14  \eta_{k+\frac12 } \left(W^1_{t_{k+\frac12}} - W^1_{t_{k}}\right)  \left(W^2_{t_{k+\frac12}} - W^2_{t_{k}} \right) \\
& + \frac14   \eta_{k+1}\left( W^1_{t_{k+1}} - W^1_{t_{k+\frac12}}\right)  \left(W^2_{t_{k+1}} -  W^2_{t_{k+\frac12}} \right),
\end{split}
\end{equation}
and
\begin{equation}
\begin{split}
z^4_k &=   - \left(\frac12 - \frac18 \left(\eta_{k+\frac12 } - \eta_{k+1} \right) \right) \left(W^1_{t_{k+1}} - W^1_{t_{k+\frac12}}\right) \left(W^2_{t_{k+\frac12}} - W^2_{t_{k}}\right)\\
& \left(\frac12 + \frac18 \left(\eta_{k+\frac12 } - \eta_{k+1} \right) \right)\left(W^1_{t_{k+\frac12}} - W^1_{t_{k}}\right) \left(W^2_{t_{k+1}} - W^2_{t_{k+\frac12}}\right)\\
& - \mu \frac{T}{4N} \left( W^1_{t_{k+1}} - 2 W^1_{t_{k+\frac12}} +  W^1_{t_{k}}\right)  + \frac14 \eta_{k+1}\left(W^1_{t_{k+\frac12}} - W^1_{t_{k}}\right)  \left(W^2_{t_{k+\frac12}} - W^2_{t_{k}} \right)
\\
&  + \frac14   \eta_{k+\frac12 } \left( W^1_{t_{k+1}} - W^1_{t_{k+\frac12}}\right)  \left(W^2_{t_{k+1}} -  W^2_{t_{k+\frac12}} \right).
\end{split}
\end{equation}
Then, one can express $Z^2$ as
\begin{equation}
\begin{split}
Z^2 & = \left(\left(\sum \limits_{k=0}^{N-1} z^1_k\right)^2 + \left(\sum \limits_{k=0}^{N-1} z^2_k\right)^2  + \left(\sum \limits_{k=0}^{N-1} z^3_k\right)^2  + \left(\sum \limits_{k=0}^{N-1} z^4_k\right)^2 - \left(\sum \limits_{k=0}^{N-1} z^0_k\right)^2 \right)^2\\
& = \Bigg(\sum \limits_{k=0}^{N-1} \left(z^1_k\right)^2 + \left( z^2_k\right)^2  + \left(z^3_k\right)^2  + \left(z^4_k\right)^2 - \left( z^0_k\right)^2 + 2 \sum \limits_{(k,l) \in \Delta_N } z^k_1z^l_1 + z^k_2z^l_2 + z^k_3z^l_3 + z^k_4z^l_4 - z^k_0 z^l_0  \Bigg)^2
\end{split}
\end{equation}
where $\Delta_N = \left\{(k,l) \in \left\{0,\ldots,N-1\right\}^2, k < l\right\}$. \\\\
\textbf{Preliminary calculus:}\\

We begin by writing $z^i_k$ in generic form:
\begin{equation}
\begin{split}
z^i_k & = \alpha^i_k \left(W^1_{t_{k+1}} - W^1_{t_{k+\frac12}}\right) \left(W^2_{t_{k+\frac12}} - W^2_{t_{k}}\right)+ \beta^i_k \left(W^1_{t_{k+\frac12}} - W^1_{t_{k}}\right) \left(W^2_{t_{k+1}} - W^2_{t_{k+\frac12}}\right)\\
&+ \gamma^i_k  \left( W^1_{t_{k+1}} - 2 W^1_{t_{k+\frac12}} +  W^1_{t_{k}}\right)  + \delta^i_k \left(W^1_{t_{k+\frac12}} - W^1_{t_{k}}\right)  \left(W^2_{t_{k+\frac12}} - W^2_{t_{k}} \right)\\
& + \omega^i_k \left( W^1_{t_{k+1}} - W^1_{t_{k+\frac12}}\right)  \left(W^2_{t_{k+1}} -  W^2_{t_{k+\frac12}} \right)
\end{split}
\end{equation}
where
\begin{center}
\begin{tabular}{|l|c|c|c|c|}
  \hline
  $\alpha_k^1 = \left(\frac12- \frac18 \left(\eta_{k+\frac12 } - \eta_{k+1} \right) \right) $ & $\alpha_k^2 = \beta_k^1$ & $\alpha_k^3 = -\beta_k^1 $ & $\alpha_k^4 = - \alpha_k^1$ & $\alpha_k^0 = 1$ \\
  
  $\beta_k^1 = - \left(\frac12+ \frac18 \left(\eta_{k+\frac12 } - \eta_{k+1} \right) \right) $ & $\beta_k^2 = \alpha_k^1 $ & $\beta_k^3 = -\alpha_k^1 $ & $\beta_k^4 =- \beta^1_k $ & $\beta_k^0 = 1$ \\
  
  $\gamma_k^1 = \mu \frac{T}{4N} $ & $\gamma_k^2 = -\gamma_k^1 $ & $\gamma_k^3 = \gamma_k^1 $ & $\gamma_k^4 =-\gamma_k^1  $ & $\gamma_k^0 = 0$ \\
  
  $\delta_k^1 = -\frac14  \eta_{k+\frac12 } $ & $\delta_k^2 =\omega_k^1 $ & $\delta_k^3 = -\delta_k^1 $ & $\delta_k^4 =-\omega_k^1 $ & $\delta_k^0 = 1$ \\
  
  $\omega_k^1 = -\frac14  \eta_{k+ 1 }  $ & $\omega_k^2 = \delta_k^1 $ & $\omega_k^3 = -\omega_k^1$ & $\omega_k^4 = -\delta_k^1 $ & $\omega_k^0 = 1$ \\
  \hline
\end{tabular}
.\end{center}

First, let us look at the expectation of $z^i_k z^j_k$:
\begin{equation}
\begin{split}
\mathbb{E}\left[z^i_k z^j_k \right] = \frac{T^2}{4N^2} \mathbb{E}\left[\alpha^i_k \alpha^j_k + \beta^i_k \beta^j_k + \delta^i_k \delta^j_k + \omega^i_k \omega^j_k\right] + \frac{T}{N}\mathbb{E}\left[\gamma^i_k \gamma^j_k\right] .
\end{split}
\end{equation}
Then, for $ i = j = 0$
\begin{equation}
\begin{split}
\mathbb{E}\left[\left(z^0_k\right)^2  \right] = \frac{T^2}{N^2}.
\end{split}
\end{equation}
For $ i = j \in \left\{1,\ldots,4\right\}$
\begin{equation}
\begin{split}
\mathbb{E}\left[\left(z^i_k\right)^2  \right] = \frac{11T^2}{64N^2} + \frac{\mu^2 T^3}{16N^3}.
\end{split}
\end{equation}
For $j = 0$ and $i \in \left\{1,\ldots,4\right\}$
\begin{equation}
\begin{split}
\mathbb{E}\left[z^i_k z^0_k \right] = 0.
\end{split}
\end{equation}
For $(i,j) \in \left\{ (1,2), (3,4) \right\}$ 
\begin{equation}
\begin{split}
\mathbb{E}\left[z^i_k z^j_k \right] = -\frac{7T^2}{64N^2} - \frac{\mu^2 T^3}{16N^3}.
\end{split}
\end{equation}
For $(i,j) \in \left\{ (1,3), (2,4) \right\}$ 
\begin{equation}
\begin{split}
\mathbb{E}\left[z^i_k z^j_k \right] = \frac{5T^2}{64N^2} + \frac{\mu^2 T^3}{16N^3}.
\end{split}
\end{equation}
For $(i,j) \in \left\{ (1,4), (2,3) \right\}$ 
\begin{equation}
\begin{split}
\mathbb{E}\left[z^i_k z^j_k \right] = -\frac{9T^2}{64N^2} - \frac{\mu^2 T^3}{16N^3}.
\end{split}
\end{equation}
Now we look at the expectation of  $\left(z^i_k\right)^2 \left(z^j_k\right)^2$:
\begin{equation}
\begin{split}
\mathbb{E}\left[\left(z^i_k\right)^2 \left(z^j_k\right)^2 \right] &= \frac{9T^4}{16N^4} \mathbb{E}\left[\left(\alpha^i_k\right)^2 \left(\alpha^j_k\right)^2 + \left(\beta^i_k\right)^2 \left(\beta^j_k\right)^2 + \left(\delta^i_k\right)^2 \left(\delta^j_k\right)^2 + \left(\omega^i_k\right)^2 \left(\omega^j_k\right)^2 \right]  \\
&+ \frac{T^4}{16N^4} \Bigg( \mathbb{E} \left[ \left(\alpha^i_k\right)^2 \left(\beta^j_k\right)^2 + \left(\alpha^j_k\right)^2 \left(\beta^i_k\right)^2 + \left(\delta^i_k\right)^2 \left(\omega^j_k\right)^2 + \left(\delta^j_k\right)^2 \left(\omega^i_k\right)^2 \right] \\
& \hspace{1.3cm} + 4 \mathbb{E}\left[ \left(\alpha_k^i \beta_k^i +  \delta_k^i \omega_k^i\right) \left(\alpha_k^j \beta_k^j +  \delta_k^j \omega_k^j \right) + \left(\alpha_k^i \beta_k^j +  \alpha_k^j \beta_k^i\right) \left(\delta_k^i \omega_k^j +  \delta_k^j \omega_k^i \right) \right] \Bigg)\\
& + \frac{3T^4}{16N^4} \Bigg( \mathbb{E}\left[ \left(\left(\alpha^i_k\right)^2 + \left(\beta^i_k\right)^2 \right) \left(\left(\delta^j_k\right)^2 + \left(\omega^j_k\right)^2 \right) + \left(\left(\alpha^j_k\right)^2 + \left(\beta^j_k\right)^2 \right) \left(\left(\delta^i_k\right)^2 + \left(\omega^i_k\right)^2 \right) \right]\\
& \hspace{1.3cm} + 4 \mathbb{E}\left[ \left(\alpha_k^i \alpha_k^j + \beta_k^i \beta_k^j\right) \left(\delta_k^i \delta_k^j + \omega_k^i \omega_k^j\right) \right] \Bigg)\\
& + \frac{T^3}{8N^3}\Bigg(  \mathbb{E}\left[ \left(\gamma_k^i\right)^2 \left(\left(\alpha_k^j\right)^2 + \left(\beta_k^j\right)^2 +\left(\delta_k^j\right)^2 + \left(\omega_k^j\right)^2 -4 \alpha_k^j \delta_k^j - 4\beta_k^j\omega_k^j\right) \right]\\
& \hspace{1.3cm} +   \mathbb{E}\left[ \left(\gamma_k^j\right)^2 \left(\left(\alpha_k^i\right)^2 + \left(\beta_k^i\right)^2 +\left(\delta_k^i\right)^2 + \left(\omega_k^i\right)^2 -4 \alpha_k^i \delta_k^i - 4\beta_k^i\omega_k^i\right) \right]\\
& \hspace{1.3cm} +   4\mathbb{E}\left[ \gamma_k^i \gamma_k^j \left(\alpha_k^i \alpha_k^j+ \beta_k^i \beta_k^j + \delta_k^i \delta_k^j + \omega_k^i \omega_k^j - 2\alpha_k^i\delta_k^j - 2\alpha_k^j\delta_k^i - 2\beta_k^i\omega_k^j - 2\beta_k^j\omega_k^i \right) \right]\Bigg)\\
& + \frac{3T^3}{8N^3}\Bigg(\mathbb{E}\left[ \left(\gamma_k^i\right)^2 \left(\left(\alpha_k^j\right)^2 + \left(\beta_k^j\right)^2 +\left(\delta_k^j\right)^2 + \left(\omega_k^j\right)^2 \right) \right]\\
& \hspace{1.3cm} +   \mathbb{E}\left[ \left(\gamma_k^j\right)^2 \left(\left(\alpha_k^i\right)^2 + \left(\beta_k^i\right)^2 +\left(\delta_k^i\right)^2 + \left(\omega_k^i\right)^2 \right) \right]\\
& \hspace{1.3cm} +   4\mathbb{E}\left[ \gamma_k^i \gamma_k^j \left(\alpha_k^i \alpha_k^j+ \beta_k^i \beta_k^j + \delta_k^i \delta_k^j + \omega_k^i \omega_k^j  \right) \right]\Bigg)\\
&+ \frac{3T^2}{N^2}\mathbb{E}\left[\left(\gamma^i_k\right)^2\left( \gamma^j_k\right)^2\right].
\end{split}
\end{equation}
Then, by straightforward calculation, for $ i = j = 0$
\begin{equation}
\begin{split}
\mathbb{E}\left[\left(z^0_k\right)^4  \right] = \frac{9T^4}{N^4}.
\end{split}
\end{equation}
For $ i = j \in \left\{1,\ldots,4\right\}$
\begin{equation}
\begin{split}
\mathbb{E}\left[\left(z^i_k\right)^4  \right] = \frac{231T^4}{1024N^4} + \frac{33\mu^2 T^5}{256N^5} + \frac{3\mu^4T^6}{256N^6}.
\end{split}
\end{equation}
For $j = 0$ and $i \in \left\{1,\ldots,4\right\}$
\begin{equation}
\mathbb{E}\left[\left(z^i_k\right)^2 \left(z^0_k\right)^2 \right] = \frac{37T^4}{64N^4} + \frac{\mu^2T^5}{16N^5}.
\end{equation}
For $(i,j) \in \left\{ (1,2), (3,4) \right\}$ 
\begin{equation}
\begin{split}
\mathbb{E}\left[\left(z^i_k\right)^2 \left(z^j_k\right)^2 \right] = \frac{127T^4}{1024N^4} + \frac{25\mu^2 T^5}{256N^5}+ \frac{3\mu^4 T^6}{256N^6}.
\end{split}
\end{equation}
For $(i,j) \in \left\{ (1,3), (2,4) \right\}$ 
\begin{equation}
\begin{split}
\mathbb{E}\left[\left(z^i_k\right)^2 \left(z^j_k\right)^2 \right] = \frac{71T^4}{1024N^4} + \frac{21\mu^2T^5}{256N^5}  + \frac{3\mu^4T^6}{256N^6}.
\end{split}
\end{equation}
For $(i,j) \in \left\{ (1,4), (2,3) \right\}$ 
\begin{equation}
\begin{split}
\mathbb{E}\left[z^i_k z^j_k \right] = \frac{159T^4}{1024N^4} + \frac{29\mu^2 T^5}{256N^5} + \frac{3\mu^4T^6}{256N^6}.
\end{split}
\end{equation}
Now, we define:
\begin{equation}
a_k = \left(z^1_k\right)^2 + \left( z^2_k\right)^2  + \left(z^3_k\right)^2  + \left(z^4_k\right)^2 - \left( z^0_k\right)^2
\end{equation}
and
\begin{equation}
b_{kl} = z^1_k z^1_l + z^2_k z^2_l + z^3_k z^3_l + z^4_k z^4_l - z^0_k z^0_l.
\end{equation}
Then $Z^2$ can be expressed as
\begin{equation}
\begin{split}
Z^2 & =\sum \limits_{k=0}^{N-1} a_k^2 + 2 \sum \limits_{(k, l) \in \Delta_N} a_k a_l + 4 \sum \limits_{(k, l),(i,j) \in \Delta_N} b_{kl} b_{ij} +4 \sum \limits_{(k, l) \in \Delta_N, j \in [\![0;N-1]\!]} a_j b_{kl}. 
\end{split}
\end{equation}
Observing that
\begin{equation}
\mathbb{E}\left[a_j b_{kl}\right] = 0
\end{equation}
\begin{equation}
\mathbb{E}\left[\sum \limits_{(k, l),(i,j) \in \Delta_N} b_{kl} b_{ij} \right] = \mathbb{E}\left[\sum \limits_{(k, l) \in \Delta_N} b_{kl}^2\right]
\end{equation}
we obtain
\begin{equation}
\begin{split}
\mathbb{E}\left[Z^2\right] & =\sum \limits_{k=0}^{N-1} \mathbb{E}\left[ a_k^2 \right]  + 2 \sum \limits_{(k, l) \in \Delta_N} \mathbb{E}\left[a_k a_l\right] + 4 \sum \limits_{(k, l) \in \Delta_N} \mathbb{E}\left[b_{kl}^2  \right].
\end{split}
\end{equation}
For $k\neq l$, by independence
\begin{equation}
\begin{split}
\mathbb{E}\left[a_k a_l\right] &= \mathbb{E}\left[a_k \right]\mathbb{E}\left[a_l\right] =  \left(\mathbb{E}\left[a_k\right]\right)^2
\end{split}
\end{equation}
this leads to
\begin{equation}
\begin{split}
\mathbb{E}\left[Z^2\right] & = N\mathbb{E}\left[ a_k^2 \right]  +  N\left(N-1\right) \left( \mathbb{E}\left[a_k\right] \right)^2 + 2 N \left(N-1\right) \mathbb{E}\left[b_{kl}^2  \right].
\end{split}
\end{equation}
To achieve our goal, we calculate the following expectations:
\begin{equation}
\mathbb{E}\left[a_k\right]  = 4 \mathbb{E}\left[\left(z^1_k\right)^2\right] - \mathbb{E}\left[\left(z^0_k\right)^2\right]  =   \frac{\mu^2 T^3}{4N^3} - \frac{5T^2}{16N^2} 
\end{equation}
\begin{equation}
\begin{split}
\mathbb{E}\left[a_k^2\right]  &=  4 \mathbb{E}\left[\left(z^1_k\right)^4\right] +  \mathbb{E}\left[\left(z^0_k\right)^4\right] - 8\mathbb{E}\left[\left(z^1_k\right)^2\left(z^0_k\right)^2\right]  + 4 \left( \mathbb{E}\left[\left(z^1_k\right)^2\left(z^2_k\right)^2\right] + \mathbb{E}\left[\left(z^1_k\right)^2\left(z^3_k\right)^2\right] + \mathbb{E}\left[\left(z^1_k\right)^2\left(z^4_k\right)^2\right]\right) \\
&= \frac{3 \mu^4 T^6}{16 N^6}+\frac{19 \mu^2 T^5}{16 N^5}+\frac{427 T^4}{64 N^4}
\end{split}
\end{equation}
\begin{equation}
\begin{split}
\mathbb{E}\left[b_{kl}^2\right]  &=  4 \left(\mathbb{E}\left[\left(z^1_k\right)^2\right]\right)^2 + \left(\mathbb{E}\left[\left(z^0_k\right)^2\right]\right)^2 - 8\left(\mathbb{E}\left[z^1_k z^0_k\right]\right)^2  + 4 \left( \left(\mathbb{E}\left[z^1_k z^2_k\right]\right)^2 + \left(\mathbb{E}\left[z^1_k z^3_k\right]\right)^2 + \left(\mathbb{E}\left[z^1_k z^4_k\right]\right)^2\right) \\
&= \frac{\mu^4 T^6}{16 N^6}+\frac{\mu^2 T^5}{4 N^5}+\frac{69 T^4}{256 N^4}.
\end{split}
\end{equation}
Combining our results, we obtain
\begin{equation}
\begin{split}
\mathbb{E}\left[Z^2\right]  &= \frac{1}{N^4} \left( \frac{3}{16} \mu^4 T^6 + \frac{9}{16} \mu^2 T^5\right) + \frac{1}{N^3} \left(\frac{11}{32} \mu^2 T^5 + \frac{1545}{256}T^4 \right) + \frac{1}{N^2} \left(\frac{163}{256}T^4 \right). 
\end{split}
\end{equation}
Then replacing $N$ by $2^{l-1}$
\begin{equation}
\begin{split}
\mathbb{E}\left[Z^2\right]  &= 2^{-4 l } \left( 3 \mu^4 T^6 + 9 \mu^2 T^5\right) + 2^{-3l} \left(\frac{11}{4} \mu^2 T^5 + \frac{1545}{32}T^4 \right) + 2^{-2l} \left(\frac{163}{64}T^4 \right).
\end{split}
\end{equation}
Dividing by $16$, we get the desired result
\begin{equation}
Y =  2^{-4 l } \left( \frac{3}{16} \mu^4 T^6 + \frac{9}{16} \mu^2 T^5\right) + 2^{-3l} \left(\frac{11}{64} \mu^2 T^5 + \frac{1545}{512}T^4 \right) + 2^{-2l} \left(\frac{163}{1024}T^4 \right). 
\end{equation}

\end{document}